\documentclass[aps,prx,article,twocolumn,floatfix, superscriptaddress,longbibliography]{revtex4-2}
\usepackage[english]{babel}
\usepackage{braket}
\usepackage{commath}
\usepackage{amsmath}
\usepackage{amsfonts}
\usepackage{graphicx}
\usepackage{dcolumn} 
\usepackage{bm}        
\usepackage{amssymb} 
\usepackage{mathrsfs}
\usepackage{epigraph}
\usepackage{braket}
\usepackage{gensymb}
\usepackage{lmodern}
\usepackage{ tipa }
\usepackage{bbold}
\usepackage{esint}
\usepackage{mathdots}
\usepackage{xcolor}
\usepackage{appendix}
\usepackage{multirow,booktabs}
\usepackage{makecell}
\usepackage{float}
\usepackage{comment}

\usepackage[colorlinks=true, linkcolor=blue, urlcolor=blue, citecolor=blue]{hyperref}
\newcommand{\kbf}{\mathbf{k}}
\newcommand{\qbf}{\mathbf{q}}

\newcommand{\rbf}      {\mathbf{r}}

\newcommand{\dbf}      {\mathbf{d}}

\newcommand{\eps} {{\bm{\varepsilon}}}
\newcommand{\kin} {{\mathbf{q}_{\rm i}}}
\newcommand{\kout} {{\mathbf{q}_{\rm s}}}
\newcommand{\dq} {{\mathbf{q}}}
\newcommand{\epsin} {{\eps_{\rm i}}}
\newcommand{\epsout} {{\eps_{\rm s}}}
\newcommand{\win} {{\omega_{\rm i}}}

\newcommand{\dw} {{\omega}}
\newcommand{\zin} {{z_{\rm i}}}
\newcommand{\dz} {{z}}
\newcommand{\matElement} {{M\mkern-2mu_{\sigma_1\sigma_2\atop\sigma_1^\prime \sigma_2^\prime}}}
\newcommand{\con} {{\rm (con)}}

\date{\today}

\begin{abstract}
Characterizing entanglement in quantum materials is crucial for advancing next-generation quantum technologies. Despite recent strides in witnessing entanglement in magnetic materials with distinguishable spin modes, quantifying entanglement in systems formed by indistinguishable electrons remains a formidable challenge. To solve this problem, we introduce a method to extract various four-fermion correlations by analyzing the nonlinearity in resonant inelastic x-ray scattering spectra. These correlations constitute the primary components of the cumulant two-particle reduced density matrix. We further derive bounds for its eigenvalues and demonstrate the linear scaling with fermionic entanglement depth, providing a reliable witness for entanglement. Using the material-relevant strongly correlated models as examples, we show how this entanglement witness can efficiently quantify multipartite entanglement across different phase regions, highlighting its advantage over quantum Fisher information.
\end{abstract}

\begin{document}

\title{Entanglement Witness for Indistinguishable Electrons Using Solid-State Spectroscopy}
\author{Tongtong Liu}
\email[\href{mailto:liutong6149@gmail.com}{liutong6149@gmail.com}]{}
\affiliation{Department of Physics, Massachusetts Institute of Technology, Cambridge, Massachusetts 02139, USA}
\affiliation{Department of Chemistry, Emory University, Atlanta, Georgia 30322, USA}
\author{Luogen Xu}
\affiliation{Department of Chemistry, Emory University, Atlanta, Georgia 30322, USA}
\author{Jiarui Liu}
\affiliation{Department of Physics and Astronomy, Clemson University, Clemson, SC 29631, USA}
\author{Yao Wang}
\email{yao.wang@emory.edu}
\affiliation{Department of Chemistry, Emory University, Atlanta, Georgia 30322, USA}
\maketitle

\section{Introduction}
Quantum materials represent a new frontier in material science, characterized by the macroscopic quantum phenomena beyond the traditional band theory\,\cite{keimer2017physics}. Though still in the early stages of exploration, these  materials have demonstrated potential for transformative applications in superconductivity, sensing, high-efficiency batteries, and quantum computing\,\cite{giustino2020the, tokura2017emergent, krelina2021quantum}. Achieving a thorough understanding and predictive simulations of quantum materials, comparable to the precision seen in semiconductors, remains a significant challenge. In quantum materials, entanglement quantifies the inseparability of a many-body wavefunction into subdivisions. It is not only fundamental to materials' collective properties, but also critical for their applications in quantum information science\,\cite{bengtsson2006geometry,horodecki2009quantum}. Therefore, detecting, quantifying, and controlling entanglement have become key objectives in the study of quantum materials in the near future.

The characterization of entanglement has been effectively demonstrated in quantum optics. One effective method involves preparing identical twin quantum states and using interferometry to detect the purity of each  partition\,\cite{islam2015measuring}. This interferometric method has been used to quantify the R\'{e}nyi entanglement entropy\,\cite{kaufman2016quantum,tubman2016measuring,linke2018measuring, brydges2019probing}, thereby providing a robust tool for entanglement analysis. Another approach in quantum simulations involves the high-fidelity measurement of connected multipoint correlations, which vanish in separable or low-entangled states\,\cite{schweigler2017experimental, hilker2017revealing, salomon2019direct, koepsell2019imaging, vijayan2020time, prufer2020experimental, zache2020extracting, koepsell2021microscopic}. These quantum optics methods have been extensively applied in the study of entanglement for quantum many-body models.

Unlike quantum simulators, solid-state materials do not allow for single-electron control or site-resolved measurements, making wavefunction tomography and interference impractical. This limitation on measurement capabilities also hinders the experimental analysis of concurrence in macroscopic materials\,\cite{wootters1981statistical,carvalho2004decoherence,mintert2005concurrence, aolita2006measuring}. Semiglobal measures, such as entanglement entropy, are suited for thermodynamic scaling, which is essential for exploring topological states through simulations\,\cite{kitaev2006topological,levin2006detecting,qi2012topological,eisert2010area}. Nonetheless, these measurements remain beyond the reach of current solid-state experimental techniques, which are confined to a narrow range of macroscopic observables.

To address the challenge of experimentally probing entanglement in materials, especially the entanglement depth in multipartite systems\,\cite{sorensen2001entanglement, guhne2005multipartite, acin2001classification,friis2019entanglement}, a practical solution known as the entanglement witness has been proposed. This approach employs correlation functions of local operators, which are accessible through solid-state experiments, to estimate the multipartite entanglement\,\cite{terhal2000bell, coffman2000distributed, amico2004dynamics, roscilde2004studying,brukner2006crucial}. Particularly for magnetic materials, spin fluctuations encoded in the dynamical spin structure factor can be translated into quantum Fisher information (QFI)\,\cite{pezze2009entanglement,hyllus2012fisher,toth2012multipartite,hauke2016measuring}, which sets a lower bound for entanglement depth\,\cite{helstrom1969quantum,braunstein1994statistical,braunstein1996generalized}. This approach has been experimentally validated in antiferromagnetic and quantum spin liquid materials using inelastic neutron scattering (INS)\,\cite{mathew2020experimental, scheie2021witnessing, laurell2021quantifying,scheie2023proximate,menon2023multipartite,fang2024amplified}. Moreover, resonant inelastic x-ray scattering (RIXS), as an alternative technique to measuring spin fluctuations, has been proposed as a promising tool to probe spin entanglement even in materials out of equilibrium\,\cite{hales2023witnessing, suresh2024electron}.

However, the effectiveness of QFI as an entanglement witness depends on selecting the appropriate local operator. For general materials formed by electrons instead of local spin moments, QFI based on spin operators is insufficient to witness entanglement. For example, spectral measurements in correlated nonmagnetic materials have identified strong non-symmetry-breaking fluctuations of charge, phonon, and Cooper pair\,\cite{schafer2001high, yokoya2005role,chatterjee2015emergence, kondo2015point,faeth2020incoherent, xu2020spectroscopic,he2021superconducting, chen2023role,chen2023anomalous}. While the notion of QFI can be generalized to other mode-based local operators\,\cite{almeida2021from,mazza2024quantum,ren2024witnessing}, it cannot depict the complexity of more general wavefunctions induced by fermionic fluctuations\,\cite{varma1997non, hilker2017revealing, koepsell2021microscopic}. Unlike distinguishable qubits or spin modes, electrons and their orbitals become independent concepts. Since orbitals (basis wavefunctions) are usually selected artificially without uniqueness, the entanglement witness for electrons should be invariant against any (single-particle) basis transformations and independent from orbital indices. Additionally, the many-body wavefunction of indistinguishable fermions is antisymmetric against exchange. Such an antisymmetric superposition already contributes ``entanglement'' in the context of qubits, whose separable counterpart is a product state, leading to the incorrect assumption that a Fermi sea is heavily entangled in terms of QFI. The entanglement witness for electrons should naturally avoid these contributions from anticommutation properties\,\cite{ghirardi2002entanglement, ghirardi2004general, plastino2009separability, amico2008entanglement, sciara2017universality, lo2016quantum}. Therefore, more sophisticated basis-independent spectral witnesses for electronic entanglement are required.

\begin{figure}[!t]
    \centering
    \includegraphics[width=\linewidth]{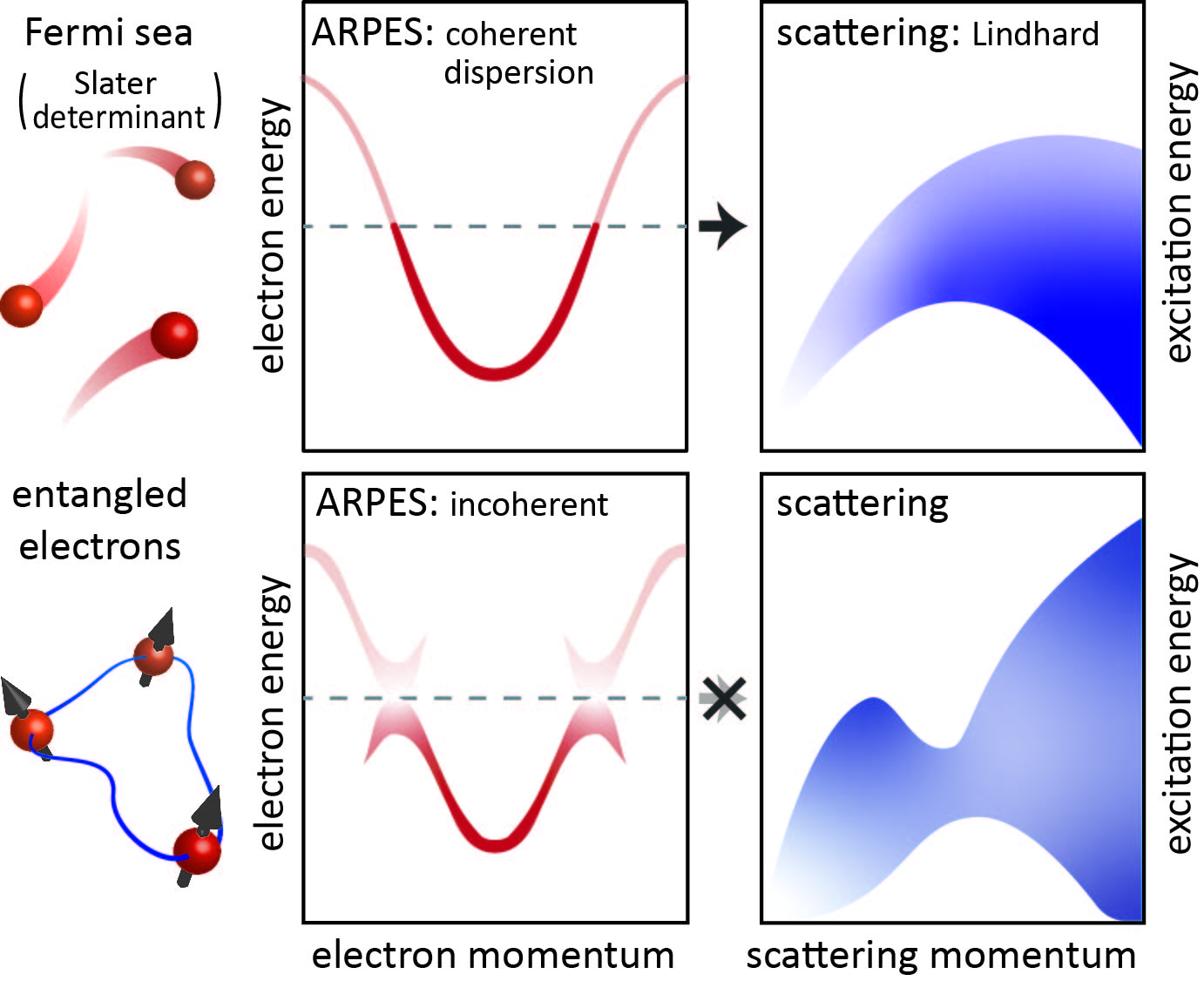}\vspace{-2mm}
    \caption{\label{fig:entanglementCartoon}
    A schematic illustrating the contrast between separable and entangled electronic systems in terms of their spectral relationships. The upper panels show a typical separable state, whose ARPES spectrum (middle) displays single or multiple electronic dispersions. The shaded area indicates unoccupied states (gray dashed line). The collective excitations measured by x-ray scattering (right) are described by the Lindhard response function, which can be inferred from the ARPES spectrum. The lower panels depict an entangled state scenario, where the ARPES spectrum appears more incoherent. The most significant difference is that the scattering spectrum cannot be directly derived from ARPES. Their distinction, especially in the form of particular energy-momentum integral, indicates entanglement.
    }
\end{figure}

Identifying a single spectral technique as a universal probe for entanglement is challenging, yet previous research indicates that the relationship between multiple spectra may reveal quantum fluctuations. As illustrated in Fig.~\ref{fig:entanglementCartoon}, the angle-resolved photoemission spectrum (ARPES) exhibits sharp quasiparticle dispersions for separable electrons in a material. The Fermi sea serves as a common example, but this principle can extend to mean-field wavefunctions with symmetry breaking. Because of the simplicity of the electronic wavefunction, particle-hole excitations can be analytically represented in the Lindhard form. Consequently, the scattering spectrum, with appropriate adjustments for matrix elements and unitary transformations, can be directly derived from the corresponding ARPES spectrum or Green's functions, essentially forming a ``bare bubble" diagram. In contrast, in many-body states beyond Gaussian representations, this direct link disappears. Deriving the scattering cross section from ARPES requires a vertex correction, which requires \textit{ad hoc} knowledge or assumptions about the interacting electron Hamiltonian and a full summation of high-order diagrams, often an impractical task. The inability to accurately reproduce scattering spectra or other multiparticle response functions (\textit{e.g.}, optical conductivity) from their single-particle counterparts is commonly viewed as an indicator of strong correlations. 

In this paper, we delve into the discrepancy between ARPES and scattering spectra and leverage it to establish an entanglement witness approach suitable for indistinguishable electrons. To achieve a basis-independent metric, we move beyond the traditional reliance on local probe operators within a scattering process. Instead, we focus on the nonlinear process involved in the intermediate state of RIXS, mediated by core-hole motion. This nonlinear process is often overlooked in standard RIXS studies, but can be identified by examining the two-momentum dependency on both incident and scattering photons. Importantly, it carries more information about electronic correlations beyond spin and charge. We successfully extract four-fermion observables that encode three-point and four-point correlations. These, along with two-point correlations derived from canonical local probes in scattering, form the primary components of a two-particle reduced density matrix (RDM). In this framework, the discrepancy between RIXS and ARPES can be quantified as a two-particle cumulant RDM (2CRDM). We further discuss the upper bounds of the 2CRDM eigenvalues $\lambda_{\max}$, which is found invariant under basis transformations, scales linearly with entanglement depth, and vanishes for Gaussian states. Thus, it qualifies as an entanglement witness for indistinguishable fermions and general materials. This spectral-based entanglement witness is then applied and validated in several representative states and models, demonstrating advantages over QFI when systems deviate from magnetic phases.

The organization of this paper is as follows. We discuss the nonlinear effect in RIXS and the resulting multipoint correlations in Sec.~\ref{sec:RIXS}. Next, we explore the connection of these correlations to the 2CRDM, discussing the upper bound of eigenvalues for various entanglement depths in Sec.~\ref{sec:entanglement}. This basis-independent, RIXS-measurable fermionic witness is then used to classify entanglement for several physically interesting models and compared with QFI in Sec.~\ref{sec:examples}. Finally, Sec.~\ref{sec:discussion} discusses specific experimental strategies and potential extensions beyond the entanglement witness. 

\section{Connected multipoint correlations from RIXS spectra}\label{sec:RIXS}
RIXS, as illustrated in Fig.~\ref{fig:RIXSCartoon}(a), is a photon-in--photon-out process to probe materials. This process uses an x-ray photon (ranging from hundreds of eVs to several keVs), precisely tuned to match a specific absorption edge. The high-energy x-ray photon induces resonant transitions of deep core-level electrons into the valence band. This process excites the material into a short-lived intermediate state with a core hole [see Fig.~\ref{fig:RIXSCartoon}(b)], typically lasting only a few femtoseconds. A valence electron subsequently recombines with this core hole, emitting another x-ray photon with slightly lower energy. Analyzing the energy and momentum differences between the two photons reveals the intrinsic collective excitations of the material\,\cite{ament2011resonant}. Its exceptional tunability enables the study of diverse excitations, including spin and charge, $d-d$ excitations, and orbital orders. 

While RIXS peak energies are often used to map collective excitations in the form of two-point correlation functions, the intensity distribution often deviates from the precise dynamical structure factors obtained through INS or EELS\,\cite{jia2016using}. These deviations arise from the finite core-hole lifetime, which cannot be ignored when the collective excitations propagate rapidly. Historically seen as a limitation in accurately representing collective excitations, {this section will demonstrate how such deviations encode} multipoint correlations, {forming a crucial framework} for the entanglement witness theory elaborated in subsequent sections.

\subsection{A brief overview of the RIXS process}\label{sec:RIXS:overview}

\begin{figure}[!t]
    \centering
    \includegraphics[width=\linewidth]{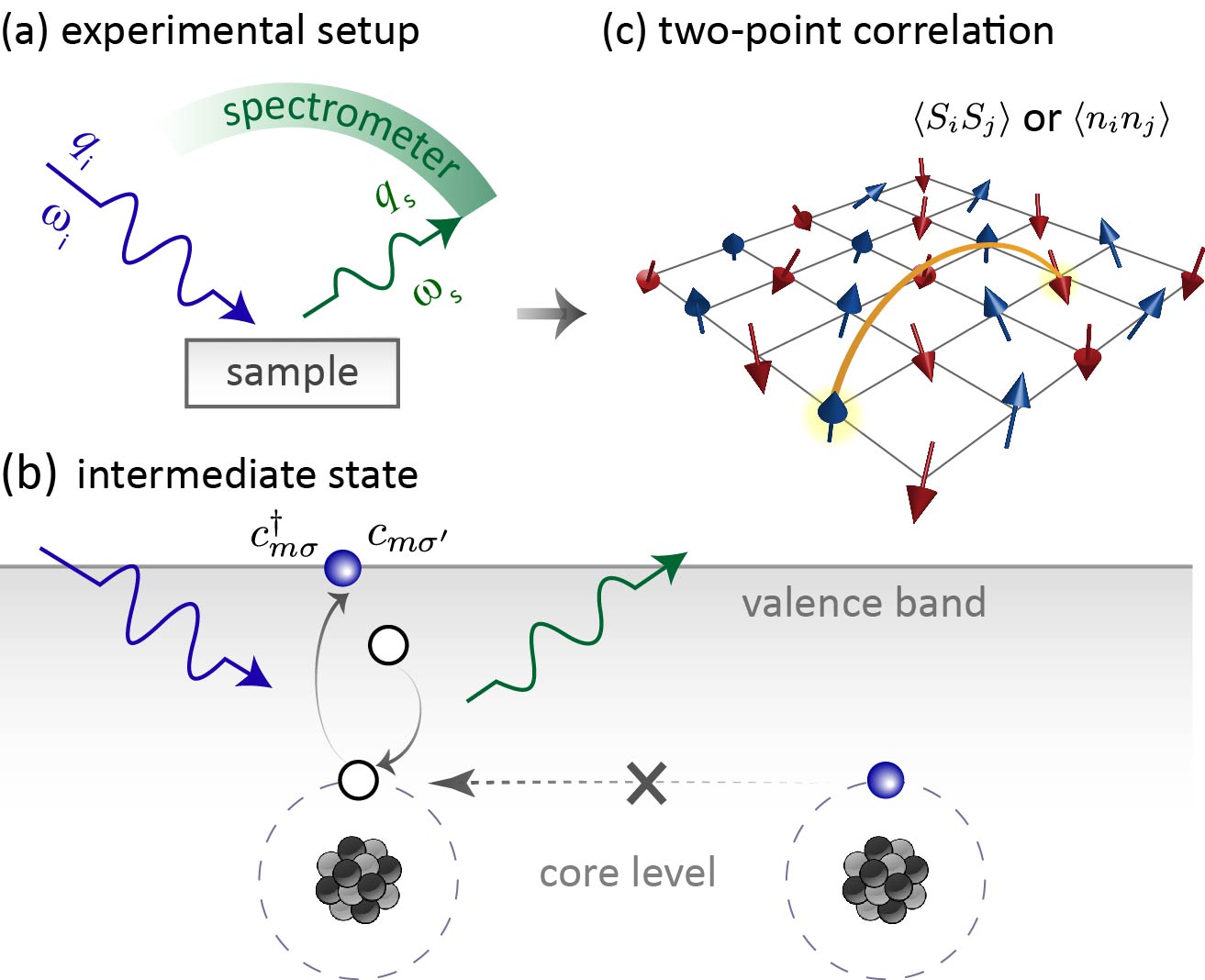}\vspace{-2mm}
    \caption{\label{fig:RIXSCartoon}
    (a) Schematic illustrating typical x-ray scattering experimental setup, where the incident beam is fixed and the spectrometer (or the sample) rotates to scan a single momentum trajectory. This arrangement is based on the underlying assumption that the spectral intensity depends solely on the momentum transfer $\dq$. (b) The intermediate state with a core hole induced by the x-ray absorption. According to the SCH assumption, the photon emission occurs at the same site as the absorption. (c) Two-point correlations,  specifically spin ($\langle S_iS_j\rangle$) and charge ($\langle n_i n_j\rangle$), probed by the spectrum.  
    }
\end{figure}

With the intermediate state, the RIXS cross section is described by the Kramers-Heisenberg formula\,\cite{wang2018theoretical}:
\begin{eqnarray}\label{eq:RIXS1}
    I\left(\kin,\kout,\win,\dw\right) &=& \frac{1}{N\pi}\sum_{m, n}e^{-i\kout\cdot (\rbf_{m}-\rbf_n)}\textrm{Im}\bra{\Psi_{\rm int}}\mathcal{D}_{n\epsout}\nonumber\\
    &&\frac1{\mathcal{H}-E_G-\omega-i0_+} \mathcal{D}_{m\epsout}^\dagger\ket{\Psi_{\rm int}}\,,
\end{eqnarray}
where $N$ is the system size, $E_G$ is the ground-state energy, $\kout$ is the momentum of the scattering photon, and $\omega$ is the energy difference between the two photons. Unlike nonresonant scattering, the state $\ket{\Psi_{\rm int}}$ in Eq.~\eqref{eq:RIXS1} represents a specific intermediate state triggered by the resonant absorption. It is determined by the momentum $\kin$ and energy $\win$ of the incident photon
\begin{equation}\label{eq:RIXS2}
    \ket{\Psi_{\rm int}(\kin,\win)}=\mkern-2mu\sum_{m^\prime} \frac{e^{i\kin\cdot \rbf_{m^\prime}}}{\sqrt{N}} \frac{1}{\mathcal{H}^\prime\mkern-1mu-\mkern-1mu E_G\mkern-1mu-\mkern-1mu\win\mkern-1mu-\mkern-1mui\Gamma}\mathcal{D}_{m^\prime\epsin}\mkern-2mu\ket{G}\,,
\end{equation}
where $\Gamma$ denotes the inverse of the core-hole lifetime and $\ket{G}$ is the ground state. For simplification, we define the scalar variables in  Eqs.~\eqref{eq:RIXS1} and \eqref{eq:RIXS2} as $\dz=E_G+\omega +i0_+$ and $\zin=E_G+\win+i\Gamma$. The absorption and emission processes involve electronic dipole transitions at a specific edge, depicted by the dipole operator 
\begin{equation}\label{eq:dipole}
\mathcal{D}_{m\bm\varepsilon}=\sum_{\alpha,\beta,\sigma} D_{\alpha\beta}^{(\bm\varepsilon)}(\qbf_{\rm i/s}) \,c_{m\beta\sigma}^{\dagger}p_{m\alpha\sigma}\,.
\end{equation}
Here, $c_{m\beta\sigma}^{\dagger}$ ($c_{m\beta\sigma}$) creates (annihilates) an electron in the $\beta$th valence bands and $p_{m\alpha\sigma}^{\dagger}$ ($p_{m\alpha\sigma}$) corresponds to the $\alpha$th core-level electron at the $m$th unit cell. The transition matrix element $D_{\alpha\beta}^{(\bm\varepsilon)}(\qbf_{\rm i/s})$ is derived from atomic orbitals. Note that both the valence bands and core levels can exhibit degeneracy, especially for the transition-metal $L$ and $M$ edges. The core-level degeneracy is crucial for probing spin-flipped excitations\,\cite{ament2009theoretical}. In contrast, the number of valence orbitals is irrelevant for our discussion here. To simplify, we assume a single valence band, using the Cu $L$ edge RIXS as an example, while maintaining general applicability. Hence, we simplify its notation to $c_{m\sigma}$, omitting orbital indices.

The intermediate-state Hamiltonian $\mathcal{H}^\prime$ in Eq.~\eqref{eq:RIXS2} includes additional terms beyond the valence-electron Hamiltonian, $\mathcal{H}$, due to the presence of the core hole:
\begin{equation}\label{eq:HubbardInter}
	\mathcal{H}^\prime = \mathcal{H} + \mathcal{H}_{\rm core} - U_c \sum_{m,\alpha}\sum_{\sigma,\sigma^\prime} c_{m \sigma}^\dagger c_{m\sigma} p_{m\alpha\sigma^\prime}p_{m \alpha\sigma^\prime}^\dagger.
\end{equation}
Here, the second term ($\mathcal{H}_{\rm core}$) in Eq.~\eqref{eq:HubbardInter} represents the core-electron Hamiltonian, while the third term characterizes the attractive interaction $U_c$ between the core hole and valence electrons. More specifically, the core-electron Hamiltonian $\mathcal{H}_{\rm core}$ is expressed as
\begin{equation}\label{eq:RIXSCoreHamiltonian}
	\mathcal{H}_{\rm core} = \sum_{m} \left(\sum_{\alpha\sigma}E_{\rm edge}p_{m\alpha\sigma}p_{m \alpha\sigma}^\dagger +  \mathcal{H}^{\rm (SOC)}_m\right)+ \mathcal{T}_c\,,
\end{equation}
where $E_{\rm edge}$ denotes the absorption edge energy and $\mathcal{H}^{\rm (SOC)}_m$ details the spin-orbit coupling (SOC) among degenerated core-level states. The last term ($\mathcal{T}_c$) represents the kinetic energy of the core hole, typically ignored due to the localized nature of core orbitals.

With spin-orbit coupling at the core level, specifically the $\mathcal{H}^{\rm (SOC)}_m$ term, the intermediate state violates spin conservation. As a result, the two spin flavors in Eq.~\eqref{eq:dipole} for the incident and scattering processes yield four combinations: one spin-conserved channel and three non-spin-conserved channels\,\cite{ament2009theoretical}. For a specific edge, the coefficients on these channels are controlled by the matrix elements $D^{(\bm\epsin)}(\kin)$ and $D^{(\bm\epsout)}(\kout)$. For specified incident and scattering x-ray beams, these four matrix elements in Eq.~\eqref{eq:RIXS1} are typically consolidated into a single $\matElement$\,\cite{ament2011theory, halasz2016resonant}. When the polarizations $\epsin$ and $\epsout$ are parallel to the scattering plane, known as the $\pi-\pi$ configuration, the coefficient $M$ simplifies to a direct product of diagonal matrices $\sigma_0\otimes \sigma_0$; in perpendicular polarization settings like the $\pi-\sigma$ configuration, $M\propto \sigma_x\otimes\sigma_x$ represents one of the non-spin-conserved channels.

In the core-electron Hamiltonian in Eq.~\eqref{eq:RIXSCoreHamiltonian}, both the potential energy and the SOC terms are spatially local. Thus, if we disregard the kinetic energy, the core hole can be treated as static during the intermediate state, as shown in Fig.~\ref{fig:RIXSCartoon}(b). This static-core-hole (SCH) assumption is common in RIXS analysis. Under this assumption, the site indices $m$ in Eq.~\eqref{eq:RIXS1} and $m'$ in Eq.~\eqref{eq:RIXS2} must be identical. Thus, the RIXS cross section simplifies to
\begin{eqnarray}\label{eq:RIXSImmobile}
    I\left(\kin,\kout,\win,\dw\right) &=& \frac{1}{N^2\pi}\sum_{m, n}e^{i\dq\cdot (\rbf_{m}-\rbf_n)}\textrm{Im}\Big\langle\mathcal{D}_{n\epsin}^\dagger  \frac{1}{\mathcal{H}^\prime-\zin^*}\nonumber\\
    &&\mathcal{D}_{n\epsout}\frac1{\mathcal{H}-z} \mathcal{D}_{m\epsout}^\dagger  \frac{1}{\mathcal{H}^\prime-\zin}\mathcal{D}_{m\epsin}\Big\rangle\,.
\end{eqnarray}
Here, the notation $\langle ...\rangle$ represents an expectation taken at the ground state $\ket{G}$. Notably, Eq.~\eqref{eq:RIXSImmobile} depends on the momentum transfer $\dq$, rather than on the individual incident or scattering momenta. Therefore, within the SCH assumption, scanning both momenta in a RIXS experiment becomes unnecessary, unless exploring dispersions with significant 3D characteristics\,\cite{lee2014asymmetry,gerber2015three,hepting2018three,bejas2024plasmon}. 

Integrating Eq.~\eqref{eq:RIXSImmobile} leads to a two-point correlation at the ultrashort core-hole lifetime (UCL) limit: 
\begin{eqnarray}\label{eq:integralImmobile}
    \mkern-4mu\iint\mkern-6mu I\mkern-2mu\left(\dq,\win,\dw\right) d\win d \dw&=& \frac{\pi }{N^2 \Gamma}\sum_{m,n}e^{i\dq\cdot\left(\rbf_{m}-\rbf_{n}\right)}\mkern-6mu\sum_{ \sigma_1, \sigma_1^\prime} \mkern-4mu\sum_{\sigma_2,\sigma_2^\prime}\mkern-4mu\matElement\nonumber\\
    && \braket{{c_{n\sigma_1^\prime}}{c^{\dagger}_{n\sigma_1} c_{m\sigma_2}}{c_{m\sigma_2^\prime}^{\dagger}}}+O\left(\frac1{\Gamma^2}\right)\,.
\end{eqnarray}
This integral can estimate the charge and spin structure factors depending on polarization settings. Specifically, the $\pi-\pi$ polarizations with $M\propto\sigma_0\otimes \sigma_0$ correspond to the (hole) charge structural factor, whereas the $\pi-\sigma$ polarizations with $M\propto \sigma_x\otimes\sigma_x$ correspond to the spin structural factor [see Fig.~\ref{fig:RIXSCartoon}(c)]. To obtain accurate results, excitations that are unrelated to the target electronic subsystem, such as the phonon excitations, must be carefully filtered out from the spectrum $I\left(\dq,\win,\dw\right)$.
    
\subsection{Impact of mobile core holes on RIXS spectra}\label{sec:RIXS:mobileCore}

\begin{figure*}[!t]
    \centering
    \includegraphics[width=\linewidth]{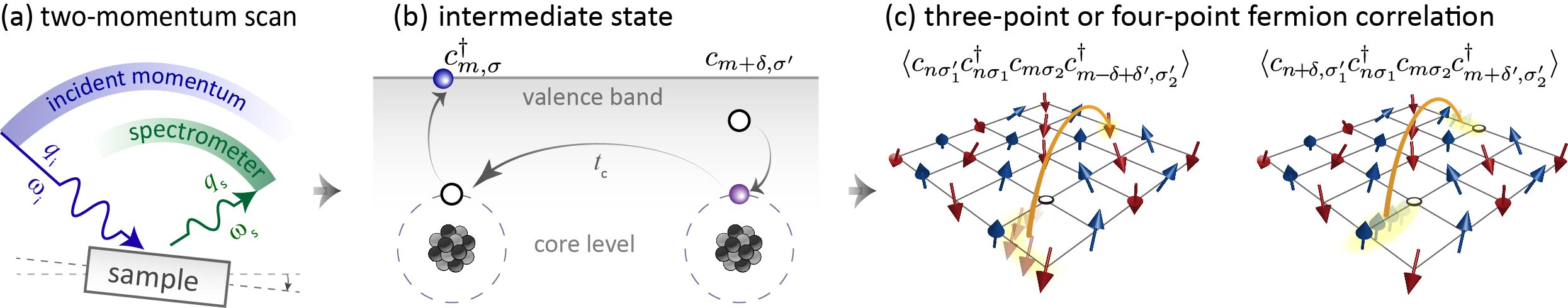}\vspace{-2mm}
    \caption{\label{fig:RIXSMobileHoleCartoon}
    (a) Schematic illustrating the proposed scattering setup, where both the sample and spectrometer rotate independently to scan two momenta, resolving nonlinear effects caused by the core hole's motion. (b) The intermediate state is affected by the mobile core hole, leading to the photon emission occurring at a different site from the absorption. (c) The three-point and four-point correlations as leading-order terms extracted from the two-momentum distribution of RIXS.\vspace{-2mm}
    }
\end{figure*}

The SCH assumption becomes questionable when considering the finite core-hole lifetime, causing deviations of RIXS from dynamical structure factors. The intermediate state with a finite lifetime enables the hopping of the core hole, significantly influencing the RIXS cross section. As illustrated by Fig.~\ref{fig:RIXSMobileHoleCartoon}, the mobility of the core hole can induce a particle-hole excitation at neighboring sites within the valence band. To analyze this, we define the regular components of Eq.~\eqref{eq:HubbardInter} that do not involve core-hole motion as $\mathcal{H}^\prime_0$. Thus, the intermediate-state Hamiltonian is rewritten as $\mathcal{H}^\prime = \mathcal{H}^\prime_0+\mathcal{T}_c$. As we show, the core-hole motion determines the spectral dependence on the incident momentum $\kin$, distinct from Eq.~\eqref{eq:RIXSImmobile}. While our derivations below are general, the momentum distribution depends on the specific core-electron band structure. Specifically, we adopt the following kinetic Hamiltonian for the core-level hoppings:
\begin{equation}\label{eq:coreBandStruct}
    \mathcal{T}_c=-t_c\sum_{\langle m,n\rangle, \sigma}\left(p_{m\alpha\sigma}^{\dagger} p_{n\alpha\sigma}+\text{H.c.}\right)\,.
\end{equation}
Here, the core-level hopping $t_c$ is significantly smaller than valence hoppings and interactions (in $\mathcal{H}$). We assume isotropic hoppings for degenerate core-level orbitals, though real materials may exhibit anisotropy or off-diagonal hoppings. In such scenarios, the hopping matrix can be diagonalized, with bandwidths expected to be comparable to $t_c$.

By treating the core-hole hopping as a perturbation, we can expand the intermediate state $\ket{\Psi}$ in Eq.~\eqref{eq:RIXS2} as 
\begin{eqnarray}\label{eq:intermediateExpansion}
    \ket{\Psi_{\rm int}(\kin,\win)}&\approx&\frac{1}{\sqrt{N}}\sum_{m^\prime} e^{i\kin\cdot \rbf_{m^\prime}} \left[\frac{1}{\mathcal{H}^\prime_0-\zin} -\frac{\mathcal{T}_c}{(\mathcal{H}^\prime_0-\zin)^2} \right.\nonumber\\
    &&\left.+\frac{\mathcal{T}_c^2}{(\mathcal{H}^\prime_0-\zin)^3}\right]\mathcal{D}_{m^\prime\epsin}\ket{G}\nonumber\\
    &=&\ket{\Psi_{\rm int}^{(0)}}+\ket{\Psi_{\rm int}^{(1)}}+\ket{\Psi_{\rm int}^{(2)}}\,,
\end{eqnarray}
where the perturbative order of each term is determined by the occurrence of $\mathcal{T}_c$. Specifically,
\begin{eqnarray}\label{eq:intermediateExpansion2}
    \ket{\Psi_{\rm int}^{(0)}(\kin,\win)}&=&\sum_{m^\prime} \frac{e^{i\kin\cdot \rbf_{m^\prime}}}{\sqrt{N}}\frac{1}{\mathcal{H}^\prime_0-\zin}\mathcal{D}_{m^\prime\epsin}\ket{G}\nonumber\\
    \ket{\Psi_{\rm int}^{(1)}(\kin,\win)}&=&-\sum_{m^\prime} \frac{e^{i\kin\cdot \rbf_{m^\prime}}}{\sqrt{N}}\frac1{(\mathcal{H}^\prime_0-\zin)^2}\mathcal{T}_c\mathcal{D}_{m^\prime\epsin}\ket{G}\nonumber\\
    \ket{\Psi_{\rm int}^{(2)}(\kin,\win)}&=&\sum_{m^\prime} \frac{e^{i\kin\cdot \rbf_{m^\prime}}}{\sqrt{N}}\frac1{(\mathcal{H}^\prime_0-\zin)^3}\mathcal{T}_c^2\mathcal{D}_{m^\prime\epsin}\mkern-2mu\ket{G}\,.
\end{eqnarray}
In these expansions, the $\mathcal{H}^\prime_0$ term is local for core holes, while $\mathcal{T}_c$ incorporates nearest-neighbor hoppings. Consequently, the combined operators $\mathcal{T}_c\mathcal{D}_{m^\prime\epsin}$ in $\ket{\Psi_{\rm int}^{(1)}(\kin,\win)}$ create a core hole at $\rbf_{m^\prime}+\rbf_{\delta}$ instead of $\rbf_{m^\prime}$, with $\rbf_{\delta}$ denoting the unit vector connecting nearest neighbors. This process is illustrated in Fig.~\ref{fig:RIXSMobileHoleCartoon}(b).

When incorporating Eq.~\eqref{eq:intermediateExpansion} into the RIXS cross section Eq.~\eqref{eq:RIXS1}, we can dissect the spectrum according to the perturbative order. The zeroth-order spectrum arises from the zeroth-order intermediate state $\ket{\Psi_{\rm int}^{(0)}}$ in Eq.~\eqref{eq:intermediateExpansion2}. Since the core hole is static in $\mathcal{H}^\prime_0$, this zeroth-order spectrum follows the form of Eq.~\eqref{eq:RIXSImmobile}:
\begin{eqnarray}\label{eq:RIXSExpansion0}
    I^{(0)}\left(\dq,\win,\dw\right) &=& \frac{1}{N^2 \pi}\sum_{m,n}e^{i\dq\cdot (\rbf_{m}-\rbf_n)}\textrm{Im}\Big\langle\mathcal{D}_{n\epsin}^\dagger  \frac{1}{\mathcal{H}^\prime_0-\zin^*}\nonumber\\
    &&\mathcal{D}_{n\epsout}\frac1{\mathcal{H}-z} \mathcal{D}_{m\epsout}^\dagger  \frac{1}{\mathcal{H}^\prime_0-\zin}\mathcal{D}_{m\epsin}\Big\rangle\,.
\end{eqnarray}
Spatial translational symmetry has been adopted in Eq.~\eqref{eq:RIXSExpansion0}. This zeroth-order spectrum is dominant in the entire cross section and is often used to represent RIXS. 

The first-order contribution of the RIXS cross section, denoted as $I^{(1)}$, arises from the cross term between the $\ket{\Psi_{\rm int}^{(0)}}$ and $\ket{\Psi_{\rm int}^{(1)}}$. This contribution is given by [see Eq.~\eqref{eq:RIXSExpansion1} in Appendix \ref{app:mobileHole} for detailed derivations]:
\begin{eqnarray}\label{eq:RIXSExpansion1}
    I^{(1)}  &&= -\frac{1}{N^2\pi}\mkern-2mu\sum_{m,n}\mkern-1mu\sum_{m^\prime, n^\prime}e^{i\kin\cdot (\rbf_{m^\prime}\mkern-2mu-\mkern-2mu\rbf_{n^\prime}\mkern-2mu)-i\kout\cdot (\rbf_{m}\mkern-2mu-\mkern-2mu\rbf_n\mkern-2mu)}\textrm{Im}\Big[\Big\langle\mkern-2mu\mathcal{D}_{n^\prime\mkern-1mu\epsin}^\dagger \mkern-4mu\mathcal{T}_c\nonumber\\
    &&  \frac{1}{(\mathcal{H}^\prime_0-\zin^*)^2}\mathcal{D}_{n\epsout}\frac1{\mathcal{H}-z} \mathcal{D}_{m\epsout}^\dagger  \frac1{\mathcal{H}^\prime_0-\zin}\mathcal{D}_{m^\prime\mkern-1mu\epsin}\mkern-2mu\Big\rangle+\mkern-4mu  \Big\langle\mathcal{D}_{n'\mkern-1mu\epsin}^\dagger \nonumber\\
    && \frac{1}{\mathcal{H}^\prime_0-\zin^*}\mathcal{D}_{n\epsout}\frac1{\mathcal{H}- z} \mathcal{D}_{m\epsout}^\dagger  \frac1{(\mathcal{H}^\prime_0-\zin)^2}\mathcal{T}_c\mathcal{D}_{m^\prime\mkern-1mu\epsin}\mkern-2mu\Big\rangle \Big]   \,.
\end{eqnarray}
Unlike the zeroth-order spectrum, $I^{(1)}$ contains a phase factor $e^{i\kin\cdot\rbf_{\delta}}$ , stemming from the inequivalence of $\rbf_{m^\prime}$ and $\rbf_{m}$ (or $\rbf_{n^\prime}$ and $\rbf_{n}$). This phase factor introduces a $\kin$ dependence in addition to the $\dq$ dependence.

\begin{figure}[!t]
   \centering \vspace*{-3mm}
    \includegraphics[width=\linewidth]{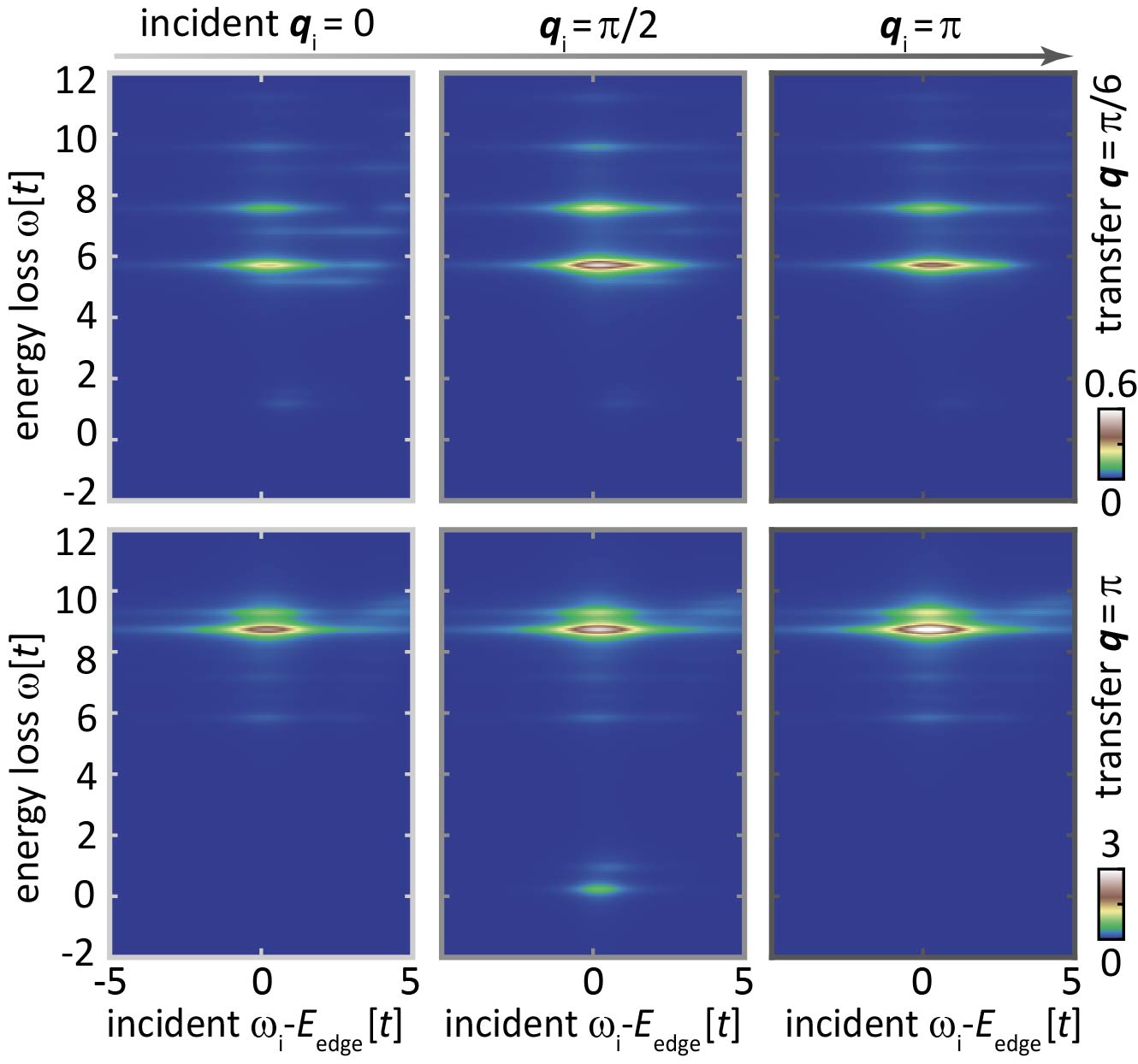}\vspace*{-3mm}
    \caption{\label{fig:RIXSspec_SC}
    RIXS spectra with the spin-conserved channel for a 12-site Hubbard model, incorporating a core-hole hopping $t_c=0.1t$. The upper panels show RIXS spectra at a momentum transfer $\dq=\pi/6$, and lower panels display spectra at $\dq=\pi$. From left to right, these panels present RIXS spectra for various incident momenta $\kin=0$, $\pi/2$, and $\pi$ (with fixed momentum transfer $\dq$). All simulations are obtained using a half-filled single-band Hubbard model for valence electrons. Parameters are chosen as Hubbard $U=8t$, core-hole interaction $U_c=4t$, and inverse core-hole lifetime $\Gamma=t$. 
    \vspace*{-2mm}
}
\end{figure}

Next, we consider the second-order RIXS cross section, denoted as $I^{(2)}$. It includes the diagonal terms of the intermediate state $\ket{\Psi_{\rm int}^{(1)}}$ and cross terms between $\ket{\Psi_{\rm int}^{(0)}}$ and $\ket{\Psi_{\rm int}^{(2)}}$. Specifically,
\begin{widetext}
\begin{eqnarray}\label{eq:RIXSExpansion2}
    &&I^{(2)} = \frac{1}{N^2\pi}\sum_{m,n\atop m^\prime, n^\prime}e^{i\kin\cdot (\rbf_{m^\prime}-\rbf_{n^\prime})-i\kout\cdot (\rbf_{m}-\rbf_n)}\textrm{Im}\Big[\Big\langle\mathcal{D}_{n'\epsin}^\dagger  \frac{1}{\mathcal{H}^\prime_0-\zin^*}\mathcal{D}_{n\epsout}\frac1{\mathcal{H}-z} \mathcal{D}_{m\epsout}^\dagger  \frac1{(\mathcal{H}^\prime_0-\zin)^3}\mathcal{T}_c^2\mathcal{D}_{m^\prime\epsin}\Big\rangle
    +\Big\langle\mathcal{D}_{n^\prime\epsin}^\dagger \mathcal{T}_c^2\nonumber\\
    &&  \frac{1}{(\mathcal{H}^\prime_0-\zin^*)^3}\mathcal{D}_{n\epsout}\frac1{\mathcal{H}-z} \mathcal{D}_{m\epsout}^\dagger  \frac1{\mathcal{H}^\prime_0-\zin}\mathcal{D}_{m^\prime\epsin}\Big\rangle
    +\Big\langle\mathcal{D}_{n^\prime\epsin}^\dagger \mathcal{T}_c \frac{1}{(\mathcal{H}^\prime_0-\zin^*)^2}\mathcal{D}_{n\epsout}\frac1{\mathcal{H}-z} \mathcal{D}_{m\epsout}^\dagger  \frac1{(\mathcal{H}^\prime_0-\zin)^2}\mathcal{T}_c\mathcal{D}_{m^\prime\epsin}\Big\rangle\Big]\,.
\end{eqnarray}
\end{widetext}
Similar to $I^{(1)}$, the second-order cross section $I^{(2)}$ introduces explicit dependencies on both the incident ($\kin$) and scattering ($\kout$) momenta. Figure~\ref{fig:RIXSspec_SC} presents an example of the full RIXS cross section, which includes contributions from all orders, for a half-filled single-band Hubbard model with $t_c=0.1t$. (The Hamiltonian and parameters are detailed in Appendix \ref{app:RIXSDetails}.)  While the overall spectral shape is mainly governed by the momentum transfer $\dq$, the distribution of spectral weight varies with different incident momenta $\kin$. Here, the matrix elements have been omitted from these presented spectra, indicating that all observed $\kin$ dependencies arise from the core-hole motion. Apart from the weight distribution, the spectra reveal several low-energy peaks appearing below 2$t$, particularly noticeable in the spectrum with $\dq=\pi$ and $\kin=\pi/2$. It is known that charge excitations are gapped in a half-filled Hubbard model. For the model parameters used in Fig.~\ref{fig:RIXSspec_SC} ($U=8t$), the Mott gap is approximately $4t$. Therefore, these low-energy excitations within the gap reflect collective fermionic modes beyond two-point charge excitations, as a nonlinear effect stemming from the intermediate state\,\cite{bisogni2014femtosecond,kumar2020spectroscopic}.

\begin{figure}[!b]
   \centering
    \includegraphics[width=\linewidth]{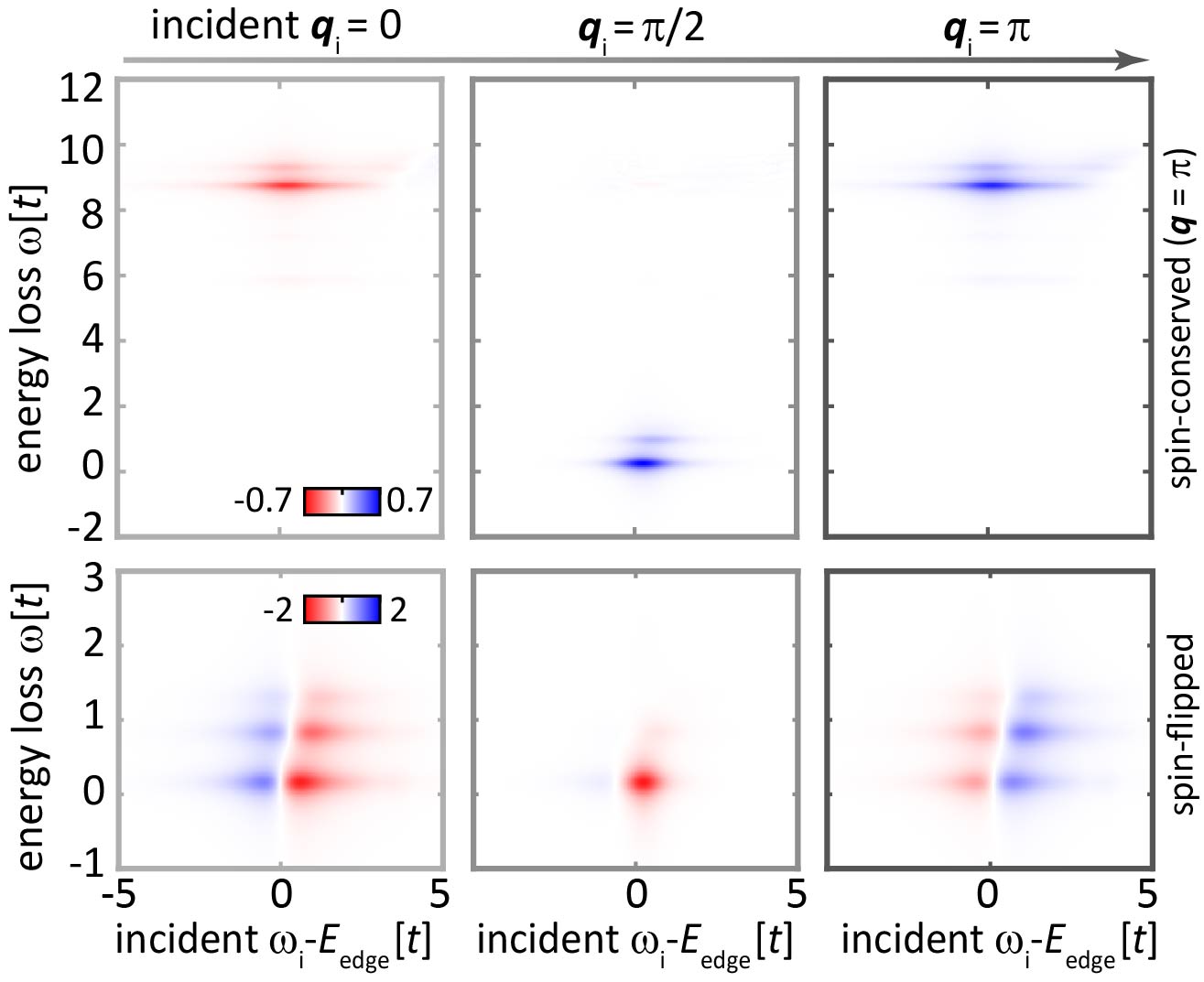}\vspace*{-3mm}
    \caption{\label{fig:RIXSspec_diff}
    The differential spectrum between RIXS spectra with the core-hole hopping $t_c=0.1t$ and without the core-hole hopping $t_c=0$, for single-band Hubbard model at momentum transfer $\dq=\pi$. The upper panels show the spectra in the spin-conserved channel, while the lower panels represent the spin-flip channel. From left to right, these panels present RIXS spectra for incident momenta $\kin=0$, $\pi/2$, and $\pi$ . All simulations are obtained in the single-band Hubbard model (for valence electrons) with the same parameters as Fig.~\ref{fig:RIXSspec_SC}. 
}
\end{figure}

The first- and second-order cross sections can be analyzed by subtracting the $\kin$-independent spectra $I^{(0)}\left(\dq,\win,\dw\right)$ with $t_c=0$. As illustrated in Fig.~\ref{fig:RIXSspec_diff}, the spin-conserved differential spectra highlight the spectral weight variation across the Mott gap and provide insightful information about the in-gap excitations attributable to the gapped charge structure. These differential spectral features, varying with different $\kin$ values, indicate the presence of dispersive multiparticle excitations. On the other hand, the spin-flipped differential spectra, particularly for $\dq=\pi$, predominantly exhibit shifts in the resonance. The presence of gapless two-spinon excitations, which form the zeroth order, implies that the $\kin$ dependency of differential spectra mainly results from shifts in the relative core-level energy compared to the spinon Fermi surface. The spinon Fermi momentum for a half-filled Hubbard model resides at $\mathbf{k}=\pi/2$. As a consequence, the low-energy excitations for $\kin=0$ and $\kin=\pi$ relate to core-level excitations for electrons at $|\mathbf{k}|>\pi/2$ and $|\mathbf{k}|<\pi/2$, respectively. Considering the positive core-hole hopping $t_c=0.1t$ used in our simulations, these incident momenta lead to a redshift and blueshift of the resonance, respectively.

\subsection{multipoint Correlations from Spectra}\label{sec:RIXS:correlatoins}
While the energy distributions in RIXS spectra reflect the dispersion of collective excitations, integrating the spectral weights yields equal-time correlation functions of ground states or thermal ensembles. These correlations provide information about the many-body electronic states in macroscopic materials, where site-resolved measurements, common in quantum optics, are infeasible. By decomposing the RIXS cross section into orders relative to the core-hole hopping $t_c$, we now analyze the correlations obtained from integrating each order of the spectrum. Although perturbative expansions cannot be performed in experiments, these correlations can be isolated according to their distinct momentum and energy dependencies, as demonstrated in this subsection.

As detailed in Eq.~\eqref{eq:integralImmobile}, integrating over the energies of the zeroth-order spectrum $I^{(0)}$ yields the spin and charge structure factors. Given the small core-hole bandwidth, even without the SCH approximation, $I^{(0)}$ dominates RIXS spectral intensity and is independent of the incident momentum $\kin$. For the first-order spectrum $I^{(1)}$, integrating Eq.~\eqref{eq:RIXSExpansion1} over $\win$ and $\dw$ leads to
\begin{eqnarray}\label{eq:I1DirectIntegral}
  &&\iint I^{(1)}(\kin,\dq,\win,\dw) d\win d\dw\nonumber\\
 &&= \frac{\pi t_c}{2N^2\Gamma^3} \sum_{ \sigma_1, \sigma_1^\prime} \mkern-4mu\sum_{\sigma_2,\sigma_2^\prime}\matElement\mathrm{Re}\Big[\sum_{m,n}\sum_{\delta}
    e^{i\dq\cdot\left(\rbf_{m}-\rbf_{n}\right)+i\kin\cdot\rbf_{\delta}}
    \nonumber\\
    &&\Big(\mkern-6mu \braket{c_{n-\delta,\sigma_1^\prime}{c^{\dagger}_{n\sigma_1} c_{m\sigma_2}}\mathcal{H}_0^\prime {c_{m\sigma_2^\prime}^{\dagger}}}\mkern-4mu-\mkern-4mu \braket{c_{n\sigma_1^\prime}\mathcal{H}_0^\prime {c^{\dagger}_{n\sigma_1} c_{m\sigma_2}}{c_{m+\delta,\sigma_2^\prime}^{\dagger}}}\mkern-4mu\Big)\Big] \nonumber\\
    &&+O\left(\Gamma^{-5}\right)\,.
\end{eqnarray}  
Although four-fermion operators appear in the expectation value, they cannot be directly utilized as elements of a RDM due to the inclusion of $\mathcal{H}_0^\prime$. Later discussions will clarify how we can segregate $I^{(1)}$ from other orders through their unique $\kin$ dependence.

Next, we examine the correlations obtained from the second-order RIXS cross section $I^{(2)}$. Directly integrating the $I^{(2)}$ over $\win$ and $\dw$ yields two primary sets of correlations, originating from the terms in Eq.~\eqref{eq:RIXSExpansion2}: 
\begin{eqnarray}\label{eq:I2DirectIntegral}
    &&\iint    I^{(2)}(\kin,\dq,\win,\dw)  d\win  d\dw\nonumber\\
    &=&\frac{\pi t_c^2}{2N^2\Gamma^3}\mkern-6mu\sum_{ \sigma\mkern-2mu_1, \sigma\mkern-2mu_1^\prime} \mkern-6mu\sum_{\sigma\mkern-2mu_2,\sigma\mkern-2mu_2^\prime}\mkern-4mu\matElement\mathrm{Re}\Big[\sum_{m,n}\sum_{\delta, \delta^\prime}\mkern-2mu
    e^{i\dq\cdot\left(\rbf\mkern-2mu_{m}-\rbf\mkern-2mu_{n}\right)+i\kin\cdot (\rbf\mkern-2mu_{\delta^\prime}-\rbf\mkern-2mu_{\delta})}
    \nonumber\\
    &&\Big(\mkern-6mu \braket{{c_{n\mkern-2mu+\mkern-2mu\delta,\sigma_1^\prime}}\mkern-2mu{c^{\dagger}_{n\sigma_1} \mkern-2muc_{m\sigma_2}}\mkern-2mu{c_{m+\delta^\prime\mkern-2mu,\sigma_2^\prime}^{\dagger}}}\mkern-4mu-\mkern-4mu\braket{{c_{n\sigma_1^\prime}}\mkern-2mu{c^{\dagger}_{n\sigma_1}\mkern-2mu c_{m\sigma_2}}\mkern-2mu{c_{m\mkern-2mu-\mkern-2mu\delta\mkern-2mu+\mkern-2mu\delta^\prime\mkern-2mu,\sigma_2^\prime}^{\dagger}}}\mkern-6mu\Big)\Big]\nonumber\\
    &&+O\left(\Gamma^{-5}\right)\,.
\end{eqnarray}
As illustrated in Fig.~\ref{fig:RIXSMobileHoleCartoon}(c), the second term represents a three-point correlation, arising from expanding one of the two intermediate states to the second order of $t_c$. The first term involves correlations among four sites, arising from expanding both intermediate states in Eq.~\eqref{eq:RIXS1} into the first order. These correlations are comparable in strength and share the same momentum dependence structure, making them inseparable from this integral. 

As discussed in Sec.~\ref{sec:entanglement}, these correlations form the RDM elements. To evaluate each of them, we can detune the incident photon energy around the resonance and design another integral. The spectral distribution in $\win$ is distinct for the first two terms (corresponding to three-point correlations) and the last term  (corresponding to four-point correlations) of Eq.~\eqref{eq:RIXSExpansion2}, reflected by the order of poles. Therefore, we consider the energy-weighted integral
\begin{eqnarray}\label{eq:I2WeightedIntegral}
    &&\iint  \frac{\win^2}{\win^2+\Gamma^2}  I^{(2)}(\kin,\dq,\win,\dw)  d\win  d\dw\nonumber\\
    &=&\frac{\pi t_c^2}{8N^2\Gamma^3}\sum_{ \sigma_1, \sigma_1^\prime} \mkern-4mu\sum_{\sigma_2,\sigma_2^\prime}\matElement\sum_{m,n}\sum_{\delta, \delta^\prime}
    e^{i\dq\cdot\left(\rbf_{m}-\rbf_{n}\right)+i\kin\cdot (\rbf_{\delta^\prime}-\rbf_{\delta})}
    \nonumber\\
    &&    \braket{{c_{n+\delta,\sigma_1^\prime}}{c^{\dagger}_{n\sigma_1} c_{m\sigma_2}}{c_{m+\delta^\prime,\sigma_2^\prime}^{\dagger}}}+O\left(\Gamma^{-5}\right)
\end{eqnarray}
This integral isolates the contributions from the four-point correlations in Eq.~\eqref{eq:I2DirectIntegral}. By substituting Eq.~\eqref{eq:I2WeightedIntegral} back into Eq.~\eqref{eq:I2DirectIntegral}, we can precisely determine the three-point correlations $\braket{{c_{n\sigma\mkern-2mu_1^\prime}}\mkern-2mu{c^{\dagger}_{n\sigma\mkern-2mu_1}\mkern-3mu c_{m\sigma\mkern-2mu_2}}\mkern-2mu{c_{m\mkern-2mu-\mkern-2mu\delta\mkern-2mu+\mkern-2mu\delta^\prime\mkern-4mu,\sigma\mkern-2mu_2^\prime}^{\dagger}}\mkern-2mu}$. Notably, this integral along the $\win$ axis requires knowledge of the inverse core-hole lifetime $\Gamma$, which can be obtained either by fitting the XAS spectral shape or by simulations.

\begin{figure}[!t]
   \centering
    \includegraphics[width=\linewidth]{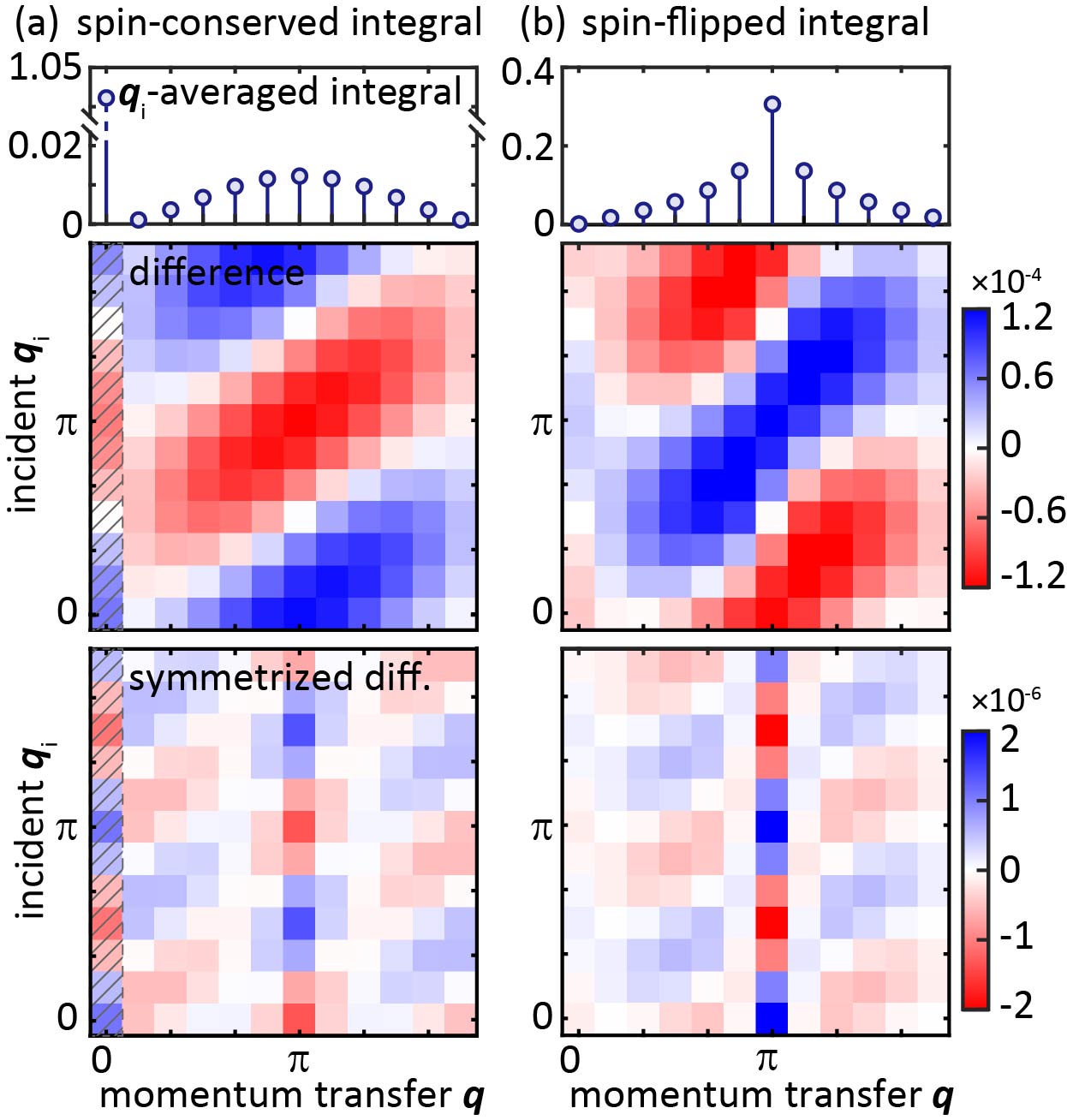}\vspace*{-3mm}
    \caption{\label{fig:integralDiff} 
    Integrated RIXS cross sections for (a) {the} spin-conserved channel and (b) {the} spin-flipped channel, obtained in a single-band Hubbard model. Top panels show $\kin$-averaged cross sections, serving as approximations to the charge and spin structure factors. Middle panels depict the difference betweeen the integrated RIXS spectra $I\left(\dq,\kin\right)$ and the averages, predominantly reflecting the $\kin$-dependent $I^{(1)}$ and $I^{(2)}$ in Eqs.~\eqref{eq:I1DirectIntegral} and \eqref{eq:I2DirectIntegral}. Bottom panels display the $t_c$-symmetrized integrals by averaging results from $t_c=0.1t$ and $-0.1t$, eliminating the $I^{(1)}$ contributions. A factor of 10 is divided from intensities in the shaded areas.\vspace*{-0mm}
}
\end{figure}

A real experiment measures the total spectral weight rather than a specific order. Therefore, to separate $I^{(0)}$, $I^{(1)}$, and $I^{(2)}$, we should leverage their distinct momentum dependence instead of their dependence on $t_c$. Using again the half-filled Hubbard model as an example, similar to those in Figs.~\ref{fig:RIXSspec_SC} and \ref{fig:RIXSspec_diff}, we analyze the $\kin$ and $\dq$ dependence of the integral of simulated RIXS cross sections for spin-conserved and spin-flipped channels [see Fig.~\ref{fig:integralDiff}]. To benchmark the correlations in the UCL limit, we extend the $\Gamma$ to 10\,$t$. As we discussed earlier, the zeroth-order term $I^{(0)}\left(\dq,\win,\dw\right)$ exhibits no $\kin$ dependence. In scenarios with minimal core-hole hopping, such as $t_c=0.1t$ used here, it can be filtered out by averaging over the incident momentum $\kin$. Hence, the $\kin$-averaged integrals in the top panels of Fig.~\ref{fig:integralDiff} reflect the charge and spin structure factors, as described in Eq.~\eqref{eq:integralImmobile}. In our half-filled Hubbard model example, the gapped charge excitations lead to minimal spin-conserved integrals at finite $\dq$. Notably, the integral at $\dq=\pi$ is relatively more pronounced due to nearest-neighbor doublon-hole fluctuations. Conversely, the spin-flipped integrals diverge logarithmically as $\dq$ approaches the nesting wavevector $\pi$ at zero temperature, a consequence of the quasi-long-range order\,\cite{voit1995one}, which is bounded in a finite cluster. 

The middle panels in Fig.~\ref{fig:integralDiff} highlight the two-momentum dependency of the integrated cross sections, contrasting with the $\kin$-averaged integrals. Notably, a single-cycle oscillation along the $\kin$ axis is evident, primarily arising from the integral of the first-order spectrum $I^{(1)}(\kin,\dq,\win,\dw)$ as detailed in Eq.~\eqref{eq:I1DirectIntegral}. This integral, consisting of several terms each marked by a nearest-neighbor index $\delta$, contains a phase factor $e^{i\kin\cdot \rbf_{\delta}}$. Although the zeroth-order spectrum $I^{(0)}$ is the primary contributor to the overall cross section, this distinct $\kin$-periodicity of $I^{(1)}$ and higher-order contributions allows us to isolate them from the $\kin$-independent background.

The integral of the second-order spectrum $I^{(2)}$ in Eq.~\eqref{eq:I2DirectIntegral}, while {it} also exhibits a $\kin$ dependency, involves a two-step core-hole hopping process and is dictated by a phase factor $e^{i\kin\cdot \left(\rbf_{\delta^\prime}-\rbf_{\delta}\right)}$ distinct from $I^{(1)}$. The phase factor involves a difference of two nearest-neighbor indices $\delta$ and $\delta^\prime$. If the lattice is bipartite, it does not overlap with that in $I^{(1)}$ and can be separated by momentum modulation. To enhance visualization of $I^{(2)}$'s momentum dependency, we present symmetrized RIXS integrals, averaging over spectra with $t_c=\pm0.1t$. This symmetrization effectively nullifies contributions from $I^{(1)}$ [see Eq.~\eqref{eq:I2DirectIntegral}]. (Note that this symmetrization can be executed only in simulations for illustrative purposes.) As shown in the bottom panels in Fig.~\ref{fig:integralDiff}, these symmetrized differential integrals exhibit a $\kin$-{periodicity} dominated by $\pi$, corresponding to the unique phase factor in $I^{(2)}$.

Therefore, to evaluate those three- and four-point correlations from a RIXS cross section, it is natural to employ Fourier factors like $e^{i\kin\cdot\left(\rbf_{\delta}-\rbf_{\delta^\prime}\right)}$ to isolate $I^{(2)}$ while filtering out the more prominent $I^{(0)}$ and $I^{(1)}$. Focusing on terms where $\rbf_{\delta}\neq \rbf_{\delta'}$ in integral Eq.~\eqref{eq:I2WeightedIntegral}, we can extract the real-space four-fermion correlation functions using a Fourier transform over both $\dq$ and $\kin$:
\begin{eqnarray}\label{eq:4pCorrelation}
        &&\frac{2\Gamma^3}{\pi t_c^2}\mkern-4mu\sum_{\dq, \kin}\mkern-3mu e^{i\dq \cdot\dbf-i \kin\cdot (\rbf_{\delta'}-\rbf_{\delta})}\mkern-7mu\iint\mkern-4mu  \frac{\win^2}{\win^2\mkern-2mu+\mkern-2mu\Gamma^2}  I(\kin,\dq,\win,\dw)  d\win  d\dw\nonumber\nonumber\\
        &\approx&\mkern-2mu\sum_{ \sigma_1, \sigma_1^\prime} \mkern-4mu\sum_{\sigma_2,\sigma_2^\prime}\mkern-2mu \matElement\mkern-2mu\braket{{c_{n+\delta,\sigma_1^\prime}}{c^{\dagger}_{n\sigma_1} c_{n+d,\sigma_2}}{c_{n+d+\delta^\prime,\sigma_2^\prime}^{\dagger}}}\,.
\end{eqnarray}
Here, $n$ can represent any site index in a translationally invariant system, and $\dbf = \rbf_m-\rbf_n$ (with corresponding index $d=m-n$) denotes the distance between the two pairs of creation-annihilation operators. Following the same Fourier transform, we then extract the sum of all second-order terms from the direct integral Eq.~\eqref{eq:I2DirectIntegral}:
\begin{eqnarray}\label{eq:4pCorrelation2}
    &&\frac{2\Gamma^3}{\pi t_c^2}\textrm{Re}\sum_{\dq, \kin} e^{i\dq \cdot\dbf-i \kin\cdot (\rbf_{\delta'}-\rbf_{\delta})}\iint    I(\kin,\dq,\win,\dw)  d\win  d\dw\nonumber\\
    &\approx&\sum_{ \sigma_1, \sigma_1^\prime} \mkern-2mu\sum_{\sigma_2,\sigma_2^\prime}\mkern-2mu \matElement\mkern-4mu \Big[ \mkern-4mu\braket{{c_{n+\delta,\sigma_1^\prime}}\mkern-2mu{c^{\dagger}_{n\sigma_1}\mkern-2mu c_{n\mkern-1mu+\mkern-1mud,\sigma_2}}\mkern-2mu{c_{n\mkern-1mu+\mkern-1mud\mkern-1mu+\mkern-1mu\delta^\prime,\sigma_2^\prime}^{\dagger}}\mkern-1mu}\nonumber\\
    &&-\braket{{c_{n\sigma_1^\prime}}\mkern-2mu{c^{\dagger}_{n\sigma_1}\mkern-2muc_{n\mkern-1mu+\mkern-1mud,\sigma_2}}\mkern-2mu{c_{n\mkern-1mu+\mkern-1mud\mkern-1mu-\mkern-1mu\delta\mkern-1mu+\mkern-1mu\delta\mkern-2mu^\prime\mkern-4mu,\sigma_2^\prime}^{\dagger}}}\mkern-2mu\Big]\,\mkern-2mu,
\end{eqnarray}
Substituting Eq.~\eqref{eq:4pCorrelation} into Eq.~\eqref{eq:4pCorrelation2}, we obtain the three-point correlations. Special consideration is necessary when $\rbf_{\delta}=\rbf_{\delta^\prime}$ or $\rbf_{\delta}-\rbf_{\delta^\prime}$ aligns with the nearest-neighbor vectors in a nonbipartite lattice. In these cases, parts of $I^{(2)}$ become indistinguishable from $I^{(0)}$ and $I^{(1)}$ in terms of the $\kin$-dependent phase factor. These scenarios limit the measurability of correlations like $ \braket{{c_{n+\delta,\sigma_1^\prime}}{c^{\dagger}_{n\sigma_1} c_{m\sigma_2}}{c_{m+\delta^\prime,\sigma_2^\prime}^{\dagger}}}$ through RIXS spectrum expansion in the small $t_c$ limit. 

{Notably, since $I^{(1)}$ and $I^{(2)}$ depend explicitly on $\kin$, preventing free tuning of $\win$ for a fixed $\kin$, unlike Eq.~\eqref{eq:integralImmobile}. To experimentally implement the integrals in Eqs.~\eqref{eq:I1DirectIntegral}--\eqref{eq:4pCorrelation2}, it is necessary to assume that effective electronic states are confined to two-dimensional planes or one-dimensional chains -- a scenario commonly found in correlated materials due to the Jahn–Teller effect. In this context, $\kin$ represents the in-plane projection of the incident momentum, and the incident angle can be adjusted to independently control $\win$ and the in-plane $\kin$.}

The correlations derived in Eqs.~\eqref{eq:4pCorrelation} and \eqref{eq:4pCorrelation2} mix various spin configurations, whose weights are determined by the matrix element $\matElement$ associated with specific polarization geometries, as discussed in Sec.~\ref{sec:RIXS:overview}. To discern correlations with particular spin configurations, we consider the linear combination of multiple scattering channels. Especially for systems preserving the spin $\mathrm{SU}(2)$ symmetry, the three spin-flip channels can be treated equivalently. Therefore, spin-specific correlations for these high-symmetry systems can be isolated using only two polarization configurations. For example, correlations with all aligned spins can be obtained by combining spin-conserved and spin-flipped channels:
\begin{eqnarray}\label{eq:spinchannels1}
    &&\braket{{c_{\cdot\uparrow}}{c^{\dagger}_{\cdot\uparrow} c_{\cdot\uparrow}}{c_{\cdot\uparrow}^{\dagger}}}\mkern-2mu=\mkern-2mu\braket{{c_{\cdot\downarrow}}{c^{\dagger}_{\cdot\downarrow} c_{\cdot\downarrow}}{c_{\cdot\downarrow}^{\dagger}}}\mkern-2mu\nonumber\\
    &=&\mkern-2mu\frac{1}{4}\mkern-4mu\sum_{\sigma_1,\sigma_2}\mkern-2mu\sum_{ \sigma_1^\prime,\sigma_2^\prime}\mkern-2mu\mkern-2mu\left(\sigma_0\mkern-3mu\otimes\mkern-3mu\sigma_0\mkern-2mu+\mkern-2mu\sigma_z\mkern-3mu\otimes\mkern-3mu\sigma_z\right)\braket{{c_{\cdot\sigma_1^\prime}}{c^{\dagger}_{\cdot\sigma_1} c_{\cdot\sigma_2}}{c_{\cdot\sigma_2^\prime}^{\dagger}}}.
\end{eqnarray}
Here, for brevity, spatial coordinates in Eqs.~\eqref{eq:4pCorrelation} and \eqref{eq:4pCorrelation2} are omitted, simplifying the notation, such as $c_{n\uparrow}$ to $c_{\cdot\uparrow}$. Similarly, the off-diagonal spin-conserved correlations are
\begin{eqnarray}\label{eq:spinchannels2}
    &&\braket{{c_{\cdot\uparrow}}{c^{\dagger}_{\cdot\uparrow} c_{\cdot\downarrow}}{c_{\cdot\downarrow}^{\dagger}}}\mkern-2mu=\mkern-2mu\braket{{c_{\cdot\downarrow}}{c^{\dagger}_{\cdot\downarrow} c_{\cdot\uparrow}}{c_{\cdot\uparrow}^{\dagger}}}\mkern-2mu\nonumber\\
    &=&\mkern-2mu\frac{1}{4}\mkern-4mu\sum_{\sigma_1,\sigma_2}\mkern-2mu\sum_{ \sigma_1^\prime,\sigma_2^\prime}\mkern-4mu\left(\sigma_0\mkern-3mu\otimes\mkern-3mu\sigma_0\mkern-2mu-\mkern-2mu\sigma_z\mkern-3mu\otimes\mkern-3mu\sigma_z\right)\braket{{c_{\cdot\sigma_1^\prime}}{c^{\dagger}_{\cdot\sigma_1} c_{\cdot\sigma_2}}{c_{\cdot\sigma_2^\prime}^{\dagger}}},
\end{eqnarray}
and spin-flipped correlations are
\begin{eqnarray}\label{eq:spinchannels3}
    &&\braket{{c_{\cdot\uparrow}}{c^{\dagger}_{\cdot\downarrow} c_{\cdot\downarrow}}{c_{\cdot\uparrow}^{\dagger}}}\mkern-2mu=\mkern-2mu\braket{{c_{\cdot\downarrow}}{c^{\dagger}_{\cdot\uparrow} c_{\cdot\uparrow}}{c_{\cdot\downarrow}^{\dagger}}}\mkern-2mu\nonumber\\
    &=& \mkern-2mu\frac12\mkern-4mu\left[\mkern-2mu\braket{{c_{\cdot\mkern-1mu\uparrow}}\mkern-1mu{c^{\dagger}_{\cdot\mkern-1mu\downarrow}\mkern-1mu c_{\cdot\mkern-1mu\downarrow}}\mkern-1mu{c_{\cdot\mkern-1mu\uparrow}^{\dagger}}\mkern-2mu}\mkern-2mu
    +\mkern-2mu\braket{{c_{\cdot\mkern-1mu\downarrow}}\mkern-1mu{c^{\dagger}_{\cdot\mkern-1mu\uparrow} \mkern-1muc_{\cdot\mkern-1mu\uparrow}}\mkern-1mu{c_{\cdot\mkern-1mu\downarrow}^{\dagger}}\mkern-2mu} \mkern-2mu
    + \mkern-2mu\braket{{c_{\cdot\mkern-1mu\uparrow}}\mkern-1mu{c^{\dagger}_{\cdot\mkern-1mu\downarrow} \mkern-1muc_{\cdot\mkern-1mu\uparrow}}\mkern-1mu{c_{\cdot\mkern-1mu\downarrow}^{\dagger}}\mkern-2mu}\mkern-2mu
    +\mkern-2mu\braket{{c_{\cdot\mkern-1mu\downarrow}}\mkern-1mu{c^{\dagger}_{\cdot\mkern-1mu\uparrow} \mkern-1muc_{\cdot\mkern-1mu\downarrow}}\mkern-1mu{c_{\cdot\mkern-1mu\uparrow}^{\dagger}}\mkern-2mu}\mkern-2mu\right] \nonumber\\
    &=&\mkern-2mu\frac{1}{2}\mkern-4mu\sum_{\sigma_1,\sigma_2}\mkern-2mu\sum_{ \sigma_1^\prime,\sigma_2^\prime}\mkern-2mu\left(\sigma_x\mkern-2mu\otimes\mkern-2mu\sigma_x\right)\braket{{c_{\cdot\sigma_1^\prime}}{c^{\dagger}_{\cdot\sigma_1} c_{\cdot\sigma_2}}{c_{\cdot\sigma_2^\prime}^{\dagger}}}\,.
\end{eqnarray}
Note that the last two terms in the second row in Eq.~\eqref{eq:spinchannels3} vanish for systems with time-reversal symmetry and conserved particle number. In systems with less symmetry, it is necessary to consider three distinct channels of the spin-flipped matrix elements. 

\begin{figure}[!b]
   \centering
    \includegraphics[width=\linewidth]{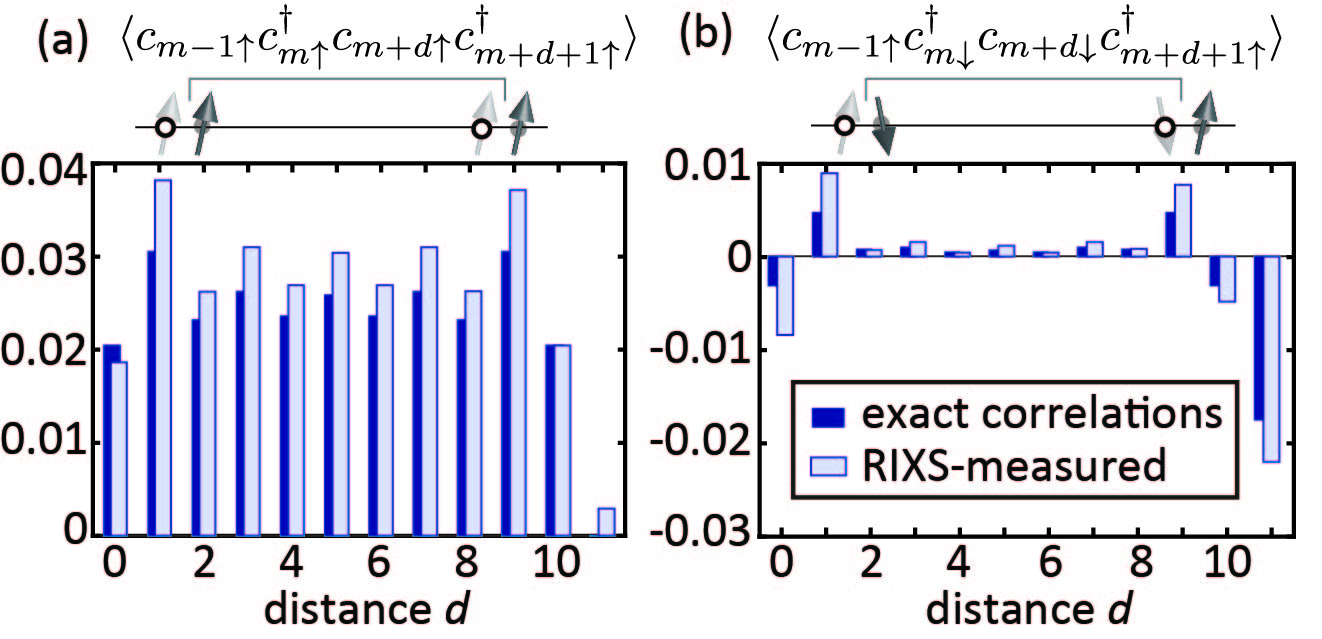}\vspace*{-3mm}
    \caption{\label{fig:realSpaceCorrDistr} 
    Real-space spin-specific correlations for (a) spin-conserved and (b) spin-flipped configurations as a function of spatial distance $d$. The dark blue bars represent correlations directly obtained from the ground-state wavefunction via ED simulations, while the light-blue bars represent correlations evaluated using the integrated RIXS cross section specified in  Eqs.~\eqref{eq:4pCorrelation} and \eqref{eq:4pCorrelation2} and the superposition of polarization geometries outlined in Eqs.~\eqref{eq:spinchannels1}--\eqref{eq:spinchannels3}. 
}
\end{figure}

By applying the above procedure, we can extract real-space, spin-specific correlation functions from the integrated RIXS spectra. As shown by the light-blue bars in Fig.~\ref{fig:realSpaceCorrDistr}, the spin-conserved correlations are positive semidefinite, while the spin-flipped ones exhibit negative components and are less pronounced for the $S=0$ ground state. To assess the accuracy of these RIXS-derived correlations, we benchmark them against the exact four-point correlations computed directly from the ground-state wavefunctions. This comparison shows a high level of consistency across all distances, with an average deviation of 15\%. This deviation arises from the finite core-hole lifetime and broadening in RIXS. As discussed in Sec.~\ref{sec:examples}, this overshooting deviation does not compromise the accuracy of the entanglement witness. It is noteworthy that the exact four-point correlations presented in Fig.~\ref{fig:realSpaceCorrDistr} exhibit a reflection symmetry about $d=N-1$, a consequence of the particle-hole symmetry in the ground state of a half-filled Hubbard model with a periodic boundary. However, this symmetry is not perfectly replicated in the correlations obtained from the RIXS integrals. This discrepancy is attributed to the inclusion of excited states in RIXS, which introduce an additional electron into the valence band, thereby disrupting the particle-hole symmetry.

\subsection{Connected correlations by differentiating RIXS and ARPES}\label{sec:RIXSconnected}

In the previous subsection, we demonstrated that RIXS can extract high-order (three-point and four-point) correlations in systems of indistinguishable fermions. While these correlations are useful for analyzing order instabilities, they do not directly indicate fermion entanglement, which represents the intrinsic complexity of the many-body wavefunction. In the context of indistinguishable fermions, the Slater determinant wavefunction -- also known as the Gaussian state -- serves as the baseline for ``separable'' states without entanglement (see detailed definitions and discussions in Sec.~\ref{sec:correlations:fermions})\,\cite{ghirardi2002entanglement, ghirardi2004general, plastino2009separability, amico2008entanglement}. These multipoint correlations are typically nonzero in Slater determinants and can be relatively large when a mean-field order is established. Thus, it is necessary to subtract the lower-order disconnected parts from the multipoint correlations. According to the Wick's theorem, a general four-point correlation $\braket{c_i c_j c_k^{\dagger} c_l^{\dagger}}$ reduces to $\braket{c_i c_l^{\dagger}}\braket{c_j c_k^{\dagger}}-\braket{c_i c_k^{\dagger}} \braket{c_j c_l^{\dagger}}$ for a Slater determinant in the canonical ensemble. Therefore, by subtracting the disconnected parts from the correlations, we obtain the connected (cumulant) correlations: 
\begin{eqnarray}\label{eq:connected3p}
    &&\braket{{c_{n\sigma_1^\prime}}c_{n+d,\sigma_2}{c^{\dagger}_{n\sigma_1}}{c_{n+d-\delta+\delta^\prime,\sigma_2^\prime}^{\dagger}}}^\con\nonumber\\&=&\left(\delta_{0,d}\delta_{\sigma_1,\sigma_2}-\braket{c_{n+d,\sigma_2}{c^{\dagger}_{n\sigma_1}}}\right)\braket{{c_{n\sigma_1^\prime}}{c_{n+d-\delta+\delta^\prime,\sigma_2^\prime}^{\dagger}}}\nonumber\\
    &-&\braket{{c_{n\sigma_1^\prime}}{c^{\dagger}_{n\sigma_1}c_{n+d,\sigma_2}}{c_{n+d-\delta+\delta^\prime,\sigma_2^\prime}^{\dagger}}}\nonumber\\
    &+&\braket{{c_{n\sigma_1^\prime}}{c^{\dagger}_{n\sigma_1}}}\braket{c_{n+d,\sigma_2}{c_{n+d-\delta+\delta^\prime,\sigma_2^\prime}^{\dagger}}}
\end{eqnarray}
and
\begin{eqnarray}\label{eq:connected4p}
    &&\braket{{c_{n+\delta,\sigma_1^\prime}} c_{n+d,\sigma_2}{c^{\dagger}_{n\sigma_1}} {c_{n+d+\delta^\prime,\sigma_2^\prime}^{\dagger}}}^\con\nonumber\\
    &=& \left(\delta_{0,d}\delta_{\sigma_1,\sigma_2}-\braket{c_{n+d,\sigma_2}{c^{\dagger}_{n\sigma_1}}}\right)\braket{{c_{n+\delta,\sigma_1^\prime}}{c_{n+d+\delta^\prime,\sigma_2^\prime}^{\dagger}}}\nonumber\\&-&\braket{{c_{n+\delta,\sigma_1^\prime}}c^{\dagger}_{n\sigma_1} c_{n+d,\sigma_2}{c_{n+d+\delta^\prime,\sigma_2^\prime}^{\dagger}}}\nonumber\\
    &+&\braket{{c_{n+\delta,\sigma_1^\prime}} {c^{\dagger}_{n\sigma_1}}}\braket{c_{n+d,\sigma_2}{c_{n+d+\delta^\prime,\sigma_2^\prime}^{\dagger}}}
\end{eqnarray}
These connected correlations vanish for any Slater determinants. Therefore, their strengths can be used to measure entanglement, as discussed in Sec.~\ref{sec:entanglement}. Note that we employ the antinormal order for the connected correlations in Eqs.~\eqref{eq:4pCorrelation} and \eqref{eq:4pCorrelation2}. 

Connected multipoint correlations have been extensively utilized in ultracold atoms to discern entanglement and properties of many-body wavefunctions\,\cite{schweigler2017experimental, hilker2017revealing,   salomon2019direct, koepsell2019imaging, vijayan2020time, prufer2020experimental, zache2020extracting, koepsell2021microscopic,wang2021higher,bohrdt2021dominant}. However, the nonlinear parts of x-ray scattering yield the bare multipoint correlations, instead of connected correlations. To determine their disconnected counterparts, we turn to another solid-state spectroscopy technique -- ARPES. The zero-temperature ARPES spectrum for a specific band is given as
\begin{eqnarray}
\mkern-18muA\mkern-3mu\left(\kbf,\mkern-1mu\omega \right)\mkern-2mu=\mkern-12mu\sum_{m,n,\sigma}\mkern-8mu\frac{e^{i\kbf\cdot (\rbf\mkern-2mu_n\mkern-2mu-\mkern-2mu\rbf\mkern-2mu_m)}}{\pi N^2} \mathrm{Im} \Big\langle \mkern-4muc^\dagger_{n\sigma}\mkern-2mu\frac{1}{\mkern-1muE_G\mkern-3mu-\mkern-3mu\mathcal{H}\mkern-3mu-\omega\mkern-3mu-\mkern-3mui0_+\mkern-3mu}\mkern-2muc_{m\sigma} \mkern-5mu\Big\rangle.
\end{eqnarray}
Here, photoemission matrix elements are omitted for brevity. When necessary, the spin flavors can be measured separately through spin ARPES. Using (spin) ARPES, it is easy to show that
\begin{eqnarray}
\sum_\kbf e^{i\kbf\cdot\dbf}\int d\dw A_\sigma\left(\kbf,\mkern-1mu\omega \right) = \braket{c_{n\sigma}^\dagger c_{n+d,\sigma}}\,.
\end{eqnarray}
These integrals form the disconnected parts in Eqs.~\eqref{eq:connected3p} and \eqref{eq:connected4p}. In systems preserving SU(2) symmetry, the disconnected parts are identical for both spin flavors and can be directly evaluated from ARPES without spin resolution.

\section{Entanglement extracted from correlations}\label{sec:entanglement}

Many-body entanglement can be witnessed by two-point correlations such as the two-tangle\,\cite{wong2001potential} and spin QFI\,\cite{hyllus2012fisher}, which are accessible via solid-state spectroscopy techniques\,\cite{hauke2016measuring}. These methods, extendable to local operators in fermionic modes\,\cite{almeida2021from}, utilize fluctuations to estimate a lower bound for the entanglement depth of a many-body wavefunction\,\cite{pezze2009entanglement}. However, these boundaries are determined by mapping correlations to isolated qubits with separable modes and rely on \textit{a priori} knowledge of the dominant bosonic excitations in the material. In general many-electron systems, especially those without local magnetic moments or charge densities, the orbital modes that can be occupied or unoccupied are chosen as bases without any preference. Entanglement among electrons cannot be properly measured based on these arbitrarily defined modes. Moreover, the entanglement of indistinguishable fermions should be evaluated in a manner such that the inherent anticommutation relations of fermions do not contribute to entanglement\,\cite{ghirardi2002entanglement, ghirardi2004general, plastino2009separability, amico2008entanglement}. These conditions necessitate an entanglement witness that is invariant under unitary basis transformations and vanishes for Slater determinants. Previous studies have suggested that the Slater rank \cite{schliemann2001quantum,eckert2002quantum,hill1997entanglement} and concurrence\,\cite{rungta2001universal,eckert2002quantum,carvalho2004decoherence} are basis-invariant measures for entanglement in two-particle systems. However, they cannot be directly measured by solid-state spectroscopy, similar to entanglement entropy. Additionally, their computational complexity scales exponentially with system size.

The identification of connected three-point and four-point correlation functions using the nonlinear effects in RIXS provides a potential avenue for witnessing entanglement in indistinguishable electrons. As we will show in this section, these multipoint correlations provide major elements in the 2CRDM, whose maximal eigenvalues can be used as a basis-independent entanglement witness to quantify the boundaries of entanglement depth.

\subsection{Entanglement in indistinguishable fermions}\label{sec:correlations:fermions}

For spin systems, a pure many-body state $\ket{\Psi}$ is defined as separable if it can be expressed as a direct product of single-spin states:
\begin{equation}\label{eq:pureproductstate}
    \ket{\Phi^{\rm (spin)}}=\ket{\phi_1}\otimes \ket{\phi_2}\otimes\cdots\otimes \ket{\phi_N},
\end{equation}
where $\ket{\phi_i}$ denotes the single-particle state for the $i$th mode. In fermionic many-body systems composed of indistinguishable particles, a many-body state must obey antisymmetry under permutations of modes (orbitals). Thus, a separable fermionic state can be expressed as a Slater determinant, or a Gaussian state\,\cite{schliemann2001quantum,eckert2002quantum}:
\begin{equation}\label{eq:SlaterDeterminant}
    \ket{\Psi}=e^{-i\sum_{ij} c_i^\dagger \xi_{ij} c_j} \prod_{j=1}^{N_e} c_j^\dagger\ket{0}= \prod_{j=1}^{N_e} \sum_k U_{jk}c_k^\dagger\ket{0}\,,
\end{equation}
where $\xi$ is a Hermitian matrix, and $U=e^{i\xi}$ is the unitary transformation acting on the fermionic basis. For convenience, the spin indices are absorbed into the orbital indices within this section. Here, we exclusively consider states with a conserved particle number $N_e$, although Gaussian states are generally definable without distinguishing between particles and holes\,\cite{weedbrook2012gaussian}.

\begin{figure}[!t]
    \centering
   \includegraphics[width=8.5cm]{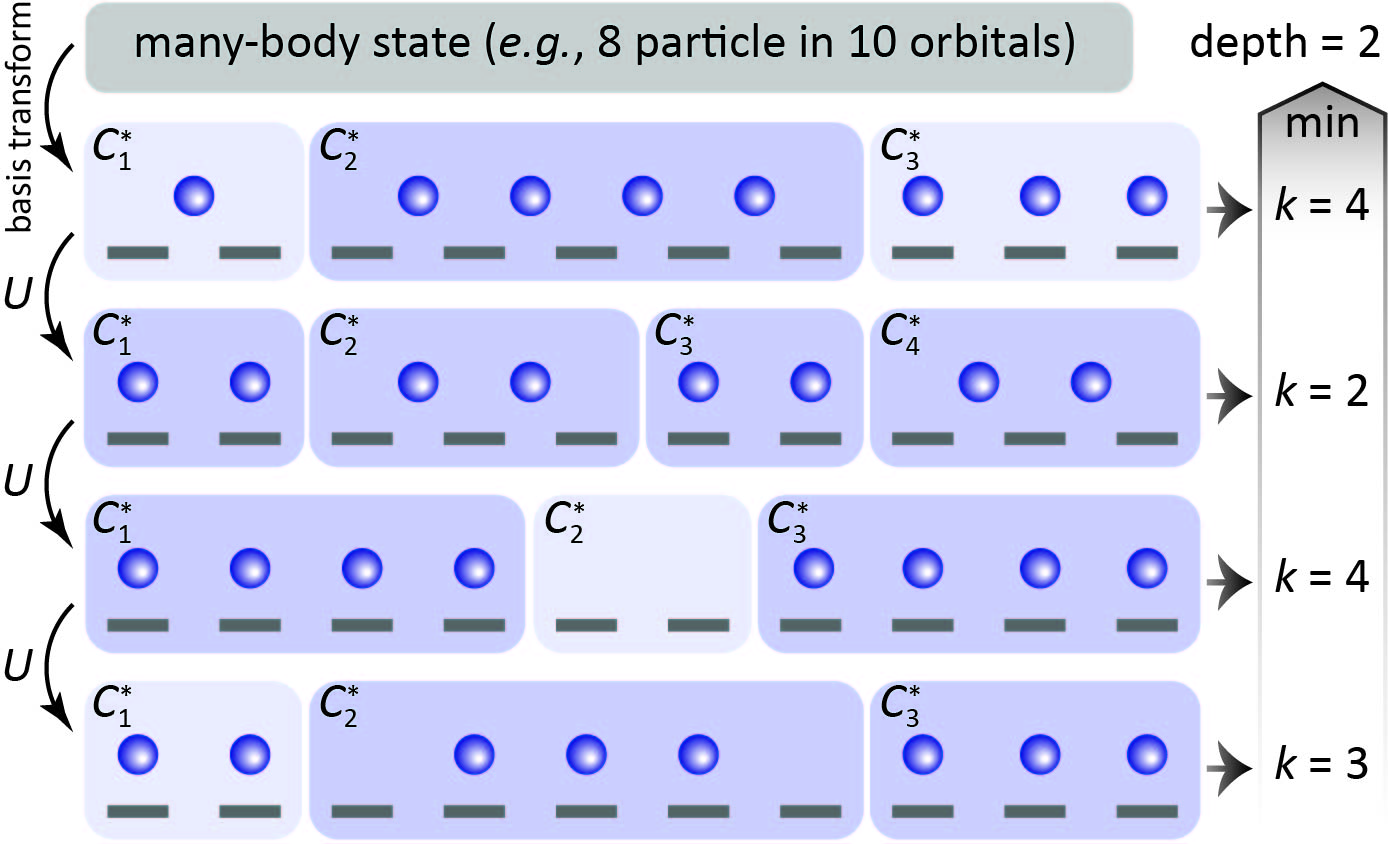}
    \caption{\label{fig:entanglementDepth} Schematic illustrating the genuinely entangled partitioning for fermionic states, highlighting the ambiguity in basis selection. For each chosen basis, the producibility is determined by identifying the maximum particle numbers among all irreducible blocks. The entanglement depth of the many-body state is subsequently defined as the minimum value among all basis selections, according to Eq.~\eqref{eq:entanglementDepth}.}    
\end{figure}

When a state cannot be expressed as a separable state through any single-particle basis transformations, it is an entangled state. To further classify different entanglement depths, the notion of $k$-producibility has been widely used in spin states and more broadly in quantum states with distinguishable modes\,\cite{toth2012multipartite, hyllus2012fisher, hauke2016measuring}. Specifically, a $k$-producible spin state is expressed in the same direct-product form as Eq.~\eqref{eq:pureproductstate}, but each block wavefunction $\ket{\phi_i}$ includes no more than $k$ spin modes. 

To generalize this notion to indistinguishable fermions, we adopt the framework introduced in Ref.~\onlinecite{almeida2021from} and consider the partitioning of second-quantized fermionic operators. Additionally, we allow arbitrary (single-particle) basis transformations on top of this framework to address the ambiguity of the orbital basis. In this context, any fermionic many-body state can be written as:
\begin{equation}\label{eq:entangledState}
    \ket{\Psi} = e^{-i\sum_{ij} c_i^\dagger \xi_{ij} c_j} 
 C_1^\star C_2^\star\cdots C_M^\star\ket{0}\,.
\end{equation}
For a fixed orbital basis determined by the Gaussian operator, the wavefunction decomposes into several block creation operators. Each $C_p^\star$ contains irreducible $N_p$-electron creation operators in a single-particle subspace $\mathcal{M}_p$, formed by a partition of fermionic orbitals: 
\begin{equation}\label{eq:gmeblock}
    C_p^\star = \sum_{\eta_p} \phi_p^\star(\eta_p) \prod_{m \in \mathcal{M}_p} (c_m^\dagger)^{\eta_p(m)}\,,
\end{equation}
where $\eta_p(m)$ denotes the occupation number (0 or 1) of the orbital $m$, with irreducible coefficients $\phi_p ^\star (\eta_p)$. Irreducibility here means that $C_p$ cannot be factorized into a product of two blocks of creation operators with nonzero fermions. All these partition subspaces constitute the entire single-particle Hilbert space, i.e., $\mathcal{M}_1\bigoplus \mathcal{M}_2\bigoplus \cdots \bigoplus \mathcal{M}_M$. The entanglement depth of a fermionic state is then defined as:
\begin{equation}\label{eq:entanglementDepth}
    \min_{\xi}\big\{ \max(N_1,N_2,\cdots, N_M)  \big\}\,
\end{equation}
where $N_p$ is the number of particles in the $p$th partition $\mathcal M_p$. This minimax definition of Eq.~\eqref{eq:entanglementDepth} prevents the misclassification of a state due to an inappropriate basis selection, such as the Fermi sea in a real-space basis. Notably, different from Ref.~\onlinecite{almeida2021from}, we define the depth of each block using its particle number (or hole number, whichever is smaller) instead of the orbital number, as a finite number of fermions can occupy an infinite number of bases. Figure~\ref{fig:entanglementDepth} shows an example of determining entanglement depth using the minimax definition. A state is called $k$-producible if $k$ is no less than the entanglement depth defined in Eq.~\eqref{eq:entanglementDepth}. Obviously, a Gaussian state in Eq.~\eqref{eq:SlaterDeterminant} is a 1-producible state in this context.

\subsection{Basis-invariant measure based on reduced density matrix}\label{sec:entanglement:RDM}
Directly identifying entanglement and quantifying the entanglement depth using Eq.~\eqref{eq:entanglementDepth} requires traversing all single-particle basis transformations via $U=e^{i\xi}$, making it a challenging task for most many-body states. A more practical approach involves considering observables that are invariant under basis transformations. An observable that effectively quantifies the difference between a state and the complete set of separable states can serve as a measure of entanglement\,\footnote{Here, we are not referring to the entanglement measure defined in the context of quantum teleportation, which requires monotonicity under local operations and classical communication\,\cite{horodecki2009quantum}.}.

The RDM is a crucial tool for characterizing orbital entanglement entropy and quantum mutual information\,\cite{rissler2006measuring,boguslawski2015orbital,mukherjee2001irreducible, kutzelnigg2002irreducible, juhasz2006cumulant}. In particular, the two-particle RDM $O_{ijkl}=\braket{c_i c_j (c_k c_l)^{\dagger}}$ acts as a tensor metric for correlations in quantum many-body systems\,\cite{mazziotti1998contracted, mazziotti2012two, mazziotti2000complete, li2021geometric, mazziotti2016pure}, exhibiting the symmetry $O_{ijkl}=-O_{jikl}=-O_{ijlk}=O_{klij}^*$. Under a Gaussian transformation, which is equivalent to a unitary basis transformation $\tilde{c}_j=\sum_k U_{jk}c_k$, the two-particle RDM of a many-body state transforms as:
\begin{equation}\label{eq:RDMtransformation}
    O_{ijkl} = \sum_{m,n}\sum_{p,q}U_{im}U_{jn}O_{mnpq}U_{pk}^* U_{ql}^*\,.
\end{equation}
The RDM has been utilized to characterize pairing in two-particle systems, with its maximal eigenvalue bounded by 2 for unpaired states\,\cite{kraus2009pairing}.

To efficiently detect entanglement, an observable should yield zero for any separable state as defined in Eq.~\eqref{eq:SlaterDeterminant} and nonzero for any entangled state. Utilizing the connected part of four-point correlations derived in Sec.~\ref{sec:RIXSconnected}, one can further construct the 2CRDM\,\cite{mazziotti1998approximate, kutzelnigg1999cumulant}
\begin{equation}\label{eq:2crdm}
    O_{ijkl}^\con = \braket{ c_i c_j (c_k c_l)^{\dagger}}
    -\braket{c_i c_k^{\dagger}}\braket{c_j c_l^{\dagger}}+\braket{ c_i c_l^{\dagger}}\braket{ c_j c_k^{\dagger}}.
\end{equation}
The 2CRDM transforms in the same way as the RDM under Eq.~\eqref{eq:RDMtransformation}. The 2CRDM isolates the portion of the two-particle (four-point) RDM that cannot be reduced to products of separable one-particle (two-point) contributions. According to Wick's theorem, for a Gaussian state that conserves particle number, $O_{ijkl}^\con\equiv 0$, making it an efficient indicator of entangled fermionic states\,\cite{skolnik2013cumulant}. This is consistent with experimental intuitions that the difference between scattering spectra and the bare-bubble response function derived from single-particle spectra signals correlations.

The 2CRDM not only indicates the presence of entanglement but also connects to the irreducible partitions in Eq.~\eqref{eq:gmeblock}. Notably, the 2CRDM $O_{ijkl} ^\con$ is nonzero only if all indices ($i$, $j$, $k$, and $l$) belong to the same partition, owing to the subtraction of disconnected correlations. For example, if $i,l \in \mathcal{M}_p$ and $j,k \in \mathcal{M}_q\neq\mathcal{M}_p$, then we have $\braket{c_i c_j (c_k c_l)^{\dagger}}=-\braket{c_i c_l^{\dagger}}\braket{c_j c_k^{\dagger}}$, with $\braket{c_ic_k^{\dagger}}=\braket{c_j c_k^{\dagger}}=0$ and, thus,  $O^\con_{ijkl}=0$. This property holds true for all the other cases except when all indices belong to the same irreducible partition. For any partitions of fermionic orbitals, the tensor 2CRDM is composed of individual tensors within each partition:
\begin{equation}\label{eq:RDMblockDiagonal}
    O_{ijkl}^\con = \sum_p o^{(p)}_{ijkl}\  \mathbb{1}_{i,j,k,l\in\mathcal{M}_p}.
\end{equation}
Therefore, the 2CRDM provides insight into the sizes of partitions in Eq.~\eqref{eq:entangledState} under a specific basis selection, although the values within each block $o^{(p)}_{ijkl}$ depend on the expression of each $C_p^\star$ in Eq.~\eqref{eq:gmeblock}.

To quantify entanglement depth considering all possible single-particle basis transformations, as defined in Eqs.~\eqref{eq:entangledState}-\eqref{eq:entanglementDepth}, the metric observable derived from the 2CRDM must be invariant under unitary transformations. Unitary invariants of a general tensor can be represented as functions of high-order singular values\,\cite{tucker1966some,de2000multilinear}. In practice, we examine the eigenvalues of the matrices $O^\con_{(ij)(kl)}$ and $O^\con_{(ik)(jl)}$ by pairing the RDM indices. While the eigenvalues of these flattened matrices do not precisely match the singular values of the tensor, they remain unitary-invariant metrics under the transformation in Eq.~\eqref{eq:RDMtransformation} and are, thus, functions of the tensor's singular values. These eigenvalues, especially the maximal eigenvalue denoted as $\lambda_{\rm max}$, serve as a basis-invariant metric to quantify the strength of the 2CRDM or, equivalently, the distance from the nearest separable state (Gaussian states or Slater determinants)\,\cite{raeber2015large}. Because of the better properties regarding the upper bounds, we choose the matrix $O^\con_{(ik)(jl)}$ and its maximal eigenvalue $\lambda_{\rm max}$ as the entanglement witness. Discussions on the other form of the flattened matrix can be found in Appendix \ref{app:flattenedMatrix}. As a benchmark, we also compare $\lambda_{\rm max}$ with the high-order singular values obtained from the canonical polyadic decomposition and Tucker decomposition of the tensor form of 2CRDM in Appendix \ref{app:tensorDecomposition}, achieving qualitative consistency.

\begin{figure}[!t]
    \centering
   \includegraphics[width=8.5cm]{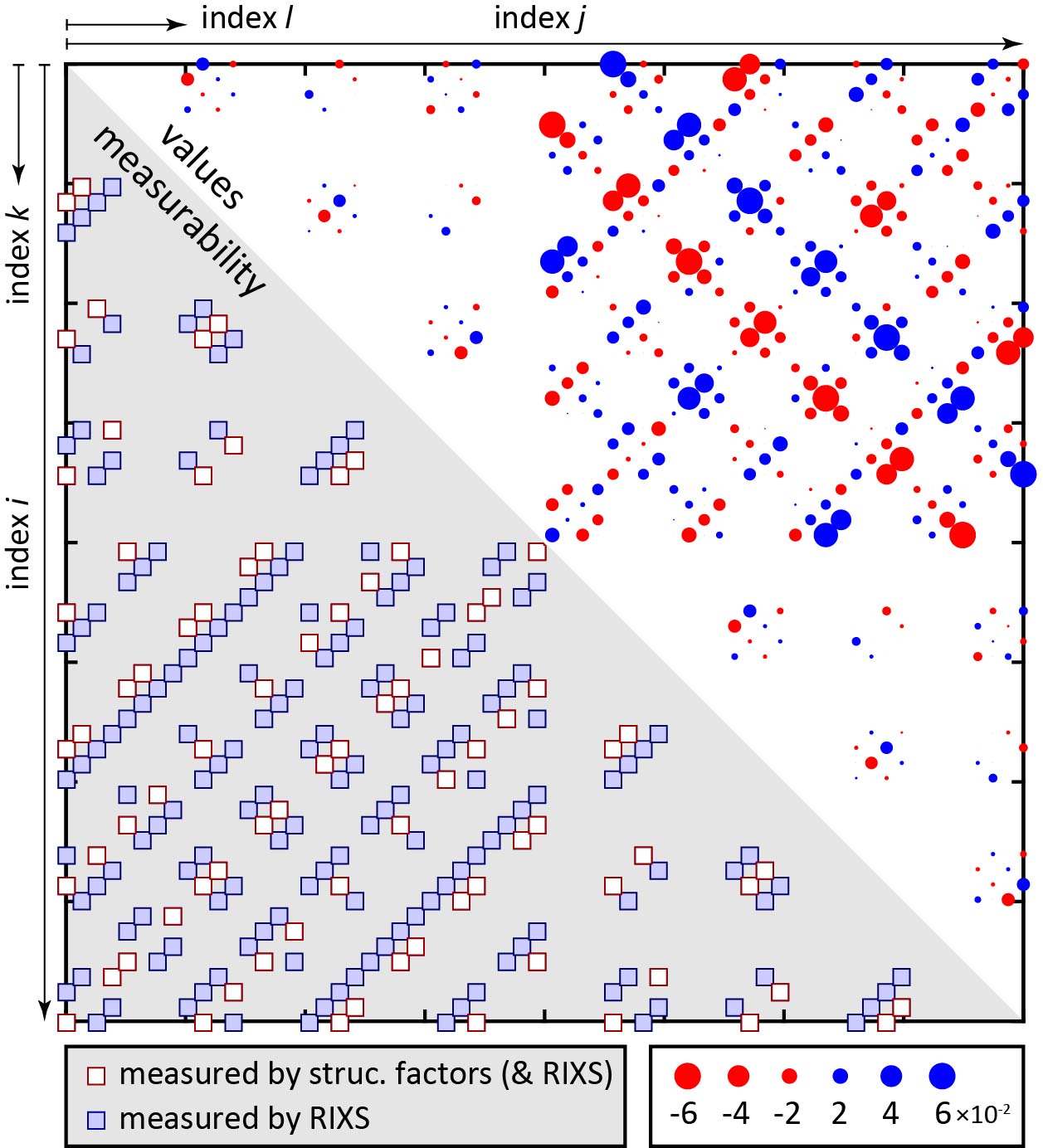}\vspace{-2mm}
    \caption{\label{fig:flattenedMatrix} Upper triangle: values and distribution of elements in the flattened matrix of the 2CRDM, with point sizes corresponding to the scale of the values. Lower triangle: matrix elements that can be measured by spin or charge structure factors (red) and by RIXS (red and blue). The matrix and its measurability are symmetric when transposing the rows ($i$ and $k$ indices) and columns ($j$ and $l$ indices). This example is drawn from the central four sites of a half-filled 1D Hubbard model in a 128-site chain.}    
\end{figure}

Figure \ref{fig:flattenedMatrix} presents an example of the flattened matrix $O^\con_{(ik)(jl)}$, obtained from a 1D extended Hubbard model, which will be discussed in detail in Sec.~\ref{sec:1DEHM}. Because of the anticommutation relations, the diagonal elements ($i=j$ and $k=l$) vanish. Moreover, only a small fraction of matrix elements hold significant values, resulting in an effectively sparse matrix, because correlations decay rapidly with distance in this system. 

While the full matrix can be computed using exact ground-state wavefunctions obtained from numerical simulations, its elements have different levels of accessibility in spectral measurements. Elements in certain rows and columns, like $i=k$ and $j=l$, correspond to two-point correlations, which can be derived from spin and charge structure factors. However, these elements alone are insufficient to cover all significant matrix elements. The leading-order expansion of RIXS spectral integrals evaluates the four-fermion correlations $\braket{{c_{n}}c_{n+d}{c^{\dagger}_{n}}{c_{n+d-\delta+\delta^\prime}^{\dagger}}}^\con$ and $\braket{{c_{n+\delta}} c_{n+d}{c^{\dagger}_{n}} {c_{n+d+\delta^\prime}^{\dagger}}}^\con$, in addition to these two-point correlations, as discussed in Sec.~\ref{sec:RIXS:correlatoins}. As shown in Fig.~\ref{fig:flattenedMatrix}, these matrix elements accessible from RIXS can cover most of the significant matrix elements in $O_{ijkl}^\con$. We further find that they provide a good approximation of the maximal eigenvalue $\lambda_{\rm max}$ of the flattened matrix. Specific models are employed to quantify errors resulting from this truncation and to validate the effectiveness of RIXS-accessible matrix elements in witnessing entanglement in Sec.~\ref{sec:examples}. 

\subsection{Boundaries and entanglement witness} \label{sec:entanglement:entangleBounds}
Since $O_{ijkl} ^\con$ becomes block diagonal according to the partitions of fermionic orbitals under a specific basis selection, the maximal eigenvalue $\lambda_{\rm max}$ is determined by
\begin{equation}
\lambda_{\rm max}(O_{ijkl} ^\con) = \max_{p}\, \lambda_{\rm max}(o^{(p)}_{ijkl})\,,
\end{equation}
where $o^{(p)}_{ijkl}$ denotes the tensor blocks in Eq.~\eqref{eq:RDMblockDiagonal}. Thus, if we can identify the maximal eigenvalue for all possible wavefunctions generated by a single irreducible $C_p^\star$ within a partition $\mathcal M_p$ containing $N_p$ particles, it satisfies
\begin{equation}
    \begin{aligned}
        \lambda_{\rm max}(O_{ijkl} ^\con) \leqslant \max_{p}\, \mu(N_p)\,,\quad \textrm{if}\ k\leqslant N_p\ \textrm{for}\ \forall p\,.\\
\mu(n) = \sup\mkern-2mu\left\{\textrm{eigenvalues of 2CRDM}\mkern-2mu: \forall\ket{\Psi_{n-\textrm{prod}}}\right\}
    \end{aligned}
\end{equation}
Given that $\lambda_{\rm max}$ remains invariant under (single-particle) basis transformations, it provides a lower bound for the entanglement depth of the many-body wavefunction. Hence, if the measured $\lambda_{\rm max}$ in a system exceeds $\mu(k)$, the wavefunction is at least $(k+1)$-producible.

For the entanglement witnessing approach to be practical, the single-partition upper bound $\mu(k)$ must increase monotonically with $k$. Obviously, the property of 2CRDM ensures that $\mu(1)=0$, suggesting that any nonzero $\lambda_{\rm max}$ signifies at least a 2-producible state. In the following subsections, we derive the upper bounds for more entangled states and prove that 
\begin{equation}\label{eq:kProdUpperBound}
\mu(k)=\frac12\left(k -\frac12\right),\quad \textrm{for}\ k>1\,.
\end{equation}
Note that the bound shown in Eq.~\eqref{eq:kProdUpperBound} has been previously examined in Ref.~\onlinecite{schouten2022large}, focusing on the half-filled Greenberger-Horne-Zeilinger (GHZ) state. The derivations that follow extend this upper bound to cover arbitrary rational
fillings and more general wavefunctions with varied entanglement depths. Notably, in the $O^\con_{(ik)(jl)}$ form of matrix flattening, the maximal matrix eigenvalue $\lambda_{\rm max}$ always exceeds the absolute value of the minimal matrix eigenvalue $\lambda_{\rm min}$, serving as a good approximation to the tensor's high-order singular values [see Appendix \ref{app:tensorDecomposition}].  

\subsubsection{Upper bound for 2-producible states}
In a system with two indistinguishable fermions, the general wavefunction can be expressed as:
\begin{equation}
\ket{\Psi_{2-\textrm{prod}}}=\sum_{i,j}\omega_{ij}c_i^{\dagger}c_j^{\dagger}\ket{0}\,,
\end{equation}
where $\omega$ is an antisymmetric matrix. It has been proven that a unitary basis transformation exists such that $U\omega U^T=\text{diag}\left(Z_1,\ldots,Z_r,0,\ldots,0\right)$\,\cite{schliemann2001quantum}. Each $Z_i$ is a $2\times 2$ matrix defined by:
\begin{equation}
    Z_i=\left(\begin{matrix}0 &z_i/2\\-z_i/2 & 0\end{matrix}\right),
\end{equation}
with $z_i>0$ and $\sum_i z_i^2=1$. The number of nonvanishing $z_i$s, denoted as $r$, is known as the Slater rank, which measures the complexity of a two-particle pairing state\,\cite{schliemann2001quantum}. Using the Slater decomposition, a $2$-producible wavefunction can be expressed as
\begin{equation}\label{eq:SlaterDecomposition}
    \ket{\Psi_{2-\textrm{prod}}}=\frac{e^{-i\sum_{ij} c_i^\dagger \xi_{ij} c_j}}{\sqrt{\sum_{j=1}^r |z_j|^2}} \sum_{j=1}^r z_j c_j^\dagger c_{j+1}^\dagger\ket{0}\,,
\end{equation}
with $ U^{\dagger}=e^{i\xi}$ the unitary transformation that block diagonalize $\omega$.

To find the maximal eigenvalue of the 2CRDM for an arbitrary $\ket{\Psi_{2-\textrm{prod}}}$, we substitute Eq.~\eqref{eq:SlaterDecomposition} into the 2CRDM Eq.~\eqref{eq:2crdm}. The block diagonal terms where all indices belong to the same $Z_i$ block leads to factorized eigenvalues $\pm z_i^4$, depending only on each individual $z_i$ coefficient. The remaining eigenvalues, which may contain information about entanglement, satisfy the $r$th degree polynomial equation:
\begin{equation}\label{eq:lambdaEq}
    \lambda^r+a_{1} \lambda^{r-1}+a_{2}\lambda^{r-2}+\cdots+a_{r-1}\lambda+a_{r}=0,
\end{equation}
where the coefficients $a_m$ are given by
\begin{eqnarray}\label{eq:coeffLambdaEq}
    a_m&=&(-1)^m\mkern-10mu\sum_{\mathcal{S}=\{j_1,\ldots,j_m\}\atop\mathcal{S}\subseteq\{1,\ldots,r\}}\mkern-6mu z_{j_1}^2 z_{j_2}^2\ldots z_{j_m}^2\nonumber
    \\&&\times\Bigg[1-\sum_{p=1}^{m}(2p-1)\mkern-6mu\sum_{\{i_1,\ldots,i_p\}\subseteq\mathcal{S}}z_{i_1}^2\ldots z_{i_p}^2\Bigg]\,.
\end{eqnarray}
For example, with Slater rank $r=2$, we have
\begin{equation}
    \begin{split}
        &\lambda^2-[z_1^2(1-z_1^2)+z_2^2(1-z_2^2)]\lambda\\&+z_1^2 z_2^2(1-z_1^2-z_2^2-3z_1^2 z_2^2)=0.
    \end{split}
\end{equation}
Given constraint $z_1^2+z_2^2=1$, the solutions are $\lambda = 3z_1^2z_2^2$ or $-z_1^2z_2^2$. 

For a general rank $r$ state, it is important to note that the exchange of any pairs of $z_i$ and $z_j$ corresponds to a basis transformation. As a result, the set of eigenvalues $\{\lambda\}$ of $O_{ijkl}^\con$ must be invariant under these exchanges. Therefore, the maximum (and minimum) eigenvalue $\lambda_{\rm max}$ ($\lambda_{\min}$) must be a symmetric function of $\{z_j\}$. Because of this reason, the factorized eigenvalues independent from Eq.~\eqref{eq:lambdaEq} cannot be the extrema $\lambda_{\rm max}$ or $\lambda_{\min}$. By imposing $z_j^2 = 1/r$, Eq.~\eqref{eq:coeffLambdaEq} becomes
\begin{eqnarray}\label{eq:symmetrizedCoefficients}
    a_m\mkern-2mu&=&\mkern-2mu{r\choose m}\frac{(-1)^m}{r^m} \mkern-6mu\left[1-\sum_{p=1}^n {m\choose p}\frac{2p-1}{r^p} \right]\nonumber\\
    &=&\mkern-2mu{r\choose m}\frac{(-1)^m}{r^m}\mkern-6mu \left[1\mkern-2mu+\mkern-4mu  \sum_{p=1}^m \mkern-4mu {m\choose p} \frac{1}{r^p}-\frac{2}{r}\mkern-2mu\sum_{p=1}^m \mkern-4mu{m\mkern-2mu-\mkern-2mu1 \choose p\mkern-2mu-\mkern-2mu1} \frac{m}{r^{p-1}}  \right]\nonumber\\
       &=&\mkern-2mu{r\choose m}\frac{(-1)^m}{r^m}\mkern-6mu\left[\left(1+\frac{1}{r}\right)^m-\frac{2m}{r}\left(1+\frac{1}{r}\right)^{m-1}  \right]\nonumber\\
       &=&\mkern-2mu(-1)^m{r\choose m}\frac{(r+1)^m}{r^{2m}}\left(1-\frac{2m}{r+1}\right)\,.
\end{eqnarray}
Substituting Eq.~\eqref{eq:symmetrizedCoefficients} into Eq.~\eqref{eq:lambdaEq}, we then obtain:
\begin{eqnarray}
    \left(\mkern-2mu\lambda_{\rm max/min} +\frac1{r}-\frac1{r^2}\mkern-2mu\right)\mkern-3mu\left(\mkern-2mu\lambda_{\rm max/min} -\frac1{r}-\frac1{r^2}\mkern-2mu\right)^{\mkern-5mur\mkern-1mu-\mkern-1mu1} \mkern-20mu = 0\,.
\end{eqnarray}
The solutions are $\lambda_{\rm max/min} = 1/r^2 \pm 1/r$ for $r\geqslant2$ and $\lambda_{\rm max/min} = 0$ for $r=1$ (the case of a Slater determinant). Therefore,
\begin{equation}
\mu(2) = \sup_{r}\mkern-2mu\left\{\lambda\right\} = \frac{3}{4}\,,
\end{equation}
which corresponds to the $k=2$ case for Eq.~\eqref{eq:kProdUpperBound}. This upper bound is reached for half-filled systems with $r=2$, consistent with the intuition that half-filled electronic states should form more entangled states.

\subsubsection{Upper bound for 3-producible states}
The general form of a 3-producible fermionic state can be expressed as:
\begin{equation}
    \ket{\Psi_{3-\textrm{prod}}}\propto\sum_{i,j,k}\omega_{ijk}c_{i}^{\dagger} c_{j}^{\dagger} c_{k}^{\dagger}\ket{0}\,,
\end{equation}
where $\omega_{ijk}$ is an antisymmetric tensor with dimension $N$, the number of orbitals. This state cannot be transformed into a sum of Slater determinants in the same manner as Eq.~\eqref{eq:SlaterDecomposition}, since the $N^3$ degrees of freedom in the $\omega_{ijk}$ tensor exceed the $N^2$ parameters available in a unitary basis transformation for solving the corresponding set of linear equations\,\cite{eckert2002quantum}. Therefore, instead of proving the maximal eigenvalues among all states, we use the generalized GHZ state and W state {[defined later in Eqs.~\eqref{eq:GHZ3} and ~\eqref{eq:Wstate3}]} to derive the upper bound $\mu(3)$. These two classes of states are known as the maximally multipartite entangled states in tripartite systems\,\cite{sarosi2014entanglement,jimenez2024generation}.

In a GHZ-like three-particle state, the particles occupy nonoverlapping sets of orbitals in the similar way as the Slater decomposition in Eq.~\eqref{eq:SlaterDecomposition}, described by\,\cite{majtey2016multipartite}:
\begin{equation}\label{eq:GHZ3}
\ket{\text{GHZ}_3}=\frac{1}{\sqrt{\sum_{j=1}^r |z_j|^2}} \sum_{j=1}^r z_j c_j^\dagger c_{j+1}^\dagger c_{j+2}^\dagger\ket{0}\,.
\end{equation}
We again ignore the factorized eigenvalues depending on individual $z_i$s. Substituting Eq.~\eqref{eq:SlaterDecomposition} into the 2CRDM Eq.~\eqref{eq:2crdm} leads to the $r$th order eigenvalue equation:
\begin{equation}\label{eq:lambdaEq3}
    \lambda^r+a_{1} \lambda^{r-1}+a_{2}\lambda^{r-2}+\ldots+a_{r-1}\lambda+a_{r}=0\,,
\end{equation}
with coefficients
\begin{eqnarray}\label{eq:coeffLambdaEq3}
    a_m&=&(-1)^m\mkern-10mu\sum_{\mathcal{S}=\{j_1,\ldots,j_m\}\atop\mathcal{S}\subseteq\{1,\ldots,r\}}\mkern-6mu z_{j_1}^2 z_{j_2}^2\ldots z_{j_m}^2\nonumber
    \\&&\times\sum_{p=0}^{m}2^{m-p}(1-3p)\mkern-6mu\sum_{\{i_1,\ldots,i_p\}\subseteq\mathcal{S}}z_{i_1}^2\ldots z_{i_p}^2\,.
\end{eqnarray}
Here, we use the symmetry among all coefficients to find the extrema of $\lambda$. With $z_j^2 = 1/r$, Eq.~\eqref{eq:coeffLambdaEq3} becomes
\begin{eqnarray}\label{eq:symmetrizedCoefficients3}
    a_m\mkern-2mu&=&\mkern-2mu{r\choose m}\frac{(-1)^m}{r^m} \mkern-6mu\left[\sum_{p=0}^n {m\choose p}\frac{1-3p}{r^p}(k-1)^{m-p} \right]\nonumber\\
    &=&\mkern-2mu{r\choose m}\frac{(-1)^m}{r^m}\mkern-6mu\left[\left(2+\frac{1}{r}\right)^m-\frac{3m}{r}\left(2+\frac{1}{r}\right)^{m-1}  \right]\nonumber\\
       &=&\mkern-2mu(-1)^m{r\choose m}\frac{(2r+1)^m}{r^{2m}}\left(1-\frac{3m}{r+1}\right)\,.
\end{eqnarray}
Substituting Eq.~\eqref{eq:symmetrizedCoefficients3} into Eq.~\eqref{eq:lambdaEq3}, we then obtain
\begin{eqnarray}\label{eq:GHZ3eigenEquation}
    \left(\mkern-2mu\lambda_{\rm max/min} +\frac1{r}-\frac1{r^2}\mkern-2mu\right)\mkern-3mu\left(\mkern-2mu\lambda_{\rm max/min} -\frac{2}{r}-\frac1{r^2}\mkern-2mu\right)^{\mkern-5mur\mkern-1mu-\mkern-1mu1} \mkern-20mu = 0\,.
\end{eqnarray}
The solutions are $\lambda_{\rm max} = 2/r + 1/r^2\leqslant5/4$ and $\lambda_{\min} = -1/r + 1/r^2\geqslant-1/4$ for $r\geqslant2$. 

We further consider the W state as another example of a maximally entangled three-particle state, whose expression in indistinguishable fermionic systems is\,\cite{dur2000three,sarosi2014entanglement,jimenez2024generation}:
\begin{equation}\label{eq:Wstate3}   \ket{\text{W}}=\frac1{\sqrt{3}}\left(c_1^{\dagger}c_5^{\dagger}c_6^{\dagger}+ c_4^{\dagger}c_5^{\dagger}c_3^{\dagger}+ c_4^{\dagger}c_2^{\dagger}c_6^{\dagger}\right)\ket{0}\,.
\end{equation}
It is straightforward to show that the maximal eigenvalue $\lambda_{\max}$ for this state is 8/9, which is less than 5/4. To explore a more general state beyond the equal-coefficient W state, we further examine the generalized spin state, defined as\,\cite{sarosi2014entanglement}:
\begin{equation}\label{eq:spin3state}
    \ket{\text{SPIN}_3}=(z_1 c_1^{\dagger}c_2^{\dagger}c_3^{\dagger}+z_2 c_1^{\dagger}c_5^{\dagger}c_6^{\dagger}+z_3 c_4^{\dagger}c_5^{\dagger}c_3^{\dagger}+z_4 c_4^{\dagger}c_2^{\dagger}c_6^{\dagger})\ket{0}\,.
\end{equation}
When the first three and last three indices are conceptually regarded as up and down spins, Eq.~\eqref{eq:spin3state} represents all possible configurations of a singly occupied spin state. The W state is a specific case within the class of $\ket{\text{SPIN}_3}$ states, corresponding to the parameters $z_1=0$ and $z_2=z_3=z_4=1/\sqrt{3}$. The $\ket{\text{GHZ}_3}$ state with $r=2$ is also a special case of $\ket{\text{SPIN}_3}$ by setting $z_1=z_2=z_3=z_4=1/2$ and applying Hadamard basis transformations for each pairs of orbitals: 
\begin{eqnarray}
     &&\frac{1}{2}\left( c_1^{\dagger} c_2^{\dagger} c_3^{\dagger}+c_1^{\dagger} c_5^{\dagger} c_6^{\dagger} + c_4^{\dagger} c_5^{\dagger} c_3^{\dagger}+ c_4^{\dagger} c_2^{\dagger} c_6^{\dagger}\right)\ket{0}\nonumber\\
     &=&\frac{1}{\sqrt{2}}\Bigg(\frac{c_1^{\dagger}+c_4^{\dagger}}{\sqrt{2}}\frac{c_2^{\dagger}+c_5^{\dagger}}{\sqrt{2}}\frac{c_3^{\dagger}+c_6^{\dagger}}{\sqrt{2}}\ket{0}+\frac{c_1^{\dagger}-c_4^{\dagger}}{\sqrt{2}}\frac{c_2^{\dagger}-c_5^{\dagger}}{\sqrt{2}}\nonumber
     \\&&\frac{c_3^{\dagger}-c_6^{\dagger}}{\sqrt{2}}\ket{0}\Bigg)\,.
\end{eqnarray}
A nice property of the generalized spin state is that the class of Eq.~\eqref{eq:spin3state} is closed under basis exchanges.

Substituting Eq.~\eqref{eq:spin3state} into the flattened matrix of the 2CRDM, the eigenproblem factorizes into three groups of eigenequations: 12 second-order equations, two third-order equations, and one sixth-order equation. First, the second-order equations are formulated as:
\begin{equation}
    \begin{split}
        &\lambda^2\pm 2(z_{a}^2 z_{b}^2-z_{c}^2 z_{d}^2)\lambda+z_{a}^4 z_{b}^4\\
        &-z_{c}^2 z_{d}^2+z_{c}^4 z_{d}^4-2z_{a}^2 z_{b}^2 z_{c}^2 z_{d}^2=0\,,
    \end{split}
\end{equation}
where $\{a,b,c,d\}$ are different combinations of $z_j$'s indices $\{1,2,3,4\}$ in Eq.~\eqref{eq:spin3state}. As previously discussed, the extremum values of eigenvalues occur at $z_j^2=1/4$, simplifying the equation to:
\begin{equation}
    \left(\lambda_{\rm max/min}+\frac{1}{4}\right)\left(\lambda_{\rm max/min}-\frac{1}{4}\right)=0\,.
\end{equation}
Thus, $\lambda_{\rm max}=1/4$ and $\lambda_{\min}=-1/4$ for this sector of eigenspace. 

Second, the third-order equations are
\begin{eqnarray}
        &&\lambda^3\pm \mkern-4mu\sum_{j_1\neq j_2}\mkern-2muz_{j_1}^2 z_{j_2}^2 \lambda^2+\frac12\Big[\mkern-2mu\sum_{j_1\neq j_2}\mkern-4mu(z_{j_1}^4 z_{j_2}^4-z_{j_1}^2 z_{j_2}^2)+\mkern-20mu\sum_{j_1\neq j_2\neq j_3}\mkern-15muz_{j_1}^2 z_{j_2}^2 z_{j_3}^2
        \nonumber\\&&-24Z^2\mp 12 Z \Big]\lambda\pm \frac16\mkern-14mu\sum_{j_1\neq j_2\neq j_3}\mkern-15muz_{j_1}^2z_{j_2}^2z_{j_3}^2(z_{j_1}^2z_{j_2}^2+z_{j_1}^2z_{j_3}^2+z_{j_2}^2z_{j_3}^2
        \nonumber\\&& +z_{j_1}^2z_{j_2}^2z_{j_3}^2)\mp2Z^2+2 Z+( 2Z\mp Z^2)\sum_{j_1\neq j_2}z_{j_1}^2 z_{j_2}^2 =0\,.
\end{eqnarray}
Here, $Z=z_1z_2z_3z_4$ and the summations contain all permutations. By setting identical variables to find the extrema, the two equations correspond to
\begin{equation}
    \begin{split}
        &\left(\lambda_{\rm max/min}+\frac{1}{4}\right)^2\left(\lambda_{\rm max/min}-\frac{5}{4}\right)=0\\
          &\textrm{or}\quad\left(\lambda_{\rm max/min}+\frac{1}{4}\right)^3=0\,,
    \end{split}
\end{equation}
giving $\lambda_{\max}=5/4$ and $\lambda_{\min}=-1/4$, the same as the results obtained from the $\ket{\text{GHZ}_3}$ state in Eq.~\eqref{eq:GHZ3eigenEquation}.

Lastly, the remaining eigenvalues are derived from the nondegenerate, sixth-order eigenequation. Because of its complicated form as a function of $z_1$, $z_2$, $z_3$, and $z_4$, and its symmetry with respect to permutations, we directly present the simplified equation with all $z_j^2$'s set identically:
\begin{equation}
\left(\lambda_{\rm max/min}+\frac{1}{4}\right)^3\left(\lambda_{\rm max/min}-\frac{1}{4}\right)^3=0\,.
\end{equation}
Thus, the $\lambda_{\max/\min}=\pm1/4$ for the sixth-dimensional sector of the eigenspace.

In conclusion, for both $\ket{\text{GHZ}_3}$ and $\ket{\text{SPIN}_3}$ states, we find consistent result about the upper bound
\begin{equation}
\mu(3) = \sup_{r}\mkern-2mu\left\{\lambda\right\} =\frac54\,,
\end{equation}
which corresponds to the $k=3$ case for Eq.~\eqref{eq:kProdUpperBound}. This upper bound is also reached by half-filled electronic states the 3-producible class.

\subsubsection{Upper bounds for $k$-producible states}
For a general $k$-particle state, expressed as\,\cite{majtey2016multipartite}
\begin{equation}
    \ket{\Psi_{k-\textrm{prod}}}\propto\sum_{i_1,\ldots,i_k} \omega_{i_1,\ldots,i_n}c_{i_1}^{\dagger} c_{i_2}^{\dagger} \ldots c_{i_k}^{\dagger}\ket{0}\,,
    \label{Eq:kprod}
\end{equation}
we still examine its maximally entangled form as a generalized GHZ state
\begin{equation}
\ket{\text{GHZ}_k}=\frac{1}{\sqrt{\sum_{j=1}^r |z_j|^2}} \sum_{j=1}^r z_j \prod_{s=0}^{k-1} c_{j+s}^\dagger \ket{0}\,.
\label{eq:GHZk}
\end{equation}

The derivation of the extremum eigenvalues follows the same strategy as the $\ket{\text{GHZ}_3}$ states. The only difference is that the coefficients in the $r$th order eigenvalue equation are:
\begin{eqnarray}\label{eq:coeffLambdaEqk}
    a_m&=&(-1)^m\mkern-10mu\sum_{\mathcal{S}=\{j_1,\cdots,j_m\}\atop\mathcal{S}\subseteq\{1,\cdots,r\}}\mkern-6mu z_{j_1}^2 z_{j_2}^2\cdots z_{j_m}^2\nonumber
    \\&&\times\sum_{p=0}^{m}(1-kp)(k-1)^{m-p}\mkern-6mu\sum_{\{i_1,\cdots,i_p\}\subseteq\mathcal{S}}z_{i_1}^2\cdots z_{i_p}^2\,.
\end{eqnarray}
By imposing the exchange symmetry among $z_j$ and focusing on the extrema eigenvalues, we obtain
\begin{eqnarray}\label{eq:symmetrizedCoefficientsk}
    a_m\mkern-2mu&=&\mkern-2mu{r\choose m}\frac{(-1)^m}{r^m} \mkern-6mu\left[\sum_{p=0}^n {m\choose p}\frac{1-kp}{r^p}(k-1)^{m-p} \right]\nonumber\\
    &=&\mkern-2mu(-1)^m{r\choose m}\frac{(kr-r+1)^m}{r^{2m}}\left(1-\frac{km}{r+1}\right)\,,
\end{eqnarray}
which simplifies the eigenvalue equation into
\begin{eqnarray}
    \left(\mkern-2mu\lambda_{\rm max/min} +\frac1{r}-\frac1{r^2}\mkern-2mu\right)\mkern-3mu\left(\mkern-2mu\lambda_{\rm max/min} -\frac{k-1}{r}-\frac1{r^2}\mkern-2mu\right)^{\mkern-5mur\mkern-1mu-\mkern-1mu1} \mkern-20mu = 0\,.
\end{eqnarray}
The solutions are $\lambda_{\rm max} = (k-1)/r + 1/r^2$ and $\lambda_{\min} = -1/r + 1/r^2$ for $r\geqslant2$. Recall that $r=1$ always leads to a separable state with $\lambda_{\rm max/min} = 0$. Therefore,
\begin{equation}
\mu(k) = \sup_{r}\mkern-2mu\left\{\lambda\right\} =\frac12\left(k -\frac12\right)\,.
\end{equation}
{This bound is consistent with the result found in Ref.~\onlinecite{schouten2022large} on the utility of 2CRDM in characterizing exciton condensation.} Until now, we have proven the upper bounds of eigenvalues in a $k$-producible state. Notably, the proof for $k>0$ is based on assuming the maximally entangled state is the GHZ form, without traversing all possible states in the two-particle case.

\subsection{Generalization to mixed states}\label{sec:mixedState}
While all the simulations presented in this paper are conducted at zero temperature, the entanglement witness theory can be extended to mixed states of an ensemble. A mixed state $\rho_k$ is defined as $k$-separable, if it can be expressed as a convex combination of $k$-producible states:
\begin{equation}
    \rho_k=\sum_{\tau}p_{\tau} \ket{\Psi^{(\tau)}_{k-\textrm{prod}}} \bra{\Psi^{(\tau)}_{k-\textrm{prod}}}\,.
\end{equation}
Here, $p_{\tau}>0$ with $\sum_\tau p_{\tau}=1$, and $\ket{\Psi^{(\tau)}_{k-\textrm{prod}}}$ is $k$-producible as defined in Sec.~\ref{sec:correlations:fermions}. A state is $(k+1)$-particle entangled if it is not $k$-separable. The 2CRDM of a mixed-state $\rho = \sum_{\tau}p_{\tau} \ket{\Psi^{(\tau)}} \bra{\Psi^{(\tau)}}$ can be expressed using its individual pure states:
\begin{equation}
\begin{split}
    O_{ijkl}^\con (\rho) =& \sum_\tau p_\tau \Big[\bra{\Psi^{(\tau)}} c_i c_j (c_k c_l)^{\dagger}\ket{\Psi^{(\tau)}}\\
    &- \bra{\Psi^{(\tau)}} c_i c_k^{\dagger}\ket{\Psi^{(\tau)}}   \bra{\Psi^{(\tau)}} c_j c_l^{\dagger}\ket{\Psi^{(\tau)}} \\
    &+   \bra{\Psi^{(\tau)}} c_i c_l^{\dagger}\ket{\Psi^{(\tau)}} \bra{\Psi^{(\tau)}} c_j c_k^{\dagger}\ket{\Psi^{(\tau)}} \Big]\,.
\end{split}
\end{equation}
Because of the additivity of $O_{ijkl}^\con$ and its flattened matrix $O^\con_{(ij)(kl)}$, the maximal eigenvalue $\lambda_{\rm max}$ for the mixed state $\rho$ is bounded by the sum of the maximal eigenvalues obtained by each pure state:
\begin{eqnarray}
\lambda_{\rm max}\mkern-2mu\big(O_{ijkl}^\con\mkern-1mu(\rho)\big) \mkern-2mu&\leqslant&\mkern-2mu \sum_\tau p_\tau \lambda_{\rm max}\mkern-2mu\big(O_{ijkl}^\con\mkern-1mu(\ket{\Psi^{\mkern-1mu(\tau)\mkern-1mu}}\mkern-1mu\bra{\Psi^{\mkern-1mu(\tau)\mkern-1mu})}\big)\nonumber\\
&\leqslant& \mkern-2mu\max_\tau \lambda_{\rm max}\mkern-2mu\big(O_{ijkl}^\con\mkern-1mu(\ket{\Psi^{\mkern-1mu(\tau)\mkern-1mu}}\mkern-1mu\bra{\Psi^{(\tau)\mkern-1mu}})\big)\,.
\end{eqnarray}
Thus, if $\lambda_{\rm max}$ for a mixed state exceeds $\mu(k)$, it cannot be {expressed} as a convex combination of all $k$-producible pure states and must therefore be $(k+1)$-particle entangled. Hence, the entanglement witness using $\lambda_{\rm max}$ applies to mixed states.

\section{Witnessing Entanglement in Representative Systems}\label{sec:examples}
In this section, we demonstrate the effectiveness of the RIXS-derived 2CRDM eigenvalue $\lambda_{\rm max}$ as an entanglement witness by applying it to various representative quantum states and material-relevant Hamiltonians. Specifically, we focus on trial wavefunctions, 1D extended Hubbard models, and quasi-1D Hubbard models with frustrated geometry. We discuss the advantages of the RDM-based fermionic entanglement witness and assess the effectiveness of RIXS measurements in these contexts. Given the nature of the models and computational complexity, all discussions in this section are restricted to zero-temperature pure states. However, as discussed earlier, the generalization to mixed states is straightforward.

\subsection{Randomly sampled many-body states}\label{sec:examples:randomStates}
\begin{figure}[!t]
    \centering
    \includegraphics[width=\linewidth]{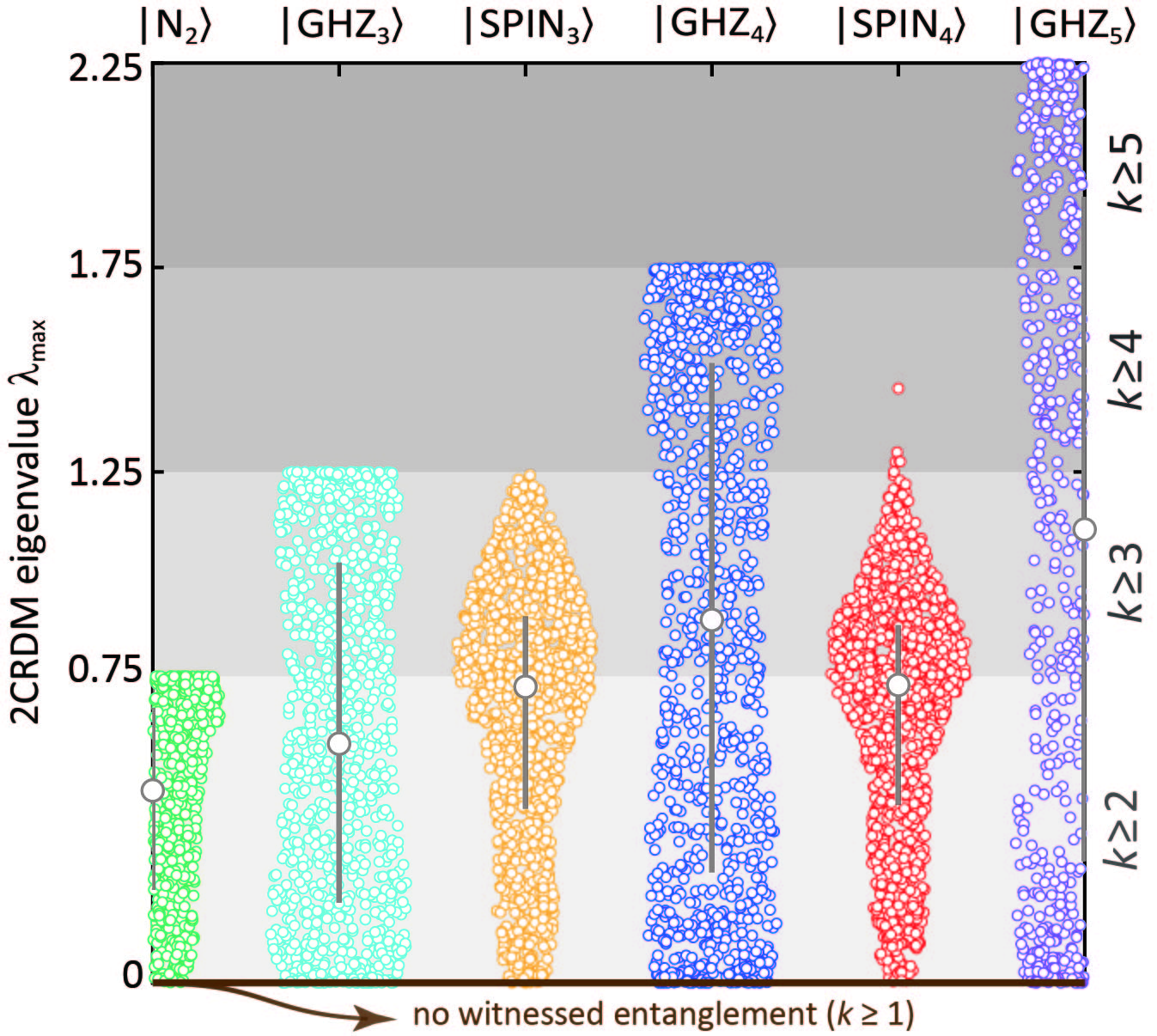}\vspace{-3mm}
    \caption{The distribution of the fermionic entanglement witness $\lambda_{\rm max}$ for random trial states across various classes of wavefunctions, evaluated using the maximal singular value for the 2CRDM. The theoretical boundaries for $k$-producible states are indicated by the darkness of the background, with the $\lambda_{\rm max}=0$ states indicating no witnessed entanglement. The white dots and gray error bars denote the means and variances within each class of random states. }
    \label{fig:randomStates}
\end{figure}

To verify the effectiveness of $\lambda_{\rm max}$ in witnessing the entanglement depth of a fermionic many-body state, we examine several classes of quantum states with known entanglement depths. One such example is the pairing state with Slater rank $r=2$, expressed by\,\cite{schliemann2001quantum}
\begin{equation}
\label{Eq:N2}
\ket{\text{N}_2 (z_1,z_2)}= \left(z_1 c_1^\dagger c_2^\dagger+z_2 c_3^\dagger c_4^\dagger\right)\ket{0}\,.
\end{equation}
Unless the two coefficients are chosen to be special values (\textit{i.e.},~$z_1z_2=0$), this state cannot be expressed as a single Slater determinant in any basis, indicating an at least bipartite entangled state. As shown in the leftmost set in Fig.~\ref{fig:randomStates}, the witness $\lambda_{\rm max}$ for randomly sampled $\ket{\text{N}_2}$ states ranges from 0 to 0.75, the upper bound $\mu(2)$ for a 2-producible state. Thus, the $\lambda_{\rm max}$ effectively captures the range of entanglement depth for $\ket{\text{N}_2}$. In other words, any states with $\lambda_{\rm max}>0.75$ cannot be represented as a $\ket{\text{N}_2}$ state. Note that the maximal $\lambda_{\rm max}$ corresponds to the 2CRDM eigenvalue, distinct from the that in the bare RDM. Here, a separable state leads to $\lambda_{\rm max}=0$ due to Wick's theorem, while it is bounded by 2 in the latter case\,\cite{kraus2009pairing}.

Similarly, we further consider the following $k$-producible fermionic GHZ states with unrestricted coefficients $z_1$ and $z_2$\,\cite{majtey2016multipartite}:
\begin{eqnarray}
\label{Eq:GHZ345}
        \ket{\text{GHZ}_3(z_1,z_2)}&=& \left(z_1 c_1^\dagger c_2^\dagger c_3^\dagger+z_2 c_4^\dagger c_5^\dagger c_6^\dagger\right)\ket{0}\nonumber\\
        \ket{\text{GHZ}_4(z_1,z_2)}&=& \left(z_1 c_1^\dagger c_2^\dagger c_3^\dagger c_4^\dagger+z_2 c_5^\dagger c_6^\dagger c_7^\dagger c_8^\dagger\right)\ket{0} \\
        \ket{\text{GHZ}_5(z_1,z_2)}&=& \left(z_1 c_1^\dagger c_2^\dagger c_3^\dagger c_4^\dagger c_5^\dagger+z_2 c_6^\dagger c_7^\dagger c_8^\dagger c_9^\dagger c_{10}^\dagger\right)\ket{0} \,.\nonumber
\end{eqnarray}
We also examine the generalized spin states with $k$ singly occupied electrons in $k$ spinful orbitals\,\cite{sarosi2014entanglement}:
\begin{equation}
\label{Eq:SPIN3}
    \begin{split}
        \ket{\text{SPIN}_3(\{z_n\}}=& \Big(z_1 c_{1\uparrow}^\dagger c_{2\uparrow}^\dagger c_{3\uparrow}^\dagger+z_2 c_{1\uparrow}^\dagger c_{2\downarrow}^\dagger c_{3\downarrow}^\dagger\\
        &+z_3 c_{1\downarrow}^\dagger c_{2\uparrow}^\dagger c_{3\downarrow}^\dagger+z_4 c_{1\downarrow}^\dagger c_{2\downarrow}^\dagger c_{3\uparrow}^\dagger\Big)\ket{0} 
    \end{split}
\end{equation}
and
\begin{equation}
\label{Eq:SPIN4}
    \begin{split}
        \ket{\text{SPIN}_4(\{z_n\})}\mkern-2mu=& \Big(\mkern-2muz_1 c_{1\uparrow}^\dagger c_{2\uparrow}^\dagger c_{3\uparrow}^\dagger c_{4\uparrow}^\dagger\mkern-2mu+\mkern-2muz_2c_{1\uparrow}^\dagger c_{2\uparrow}^\dagger c_{3\downarrow}^\dagger c_{4\downarrow}^\dagger\\
        &+\mkern-2muz_3c_{1\uparrow}^\dagger c_{2\downarrow}^\dagger c_{3\uparrow}^\dagger c_{4\downarrow}^\dagger\mkern-2mu+\mkern-2muz_4c_{1\uparrow}^\dagger c_{2\downarrow}^\dagger c_{3\downarrow}^\dagger c_{4\uparrow}^\dagger\\
        &+\mkern-2muz_5c_{1\downarrow}^\dagger c_{2\uparrow}^\dagger c_{3\uparrow}^\dagger c_{4\downarrow}^\dagger\mkern-2mu+\mkern-2muz_6c_{1\downarrow}^\dagger c_{2\uparrow}^\dagger c_{3\downarrow}^\dagger c_{4\uparrow}^\dagger\\
        &+\mkern-2muz_7c_{1\downarrow}^\dagger c_{2\downarrow}^\dagger c_{3\uparrow}^\dagger c_{4\uparrow}^\dagger\mkern-2mu+\mkern-2muz_8c_{1\downarrow}^\dagger c_{2\downarrow}^\dagger c_{3\downarrow}^\dagger c_{4\downarrow}^\dagger\Big)\mkern-4mu\ket{0}.
    \end{split}
\end{equation}
For each class of wavefunctions, we randomly sample 1000 sets of coefficients in a $p$-dimensional unit space, where $p$ is the number of coefficients in each class:
\begin{equation}
\begin{split}
(&z_1,\ldots,z_p)=\big(\cos\alpha_1,\sin\alpha_1\cos\alpha_2,\ldots,\\
&\sin\alpha_1\ldots\cos\alpha_{p-2},\sin\alpha_1\ldots\sin\alpha_{p-1}\big)\,.
\end{split}
\end{equation}
These sampled states are automatically normalized.

\begin{figure*}[!t]
    \includegraphics[width=18cm]{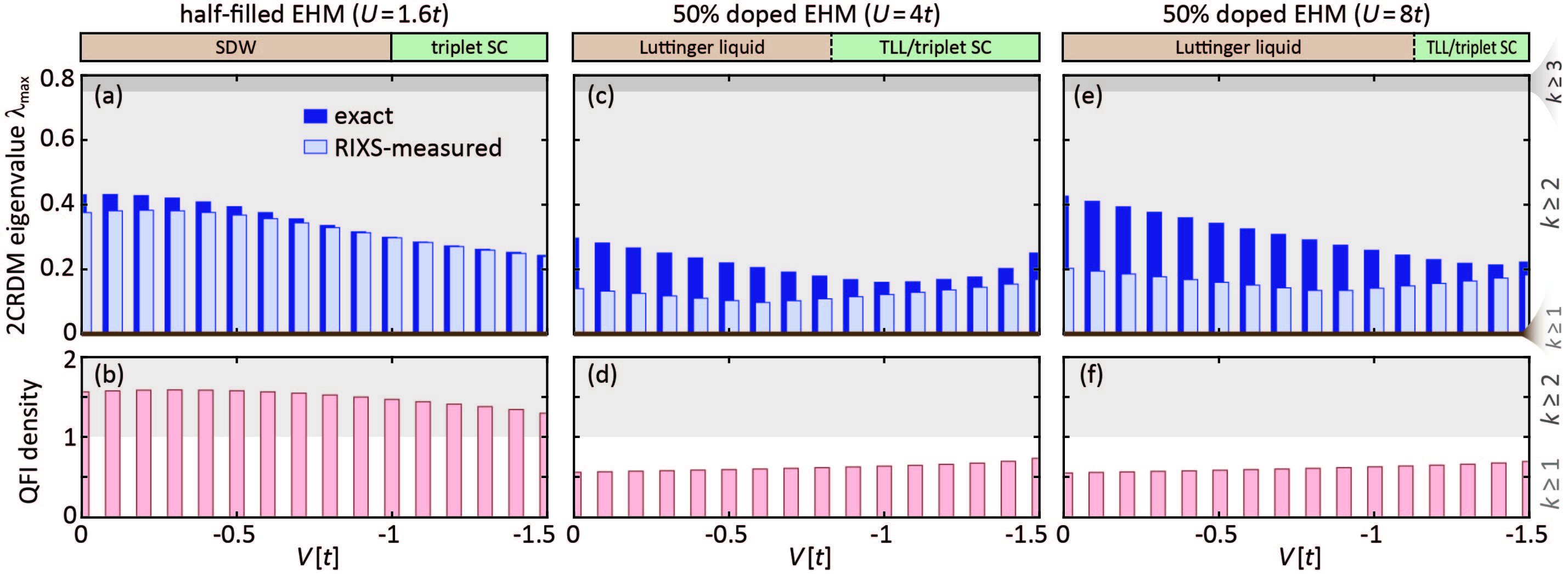}\vspace{-3mm}
    \caption{\label{fig:EHM}
    (a) The maximal eigenvalue of the 2CRDM, evaluated using the exact ground-state wavefunction (dark blue) and the RIXS-measured correlations (light blue), for a half-filled EHM with $U=1.6t$ and varying $V$ {values}. (b) The spin QFI density for the same model as in (a). The upper bar indicates the ground-state phases identified in Ref.~\onlinecite{qu2022spin}, while background darkness indicates the bounds of different $k$-producible states. (c)-(f) The $V$ dependence for the two entanglement witnesses, similar to (a) and (b), but for the 50\% doped EHM with (c), (d) $U=4t$ and (e), (f) $U=8t$. All simulations are performed using zero-temperature DMRG on a 128-site chain, with the correlation measurements restricted to the central $36$ sites. }
\end{figure*}

As shown in Fig.~\ref{fig:randomStates}, the 2CRDM eigenvalues for all samples within each class of state are shown as violin plots. For any $k$-particle entangled states, the simulated $\lambda_{\rm max}$ values fall within the bound of $(k+0.5)/2$,, thereby validating the $\mu(k)$ derived in Eq.~\eqref{eq:kProdUpperBound}. Because of the high parameter-space dimension for $\ket{\text{SPIN}_4}$, our samples do not reach its upper-bound value but clearly exceed the bound for 3-producible states. It is important to recognize that $\lambda_{\rm max}$, as an entanglement witness, indicates only the lower bound of entanglement depth for a given many-body state. For instance, a three-particle entangled $\ket{\text{GHZ}_3}$ state may exhibit $\lambda_{\rm max}<0.75$ for many sampled coefficients, where the witness may be less efficient and conclude only that the state is at least two-particle entangled. Determining the precise entanglement depth necessitates comprehensive information about the entire many-body wavefunction and cannot be inferred from 2CRDM and RIXS spectra, which is beyond the scope of this paper.

\subsection{Extended Hubbard model with mixed-sign interactions}\label{sec:1DEHM}

While the few-body quantum states provide a clear statistical distribution for the entanglement witness $\lambda_{\rm max}$ and validate its bound, electronic wavefunctions in quantum materials, especially at the thermodynamic limit, remain inaccessible by solid-state measurements. Effective electronic Hamiltonians, where band structures and interactions are codetermined by \textit{ab initio} simulations and experimental measurements, provide a widely accessible description of materials. Validating the RIXS-accessible witness also requires an electronic Hamiltonian that defines both the ground state and all excited states. Therefore, we turn to material-relevant tight-binding models and first consider interacting electrons in a 1D chain. 

The simplest description of electronic interaction is the Hubbard model, which simplifies Coulomb repulsions into an on-site $U$. Extending this model, we further include the nearest-neighbor interaction $V$, leading to the extended-Hubbard model (EHM):
\begin{equation}
\mathcal{H} \mkern-2mu=\mkern-2mu -t \mkern-4mu\sum_{\langle ij \rangle, \sigma} \mkern-4mu(c^\dagger_{i\sigma} c_{j\sigma} +  \mathrm{H.c.}) + U \mkern-3mu\sum_i n_{i\uparrow} n_{i\downarrow} + V\mkern-16mu\sum_{\langle ij \rangle,\sigma,\sigma^\prime}\mkern-10mu n_{i\sigma} n_{j\sigma^\prime},
\end{equation} 
where $n_{i\sigma} = c^\dagger_{i\sigma}c_{i\sigma}$ denotes the electron density at site $i$ with spin $\sigma$. Because of the relevance for 1D cuprate chains like Ba$_{2-x}$Sr$_x$CuO$_{3+\delta}$\,\cite{chen2021anomalously}, we focus on the mixed-sign interactions with repulsive $U>0$ and attractive $V<0$. 

We first consider the EHM at half filling, where the spin-density wave (SDW) phase dominates, except for a small region of the triplet superconductivity phase\,\cite{qu2022spin}. To better approximate the thermodynamic limit, we simulate a 128-site 1D EHM using the density matrix renormalization group (DMRG). We select the 36 sites in the center of the chain to measure the 2CRDM, which ensures approximately translational invariance. As shown in Fig.~\ref{fig:EHM}(a), the maximal eigenvalue $\lambda_{\rm max}$ obtained from the the 2CRDM of the EHM is 0.4 for the Hubbard model with $U=1.6t$, indicating an at least two-particle entangled state. Because of the relatively weak interaction, $\lambda_{\rm max}$ lies in the middle of the bounds for separable and 2-producible states. This witnessed entanglement depth aligns with the result obtained from the spin QFI at $q=\pi$. As shown in Fig.~\ref{fig:EHM}(b), the QFI density is around 1.6, also in the middle of these two bounds, witnessing a bipartite spin state. This consistency is expected in the half-filled system with an SDW ground state, where spin excitations dominate. It is worth noting that the half-filled Hubbard model exhibits a quasi-long-range SDW state, causing a logarithmic divergence of the QFI density with the system size at zero temperature\,\cite{voit1995one}. To avoid this singularity, the comparison between $\lambda_{\rm max}$ and QFI here is restricted to a finite system of the same size. This divergence is removed at any finite temperature and, therefore, irrelevant for experimental measurements. 

Including the attractive $V$ suppresses spin correlations by favoring the triplet pairing instability, which is reflected in the reduction of both the QFI and $\lambda_{\rm max}$. A previous DMRG study has demonstrated that the Luttinger parameter $K_\rho$ exceeds 1 at around $V=-t$ for this chosen $U$ value\,\cite{qu2022spin}. As a result, the triplet superconductivity becomes the dominant charge-$2e$ correlation, replacing the SDW. Although the triplet correlation also exhibits a logarithmic divergence with the system size in this phase, its absolute strength of correlations is much weaker than the spin correlations and is negligible in a finite system. The interplay between the SDW and triplet superconductivity state accounts for the decrease in both $\lambda_{\rm max}$ and QFI with the presence of $V$.

According to the derivations in Sec.~\ref{sec:RIXS:correlatoins}, the leading nonlinear effects in RIXS measure only the three- and four-point correlations in the form of Eq.~\eqref{eq:I2DirectIntegral} and do not access all elements in the 2CRDM. To test the effectiveness of witnessing entanglement with incomplete elements, we further simulate the RIXS-measured 2CRDMs and evaluate their maximal eigenvalues $\lambda_{\rm max}$. As shown in Fig.~\ref{fig:EHM}(a), the RIXS-measured $\lambda_{\rm max}$ closely approximates the exact results, demonstrating the dominant role of these short-range correlations in the eigenvalue. Only a slight discrepancy is observed for small $V$, where the EHM exhibits quasi-long-range order. 

Upon doping, the ground state of the EHM evolves into a gapless state, diverging from its SDW configuration. For $U=4t$, the 2CRDM eigenvalue $\lambda_{\rm max}$ decreases, indicating a potentially less entangled ground state [see Fig.~\ref{fig:EHM}(c)]. Despite the reduction, $\lambda_{\rm max}$ still witnesses a (at least) two-particle entangled state. This aligns with the expectation that a Luttinger liquid cannot be equated to a Fermi sea, meaning the connected part of correlations is nonvanishing. In contrast, the QFI density fails to witness any entanglement in the doped scenarios, as two-particle spin excitations cease to dominate within the Luttinger liquid state, rendering spin QFI insensitive to single-particle fermionic excitations. Moreover, as the nearest-neighbor interaction $V$ increases, we observe a dip in $\lambda_{\rm max}$ near $V=-0.9t$, coinciding with the Luttinger parameter $K_\rho$ crossing 1, indicating a shift in the dominant charge-2$e$ excitation transitions toward triplet pairing. This transition is a crossover rather than a broken-symmetry phase transition due to the system maintaining its gapless Luttinger liquid nature\,\cite{qu2022spin}. The $\lambda_{\rm max}$ reflects the strengths of either quasi-long-range correlations on the two sides of the crossover, resulting in a nonmonotonic dependence on $V$. Conversely, the QFI density remains nearly constant across varying $V$ strengths, highlighting the QFI's ineffectiveness in a doped, gapless system.

When comparing the exact $\lambda_{\rm max}$ with the approximated value derived from RIXS-measured RDM elements, we find that the latter significantly underestimates $\lambda_{\rm max}$, especially when contrasted with the half-filled system shown in Fig.~\ref{fig:EHM}(a). This discrepancy can be attributed to the slower spatial decay of single-particle correlations $\langle c_i c_j^\dagger\rangle$ in the Luttinger liquid state, while the RIXS-measured RDM elements truncate in distance. Nevertheless, despite this underestimation, the RIXS-measured $\lambda_{\rm max}$ effectively identifies an entangled state and captures crossover-induced nonmonotonicity, unlike the QFI results. Importantly, since $\lambda_{\rm max}$ indicates the lower bound of entanglement, an underestimating $\lambda_{\rm max}$ through RIXS does not compromise the validity of the bound.

We delve deeper into the impacts of correlations by examining the $U=8t$ system. As shown in Fig.~\ref{fig:EHM}(e), stronger interactions lead to a reduced Luttinger parameter $K_\rho$ and, therefore, more pronounced, longer-range spin correlations. This is evidenced by the enhanced $\lambda_{\rm max}$ values compared to those in the $U=4t$ scenario. However, these values remain well below the upper bound for two-particle entangled states and, thus, do not alter the witnessed entanglement depth. Concurrently, the QFI density remains largely unchanged and continues to fail in witnessing entanglement, highlighting its limitations in doped systems. Additionally, the discrepancies between the RIXS-measured $\lambda_{\rm max}$ and the exact values are more evident in the $U=8t$ system due to the enhanced and extended spin correlations. Despite this, the conclusion that a two-particle entangled state is witnessed remains unchanged.

\subsection{Triangular lattice Hubbard model}\label{sec:examples:triangularHubbard}
\begin{figure}[!t]
    \centering
    \includegraphics[width=\linewidth]{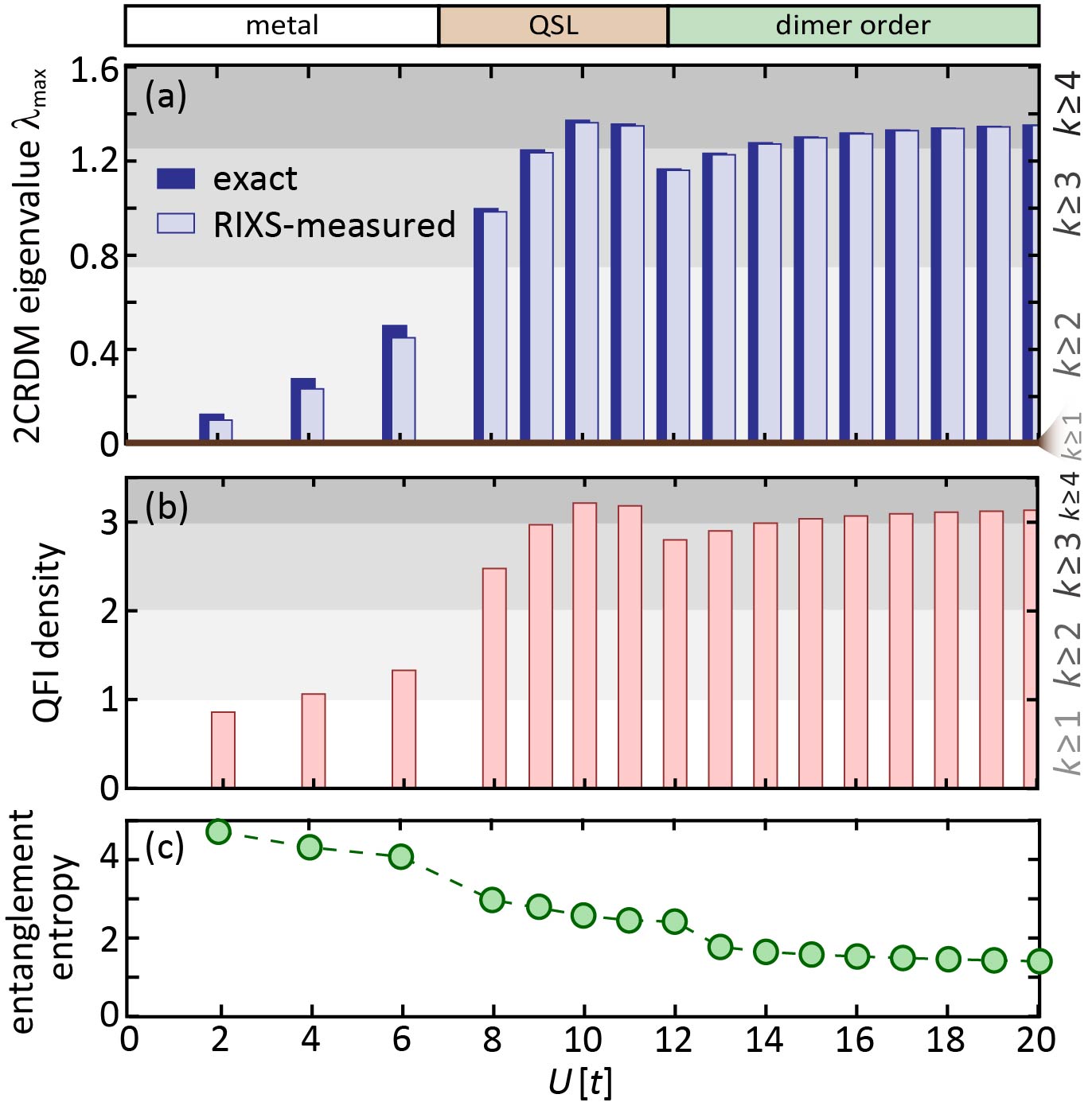}\vspace{-3mm}
    \caption{\label{fig:triangularHubbard}(a) The 2CRDM witness using the maximal eigenvalues $\lambda$, evaluated by the exact ground-state wavefunctions (dark blue) and derived from RIXS spectra (light blue) for the triangular-lattice Hubbard model in a $72\times 3$ {cylinder}. The upper bar indicates the ground-state phases for different $U$s, adapted from Ref.~\onlinecite{peng2021gapless}. (b) The spin QFI density evaluated using the ground states of the same model. The witnessed entanglement depths are indicated by the background darknesses in (a) and (b). (c) The von Neumann entanglement entropy $S(x=L_x/2)$ calculated the center of the system.}
\end{figure}

We expand our exploration to systems beyond 1D, particularly focusing on frustrated geometries where we expect to find highly entangled many-body wavefunctions. Here, we use the triangular lattice as an example, corresponding to the quantum spin liquid (QSL) candidate materials $\kappa$-(ET)$_2$Cu$_2$(CN)$_3$\,\cite{shimizu2003spin}. Figure~\ref{fig:triangularHubbard}(a) shows the results obtained from the Hubbard model on a $72\times 3$ three-leg triangular cylinder. Previous research has delineated its phase diagram, identifying metallic, QSL, and dimer-order phases\,\cite{peng2021gapless}. Our analysis reveals that $\lambda_{\rm max}$ generally increases with the interaction strength $U$ due to stronger electronic correlations. In the strong-coupling limit, the system forms a dimer-order state with period-2 spin patterns, resulting in six-site dimer supercells on the three-leg ladder lattice. Each supercell can be regarded as a partition $\mathcal M_p$ in Eq.~\eqref{eq:gmeblock}, which hosts 6 electrons and the maximally possible entanglement depth is 6, if it is irreducible. Here, the 2CRDM reaches 1.35, exceeding the upper bound for a three-particle producible state (5/4), thereby witnessing at least a four-particle entangled state. This aligns with the size of individual dimer supercells, considering that each partition may not be maximally entangled. 

As $U$ decreases to $12t$, the ground state transitions to a gapless QSL phase\,\cite{peng2021gapless}. In this regime, $\lambda_{\rm max}$ exhibits a noticeable increase, indicating the enhanced entanglement of wavefunctions due to the frustrated geometry. Notably, the entanglement witness only characterizes entanglement depth without differentiating between long-range and short-range entanglement. Consequently, the witnessed depth in the QSL phase is also at least 4, similar to that in the strong-coupling regime of the dimer-order state. Both these two correlated phases are dominated by spin excitations. As a result, the spin QFIs display the same $U$ dependence as $\lambda_{\rm max}$, and quantitatively witness the same three-particle and four-particle entangled states in these two phases [see Fig.~\ref{fig:triangularHubbard}(b)].

As the interaction strength further decreases to $U\sim 7t$, the system transitions to a metallic ground state\,\cite{peng2021gapless}. Despite the absence of a single-particle gap, this metallic state remains correlated and distinguishable from a simple Fermi sea. The 2CRDM eigenvalue $\lambda_{\rm max}$ captures this correlation, consistently witnessing an at least two-particle entangled state throughout the phase diagram for $U>0$. In stark contrast, the QFI density, evaluated from the ground state of the triangular-lattice Hubbard model, falls below 1 (the upper bound for separable states) for $U<4t$. Thus, the spin QFI fails to recognize the correlated nature of this metallic state due to the diminished spin excitations within this regime. 

We further analyze the entanglement entropy, which quantifies the entanglement of a many-body state across a given partition. The scaling behavior of entanglement entropy with system dimensions has been widely used for identifying gapless modes and topological order\,\cite{levin2006detecting, kitaev2006topological, calabrese2004entanglement}. Here, we calculate the von Neumann entanglement entropy at the center of the quasi-1D system:
\begin{equation}
    S(x=L_x/2)=-\mathrm{Tr}(\rho_{x}\log \rho_{x})\,.
\end{equation}
Here, $\rho_{x}$ is the reduced density matrix for a subsystem of length $x$, with $x$ set to $L_x/2=36$. To avoid confusion,  $\rho_{x}$ is different from, though related to, the two-particle RDM $\braket{c_i c_j (c_k c_l)^{\dagger}}$ used in other contexts of this paper\,\cite{rissler2006measuring}. As shown in Fig.~\ref{fig:triangularHubbard}(c), the entanglement entropy is low in the strong-coupling limit (dimer order) but increases rapidly as the system becomes gapless. Because entanglement entropy is specific to a particular partition in a chosen basis (typically real-space orbitals), it is less effective in depicting the basis-invariant entanglement depth. The rise in $S(x=L_x/2)$ for small $U$ indicates that electrons become less localized, making the real-space basis less effective for describing the many-body wavefunction. 

Recognizing the advantage of the 2CRDM entanglement witness over QFI and entropy, we further assess the accuracy of RIXS-derived eigenvalues. In a manner similar to the 1D EHM example in Sec.~\ref{sec:1DEHM}, RIXS precisely captures the $\lambda_{\rm max}$ for systems with relatively strong interactions, i.e., in the QSL and dimer-order phases. In these phases, electrons are localized into spins, making the three- and four-particle correlations derived from the nonlinearity of RIXS in Eq.~\eqref{eq:I2DirectIntegral} sufficiently accurate to approximate most RDM elements. The RIXS-measured $\lambda_{\rm max}$ starts to deviate from the exact value upon entering the metallic phase, where electrons become less localized. Nevertheless, unlike the doped systems discussed in Sec.~\ref{sec:1DEHM}, the finite interaction and half-filled configuration in the metallic phase restrict electron delocalization, resulting in a RIXS-measured error of just 1\%. Across all simulated model parameters, the RIXS-measured $\lambda_{\rm max}$ consistently witnesses the same entanglement depth as the exact $\lambda_{\rm max}$, further emphasizing the reliability of this fermionic entanglement witness.

\section{Summary and Outlook}\label{sec:discussion}

This work advances the field of spectral characterization of quantum entanglement by developing a robust theoretical framework that extends beyond the QFI {traditionally used for} distinguishable modes. {For indistinguishable fermions, a reliable entanglement witness must be resilient to fermionic antisymmetry, invariant under basis transformations, and exhibits monotonic scaling with entanglement depth. We propose a practical and experimentally viable entanglement witness through x-ray scattering techniques.} As detailed in Sec.~\ref{sec:RIXS}, high-precision RIXS spectra enable the measurement of four-fermion correlations $\braket{c_i c_j (c_k c_l)^{\dagger}}$, leveraging information from two-momentum distributions. By subtracting the two-fermion correlations $\braket{c_i c_j^{\dagger}}$, accessible via ARPES, we obtain the dominant elements of the 2CRDM. Derivations in Sec.~\ref{sec:entanglement} reveal that the maximal eigenvalue ($\lambda_{\max}$) of the 2CRDM {fulfills the stringent criteria for a fermionic entanglement witness}. Utilizing this witness tool, we investigate representative systems in Sec.~\ref{sec:examples}, demonstrating its capability to characterize entanglement in extended and triangular Hubbard models. Notably, we observe that the spin QFI fails to characterize entanglement in phases where spin fluctuations are not dominant, while our electronic entanglement witness remains effective across all studied phases. With the feasibility of implementation using current x-ray scattering techniques, this approach provides a versatile tool for detecting entangled states in quantum materials and advancing material design for quantum technologies.

In practice, extracting four-fermion correlations requires performing a the Fourier transformation on the incident-photon momentum {and energy. As detailed in Sec.~\ref{sec:RIXS:correlatoins}, this process requires independent control of both incident energy and the momentum relevant to electronic excitations. This is feasible under the assumption that electronic states are effectively confined to separated layers or chains, a condition often met in strongly correlated materials with significant quantum fluctuations. For these materials,} the collection of high-quality two-momentum information {is enabled} by rotating both the sample and the spectrometer. This approach is commonly used to characterize collective excitations such as plasmons and charge density waves\,\cite{lee2014asymmetry,gerber2015three,hepting2018three,bejas2024plasmon}. {For systems with 3D electronic dispersions, employing a single Lorentzian approximation for the $\win$ dependence of RIXS spectra can approximate the four-fermion correlations needed for entanglement witness. Moreover, utilizing the continuous polarization dependence of the incident and scattered photons offers an alternative strategy, potentially reducing reliance on the $\kin$ dependence and broadening applicability to molecular systems without translational symmetry. Although these strategies are beyond our scope, they highlight promising avenues for future exploration in specific experimental systems.} 

{In addition,} separating various correlations requires different weighted integrals through the detuning of incident energy, as specified in Eq.~\eqref{eq:I2WeightedIntegral}. Accurate evaluation of these integrals requires information about the magnitude of the core hole lifetime ($\Gamma$) and the core level hopping ($t_c$). In typical RIXS experiments, $\Gamma$ can be estimated through the spectral broadening along the $\win$ axis or the corresponding x-ray absorption spectrum, while $t_c$ can be estimated by various weighted integrals over $\win$. In the worst case, when the resonance detuning is unavailable, the energy integral gives a superposition of a pair of four-fermion correlations, following Eq.~\eqref{eq:I2DirectIntegral}. In this case, a compromised approach to estimate the individual elements in RDM is using the Cauchy–Schwarz inequality
\begin{eqnarray}\label{eq:CauchySchwarz}
    &&\frac12\Big(\mkern-6mu\braket{{c_{n\sigma\mkern-2mu_1^\prime}}\mkern-2mu{c^{\dagger}_{n\sigma\mkern-2mu_1} c_{m\sigma\mkern-2mu_2}}\mkern-2mu{c_{m-\delta+\delta^\prime,\sigma\mkern-2mu_2^\prime}^{\dagger}}}+\braket{{c_{n+\delta,\sigma\mkern-2mu_1^\prime}}\mkern-2mu{c^{\dagger}_{n\sigma\mkern-2mu_1} \mkern-2muc_{m\sigma\mkern-2mu_2}}\mkern-2mu{c_{m+\delta^\prime,\sigma\mkern-2mu_2^\prime}^{\dagger}}}\mkern-6mu\Big)^2\nonumber\\
    &&\leqslant \mkern-3mu\Big|\mkern-4mu\braket{{c_{n\sigma\mkern-2mu_1^\prime}}\mkern-2mu{c^{\dagger}_{n\sigma\mkern-2mu_1}\mkern-3mu c_{m\sigma\mkern-2mu_2}}\mkern-2mu{c_{m\mkern-2mu-\mkern-2mu\delta\mkern-2mu+\mkern-2mu\delta^\prime\mkern-4mu,\sigma\mkern-2mu_2^\prime}^{\dagger}}\mkern-2mu}\mkern-4mu\Big|^2\mkern-7mu    +    \mkern-4mu\Big|\mkern-4mu\braket{{c_{n\mkern-2mu+\mkern-2mu\delta\mkern-2mu,\sigma\mkern-2mu_1^\prime}}\mkern-2mu{c^{\dagger}_{n\sigma\mkern-2mu_1}\mkern-3mu c_{m\sigma\mkern-2mu_2}}\mkern-2mu{c_{m\mkern-2mu+\mkern-2mu\delta^\prime\mkern-4mu,\sigma\mkern-2mu_2^\prime}^{\dagger}}\mkern-2mu}\mkern-4mu\Big|^2   .
\end{eqnarray}
Thus, it gives the lower bounds for each element.

The extraction of four-fermion correlations from RIXS spectra depends largely on the ratio between the core-hole lifetime and the core-hole hopping timescales. In this work, we assumed a relatively small ratio, which is typical for transition-metal oxides, enabling the spectrum to be expanded to the leading order of $O({t_c^2}/{\Gamma^{{2}}})$. {For representative correlated materials—such as cuprates, nickelates, manganates, and ruthenates—this ratio is estimated to be around ${t_c^2}/{\Gamma^{{2}}}\sim 10\%$.} With future improvements in RIXS resolution and measurement precision {(to below 1\%)}, higher-order terms such as $O({t_c^4}/{\Gamma^4})$ experimentally accessible, allowing additional elements of the 2CRDM to be measured beyond those defined in  Eqs.~\eqref{eq:connected3p} and \eqref{eq:connected4p}. Additional spectroscopic techniques, including Raman scattering and pair photoemission spectra, offer complementary means of probing these elements. However, their reliance on single-momentum dependencies restricts their utility for disentangling specific four-point correlations. Instead, these techniques are best suited for constraining the bounds of otherwise inaccessible RDM elements, much like the method outlined in Eq.\eqref{eq:CauchySchwarz}.

The capability to measure the two-particle RDM and its cumulants has broader implications than just witnessing entanglement. According to Rosina's theorem, these RDMs for a nondegenerate ground state can reconstruct the many-electron wavefunction for systems with only two-particle interactions\,\cite{rosina1968reduced, mazziotti1998contracted, mazziotti2000complete}. Although the specific reconstruction method is complex and requires matrix elements beyond the reach of RIXS, it suggests that other significant observations, including energy, pairing correlation, and polarizability, can also be reconstructed using these RDMs. In terms of entanglement depth, other observables apart from maximal eigenvalues can also be used to witness entanglement, provided they are invariant under single-particle basis transformations and increase with the entanglement depth. For example, the Frobenius norm of $O_{ijkl}^\con$, or equivalently the second-order trace $\text{Tr}\big[(O^{\con})^2\big]$, scales quadratically with $k$ and is separable for disconnected partitions according to Eq.~\eqref{eq:RDMblockDiagonal}. The upper bound of these observables for a $k$-producible state can be derived similarly to Sec.~\ref{sec:entanglement}, allowing them to witness fermionic entanglement. However, as they intrinsically mix $\lambda_{\max}$ with other eigenvalues, the bounds derived from these observables are less tight than those from $\lambda_{\max}$. Moreover, in systems with specific symmetries, like spin SU(2) symmetry, the allowed form of wavefunctions can be restricted and a tighter bound achievable. We leave the exploration on specific materials for future studies.

The authors thank Paola Cappellaro, Mark Dean, Francesco Evangelista, Rosario Lo Franco, Robert Konik, Mingda Li, David Mazziotti, Matteo Mitrano, Cheng Peng, Xiao-Gang Wen, and Yuanzhe Xi for insightful discussions. This work is supported by the U.S. Department of Energy, Office of Science, Basic Energy Sciences, under Early Career Award No.~DE-SC0024524. The simulation used resources of the National Energy Research Scientific Computing Center, a U.S. Department of Energy Office of Science User Facility located at Lawrence Berkeley National Laboratory, operated under Contract No.~DE-AC02-05CH11231 using NERSC award BES-ERCAP0031226.

\appendix

\section{DETAILS OF RIXS SIMULATIONS}\label{app:RIXSDetails}

The examples of RIXS calculations discussed in Sec.~\ref{sec:RIXS} employ the single-band Hubbard model, with the valence-electron Hamiltonian defined as
\begin{equation}\label{eq:Hubbard}
{\mathcal{H}} = -t \sum_{i\sigma} \left(c_{i\sigma}^\dagger c_{i+1, \sigma} + \mathrm{H.c.} \right)  + \, U \sum_{i} n_{i\uparrow}n_{i\downarrow}
\,.
\end{equation}
The nearest-neighbor hopping amplitude $t$ governs the band structure, while the on-site Coulomb repulsion $U$ controls the electronic correlations within the model. 

To incorporate the x-ray processes, the full Hamiltonian $\mathcal{H}^\prime$ includes additional terms that account for core holes, as described in Eq.~\eqref{eq:HubbardInter} and restated here:
\begin{eqnarray}\label{eq:intermediateHamAppendix}
	\mathcal{H}^\prime &=& \mathcal{H} + \sum_{m} \left(\sum_{\alpha\sigma}E_{\rm edge}p_{m\alpha\sigma}p_{m \alpha\sigma}^\dagger +  \mathcal{H}^{\rm (SOC)}_m\right)\nonumber\\
 && - U_c \sum_{m,\alpha}\sum_{\sigma,\sigma^\prime} c_{m \sigma}^\dagger c_{m\sigma} p_{m\alpha\sigma^\prime}p_{m \alpha\sigma^\prime}^\dagger+ \mathcal{T}_c.
\end{eqnarray}
The example presented in Sec.~\ref{sec:RIXS:mobileCore} focuses on the transition-metal $L$-edge RIXS, where core levels correspond to the $2p_{x,y,z}$ orbitals of transition metal atoms. The core-level spin-orbit coupling is given by:
\begin{eqnarray}\label{eq:coreSOCHam}
	\mathcal{H}^{\rm (SOC)}_m =\lambda\sum_{\alpha\alpha^\prime}\sum_{ \sigma\sigma^\prime}p_{m\alpha\sigma}^\dagger \chi_{\alpha\alpha^\prime}^{\sigma\sigma^\prime}p_{m\alpha^\prime\sigma^\prime}.
\end{eqnarray}
In the simulations presented in the main text, the core-hole potential $U_c$ is consistently set to 4$t$\,\cite{tsutsui2000resonant,jia2016using, wang2021xray}. The edge energy $E_{\rm edge}$ is chosen as 938\,eV, corresponding to the Cu $L$-edge x-ray absorption, and the spin-orbit coupling $\lambda$ of the core states is set to 13\,eV \cite{tsutsui2000resonant}.

\section{VARIOUS INTEGRALS OF RIXS SPECTRA}\label{app:mobileHole}
This section details the spectral integrals derived in Sec.~\ref{sec:RIXS:mobileCore} and their corresponding momentum dependence. We decompose the intermediate-state Hamiltonian $\mathcal{H}^\prime$ in Eq.~\eqref{eq:intermediateHamAppendix} into $\mathcal{H}^\prime_0+\mathcal{T}_c$ and treat the core-level kinetic Hamiltonian $\mathcal{T}_c$ as a perturbation. Many of the integrals derived in this section take the form
\begin{equation}
\int_{-\mkern-1mu\infty}^{\infty}\mkern-6mu  \frac{(x^2)^\ell}{(x^2\mkern-1mu+\mkern-1mu\Gamma^2)^\ell} \frac{dx}{(x\mkern-2mu-\mkern-2muE_1\mkern-2mu-\mkern-2mui \Gamma)^{a\mkern-1mu+\mkern-1mu1}(x\mkern-2mu-\mkern-2muE_2\mkern-2mu+\mkern-2mui \Gamma)^{b\mkern-1mu+\mkern-1mu1}},
\end{equation}
where $a+1$ and $b+1$ indicate the orders of the poles in the complex plane, with $\Gamma>0$. We denote this standard integral as $\Xi^{(\ell)}_{ab}(E_1, E_2; \Gamma)$. The residue theorem is employed to derive the closed-form expression of these integrals. For $\ell=0$, the integral reduces to
 \begin{equation}
     \Xi^{(0)}_{ab}(E_1, E_2; \Gamma)     =\frac{ (a+b)!}{a!\,b!}\frac{2\pi i \,(-1)^a }{(E_1-E_2+2\Gamma i)^{a+b+1}}\,,
\end{equation}
while for nonzero $\ell$ the equation becomes complicated.

The zeroth-order spectrum in Eq.~\eqref{eq:RIXSExpansion0} excludes $\mathcal{T}_c$, with $\mathcal{H}^\prime$ in the propagator being replaced by $\mathcal{H}^\prime_0$. To evaluate its integral, we expand the intermediate state in terms of the eigenstates of $\mathcal{H}^\prime_0$, denoted as $\{\ket{\Psi}\}$. The zeroth-order integral can thus be expressed as:
\begin{eqnarray}\label{eq:zerothOrderContribution_1}
&&\iint I^{(0)}\mkern-2mu\left(\dq,\win,\dw\right) d\win d \dw\nonumber\\
    &=& \frac{1}{N^2 }\sum_{m,n}\sum_{\Psi_1, \Psi_2}e^{i\dq\cdot(\rbf_{m}-\rbf_{n})}\bra{G}\mathcal{D}_{n\epsin}^\dagger\ket{\Psi_1}\bra{\Psi_2}\mkern-2mu\mathcal{D}_{m\epsin}\mkern-2mu\ket{G}\mkern-6mu\nonumber\\
        && \bra{\Psi_1}\mathcal{D}_{n\epsout}\mathcal{D}_{m\epsout}^\dagger  \ket{\Psi_2}\Xi^{(0)}_{00}(E_{\Psi_1}^{(0)}, E_{\Psi_2}^{(0)}; \Gamma) \nonumber\\
    &=& \frac{2\pi i}{N^2 }\sum_{m,n}\sum_{\Psi_1, \Psi_2}\mkern-6mue^{i\dq\cdot(\rbf_{m}-\rbf_{n})}\textrm{Re}\bra{G}\mathcal{D}_{n\epsin}^\dagger\ket{\Psi_1}\bra{\Psi_2}\mkern-2mu\mathcal{D}_{m\epsin}\mkern-2mu\ket{G}\mkern-6mu\nonumber\\
        && \bra{\Psi_1}\mathcal{D}_{n\epsout}\mathcal{D}_{m\epsout}^\dagger  \ket{\Psi_2}\mkern-2mu \left[E_{\Psi_1}^{(0)} - E_{\Psi_2}^{(0)} +2\Gamma i\right]^{-1}\nonumber\\
   &\approx& \frac{\pi }{N^2 \Gamma}\mkern-2mu\sum_{m,n}e^{i\dq\cdot\left(\rbf_{m}-\rbf_{n}\right)}\mkern-6mu\sum_{ \sigma_1, \sigma_1^\prime} \mkern-4mu\sum_{\sigma_2,\sigma_2^\prime}\mkern-4mu\matElement\nonumber\\
    && \bra{G}{c_{n\sigma_1^\prime}}{c^{\dagger}_{n\sigma_1} c_{m\sigma_2}}{c_{m\sigma_2^\prime}^{\dagger}}\ket{G}\,.
\end{eqnarray}
The last step yields a result identical to the zeroth-order integral Eq.~\eqref{eq:integralImmobile}, which assumes immobile core holes.

Next, we proceed with the first-order spectrum in Eq.~\eqref{eq:RIXSExpansion1}, stemming from the cross term between the $\ket{\Psi_{\rm int}^{(0)}}$ and $\ket{\Psi_{\rm int}^{(1)}}$. By expanding in the eigenstates of $\mathcal{H}^\prime_0$, the integral $\iint I^{(1)}(\kin,\dq,\win,\dw) d\win d\dw$ transforms into:
\begin{widetext}
\begin{eqnarray}\label{eq:appendixIntegralFirstOrder}
     && \frac{1}{N^2}\sum_{m,n}\sum_{ m^\prime, n^\prime}\sum_{\Psi_1,\Psi_2}e^{i\kin\cdot (\rbf_{m^\prime}-\rbf_{n^\prime})-i\kout\cdot (\rbf_{m}-\rbf_n)}\bra{\Psi_1}\mathcal{D}_{n\epsout}\mathcal{D}_{m\epsout}^\dagger  \ket{\Psi_2}\nonumber\\
    &&\times \Big[ \bra{G}\mathcal{D}_{n'\epsin}^\dagger \mathcal{T}_c\ket{\Psi_1}\bra{\Psi_2}\mathcal{D}_{m^\prime\epsin}\ket{G}\Xi^{(0)}_{01}(E_{\Psi_1}^{(0)}, E_{\Psi_2}^{(0)}; \Gamma) +
    \bra{G}\mathcal{D}_{n'\epsin}^\dagger \ket{\Psi_1} \bra{\Psi_2}\mathcal{T}_c\mathcal{D}_{m^\prime\epsin}\ket{G}\Xi^{(0)}_{10}(E_{\Psi_1}^{(0)}, E_{\Psi_2}^{(0)}; \Gamma)  \Big]\nonumber\\
   &=& \frac{2\pi i}{N^2}\mkern-6mu\sum_{m,n\atop m^\prime, n^\prime}\mkern-6mu\sum_{\Psi_1,\mkern-2mu\Psi_2}\mkern-6mue^{i\kin\cdot (\rbf_{m\mkern-1mu^\prime}\mkern-1mu-\mkern-1mu\rbf_{n\mkern-1mu^\prime}\mkern-1mu)-i\kout\cdot (\rbf_{m}-\rbf_n\mkern-1mu)}\mkern-2mu\bra{\mkern-1mu\Psi\mkern-1mu_1\mkern-1mu}\mkern-2mu\mathcal{D}_{n\epsout}\mkern-2mu\mathcal{D}_{m\epsout}^\dagger \mkern-2mu \ket{\mkern-1mu\Psi\mkern-1mu_2\mkern-1mu} \mkern-2mu \frac{\bra{G}\mathcal{D}_{n\mkern-1mu^\prime\epsin}^\dagger \mkern-1mu\mathcal{T}_c\mkern-2mu\ket{\mkern-1mu\Psi\mkern-1mu_1\mkern-1mu}\bra{\mkern-1mu\Psi\mkern-1mu_2\mkern-1mu}\mathcal{D}_{m\mkern-1mu^\prime\epsin}\mkern-2mu\ket{G}-\bra{G}\mathcal{D}_{n\mkern-1mu^\prime\epsin}^\dagger \ket{\mkern-1mu\Psi\mkern-1mu_1\mkern-1mu}\bra{\mkern-1mu\Psi\mkern-1mu_2\mkern-1mu}\mkern-2mu\mathcal{T}_c\mathcal{D}_{m\mkern-1mu^\prime\epsin}\mkern-2mu\ket{G}}{(E_{\Psi_1}^{(0)}-E_{\Psi_2}^{(0)}+2\Gamma i)^2}\nonumber\\
    &\approx& \frac{\pi t_c}{2 N^2\Gamma^3}\mkern-6mu\sum_{ \sigma_1, \sigma_1^\prime}\sum_{ \sigma_2,\sigma_2^\prime}\mkern-4mu\matElement\mathrm{Re}\mkern-2mu\Big[\mkern-2mu\sum_{m,n,\delta}\mkern-4mu e^{i\dq\cdot\left(\rbf_{m}-\rbf_{n}\right)+i\kin\cdot\rbf_{\delta}} \mkern-2mu\Big(\mkern-2mu\braket{c_{n-\delta,\sigma_1^\prime}c^{\dagger}_{n\sigma_1}c_{m\sigma_2}\mathcal{H}_0^\prime c_{m\sigma_2^\prime}^{\dagger}}-\braket{c_{n\sigma_1^\prime}\mathcal{H}_0^\prime c^{\dagger}_{n\sigma_1} c_{m\sigma_2}c_{m+\delta\sigma_2^\prime}^{\dagger}}\Big)\mkern-2mu\Big],
\end{eqnarray}
\end{widetext}
It is crucial to observe that the leading-order imaginary part of the last step cancels out due to the symmetry when the two terms are exchanged. Therefore, Eq.~\eqref{eq:appendixIntegralFirstOrder} reproduces the Eq.~\eqref{eq:I1DirectIntegral} of the main text. While the correlations do not directly contribute to the RDM, this first-order spectral integral includes a phase factor $e^{i\kin\cdot\rbf_{\delta}}$, allowing it to be distinguished from the zeroth-order and second-order integrals. This distinction means that Eq.~\eqref{eq:appendixIntegralFirstOrder} does not require actual computation in practice.

Finally, the second-order spectral integral contains three terms corresponding to the integral of the first-order intermediate state $\ket{\Psi_{\rm int}^{(1)}}$ and the cross terms between the zeroth- and second-order intermediate states. Similar to the above derivations, the integral $\iint I^{(2)}(\kin,\dq,\win,\dw) d\win d\dw$ corresponds to 
\begin{widetext}
\begin{eqnarray}\label{eq:appendix:integralSecondOrder}
     && \frac{1}{N^2}\mkern-6mu\sum_{m,n\atop m^\prime, n^\prime}\mkern-6mu\sum_{\Psi_1,\Psi_2}\mkern-6mue^{i\kin\cdot (\rbf_{m^\prime}-\rbf_{n^\prime})-i\kout\cdot (\rbf_{m}-\rbf_n)}\bra{\Psi_1}\mathcal{D}_{n\epsout}\mathcal{D}_{m\epsout}^\dagger  \ket{\Psi_2} \Bigg[ \bra{G}\mathcal{D}_{n'\epsin}^\dagger \mathcal{T}_c\ket{\Psi_1}\bra{\Psi_2}\mathcal{T}_c\mathcal{D}_{m^\prime\epsin}\ket{G}\Xi^{(0)}_{11}(E_{\Psi_1}^{(0)}, E_{\Psi_2}^{(0)}; \Gamma)\nonumber\\
    && + \bra{G}\mathcal{D}_{n'\epsin}^\dagger \mathcal{T}_c^2\ket{\Psi_1}\bra{\Psi_2}\mathcal{D}_{m^\prime\epsin}\ket{G}\Xi^{(0)}_{02}(E_{\Psi_1}^{(0)}, E_{\Psi_2}^{(0)}; \Gamma) +
    \bra{G}\mathcal{D}_{n'\epsin}^\dagger \ket{\Psi_1} \bra{\Psi_2}\mathcal{T}_c^2\mathcal{D}_{m^\prime\epsin}\ket{G}\Xi^{(0)}_{20}(E_{\Psi_1}^{(0)}, E_{\Psi_2}^{(0)}; \Gamma)  \Bigg]\nonumber\\
   &=& \frac{2\pi i}{N^2}\sum_{m,n\atop m^\prime, n^\prime}\sum_{\Psi_1,\Psi_2}e^{i\kin\cdot (\rbf_{m^\prime}-\rbf_{n^\prime})-i\kout\cdot (\rbf_{m}-\rbf_n)}\bra{\Psi_1}\mathcal{D}_{n\epsout}\mathcal{D}_{m\epsout}^\dagger  \ket{\Psi_2}\nonumber\\
    &&\times  \frac{\bra{G}\mathcal{D}_{n'\epsin}^\dagger \mathcal{T}_c^2\ket{\Psi_1}\bra{\Psi_2}\mathcal{D}_{m^\prime\epsin}\ket{G}+\bra{G}\mathcal{D}_{n'\epsin}^\dagger \ket{\Psi_1}\bra{\Psi_2}\mathcal{T}_c^2\mathcal{D}_{m^\prime\epsin}\ket{G}-2\bra{G}\mathcal{D}_{n'\epsin}^\dagger \mathcal{T}_c\ket{\Psi_1}\bra{\Psi_2}\mathcal{T}_c\mathcal{D}_{m^\prime\epsin}\ket{G}}{(E_{\Psi_1}^{(0)}-E_{\Psi_2}^{(0)}+2\Gamma i)^3}\nonumber\\
    &\approx& \frac{\pi t_c^2}{2 N^2\Gamma^3}\mkern-6mu\sum_{ \sigma_1, \sigma_1^\prime\atop\sigma_2,\sigma_2^\prime}\mkern-4mu\matElement\mathrm{Re}\mkern-2mu\Bigg[\sum_{m,n}\sum_{\delta, \delta^\prime}\mkern-2mu
    e^{i\dq\cdot\left(\rbf\mkern-2mu_{m}-\rbf\mkern-2mu_{n}\right)+i\kin\cdot (\rbf\mkern-2mu_{\delta^\prime}-\rbf\mkern-2mu_{\delta})}\Big(\mkern-2mu \mkern-4mu\braket{{c_{n\mkern-2mu+\mkern-2mu\delta,\sigma_1^\prime}}\mkern-2mu{c^{\dagger}_{n\sigma_1} \mkern-2muc_{m\sigma_2}}\mkern-2mu{c_{m+\delta^\prime\mkern-2mu,\sigma_2^\prime}^{\dagger}}}\mkern-4mu
    -\braket{{c_{n\sigma_1^\prime}}\mkern-2mu{c^{\dagger}_{n\sigma_1}\mkern-2mu c_{m\sigma_2}}\mkern-2mu{c_{m\mkern-2mu-\mkern-2mu\delta\mkern-2mu+\mkern-2mu\delta^\prime\mkern-2mu,\sigma_2^\prime}^{\dagger}}}\mkern-6mu\Big)\Bigg],
\end{eqnarray}
\end{widetext}
reproducing the expression of Eq.~\eqref{eq:I2DirectIntegral} in the main text.

The phase factor $e^{i\kin\cdot(\rbf_{\delta}-\rbf_{\delta^\prime})}$ is sufficient to distinguish the second-order contribution from the zeroth- and first-order counterparts, yet it cannot separate the two types of four-fermion correlations in Eq.~\eqref{eq:appendix:integralSecondOrder}. To address this, it becomes necessary to consider their energy distributions prior to integration. It is important to note that 
 \begin{equation}
     \mathrm{Re}\,\Xi^{(1)}_{11}(E_1, E_2; \Gamma)     \approx \frac{\pi }{8\Gamma^3} \,,
\end{equation}
whereas,
\begin{equation}
 \begin{split}
     &\mathrm{Re}\,\Xi^{(1)}_{02}(E_1, E_2; \Gamma) \approx \frac{\pi }{64\Gamma^5}(4E_1^2\mkern-2mu -\mkern-2mu3E_1E_2\mkern-2mu-\mkern-2mu12E_2^2) \,,\\
     &\mathrm{Re}\,\Xi^{(1)}_{20}(E_1, E_2; \Gamma) \approx \frac{\pi }{64\Gamma^5}(4E_2^2\mkern-2mu-\mkern-2mu3E_1E_2\mkern-2mu-\mkern-2mu12E_1^2) \,.
 \end{split}
\end{equation}
Thus, by introducing an energy-weighted integral for the second-order term $I^{(2)}$, specifically with the weight ${\win^2}/{(\win^2+\Gamma^2)}$, the leading-order contribution becomes:
\begin{eqnarray}
&&\iint \frac{\win^2}{\win^2+\Gamma^2}I^{(2)}(\kin,\dq,\win,\dw) d\win d\dw\nonumber\\
&\approx&\frac{1}{N^2}\mkern-6mu\sum_{m,n\atop m^\prime, n^\prime}\mkern-6mu\sum_{\Psi_1,\Psi_2}\mkern-6mue^{i\kin\cdot (\rbf_{m^\prime}-\rbf_{n^\prime})-i\kout\cdot (\rbf_{m}-\rbf_n)}\bra{G}\mathcal{D}_{n'\epsin}^\dagger \mathcal{T}_c\ket{\Psi_1} \nonumber\\
    && \bra{\Psi_1}\mathcal{D}_{n\epsout}\mathcal{D}_{m\epsout}^\dagger  \ket{\Psi_2}\bra{\Psi_2}\mathcal{T}_c\mathcal{D}_{m^\prime\epsin}\ket{G}\Xi^{(1)}_{11}(E_{\Psi_1}^{(0)}, E_{\Psi_2}^{(0)}; \Gamma)\nonumber\\
&\approx& \frac{\pi t_c^2}{8 N^2\Gamma^3}\mkern-6mu\sum_{ \sigma_1, \sigma_1^\prime}\mkern-2mu\sum_{ \sigma_2,\sigma_2^\prime}\mkern-4mu\matElement\mathrm{Re}\mkern-2mu\sum_{m,n}\sum_{\delta, \delta^\prime}\mkern-2mu
    e^{i\dq\cdot\left(\rbf\mkern-2mu_{m}-\rbf\mkern-2mu_{n}\right)+i\kin\cdot (\rbf\mkern-2mu_{\delta^\prime}-\rbf\mkern-2mu_{\delta})} \nonumber\\
    &&\braket{{c_{n\mkern-2mu+\mkern-2mu\delta,\sigma_1^\prime}}\mkern-2mu{c^{\dagger}_{n\sigma_1} \mkern-2muc_{m\sigma_2}}\mkern-2mu{c_{m+\delta^\prime\mkern-2mu,\sigma_2^\prime}^{\dagger}}}\,.
\end{eqnarray}
It includes only the four-point correlation term, thereby reproducing Eq.~\eqref{eq:I2WeightedIntegral} from the main text.

It is also possible to consider alternative weighted integrals, such as $\Xi^{(2)}_{11}$ and $\Xi^{(2)}_{20/02}$. These integrals produce $O(t_c^2/\Gamma^{3})$ terms, but with coefficients distinct from those in Eq.~\eqref{eq:appendix:integralSecondOrder}. By combining these integrals, one can separate the contributions of three-point and four-point correlations. Practically, using multiple weighted integrals can help mitigate the uncertainties associated with the estimation of $\Gamma$ and $t_c$. In addition, while we focus the integrals for $I^{(0)}$, $I^{(1)}$, and $I^{(2)}$, we have confirmed that all higher-order integrals (\textit{i.e.}, $I^{(m)}$ for $m>2$) produce correlations with prefactors $O(\Gamma^{-5})$ or smaller. Therefore, they can be neglected in practice, as the focus is on the leading nonlinear term $O(t_c^2/\Gamma^{3})$ in RIXS.

\section{AN ALTERNATIVE FORM OF THE FLATTENED MATRIX FROM THE 2CRDM}\label{app:flattenedMatrix}

\begin{figure}[!t]
    \centering
   \includegraphics[width=8.5cm]{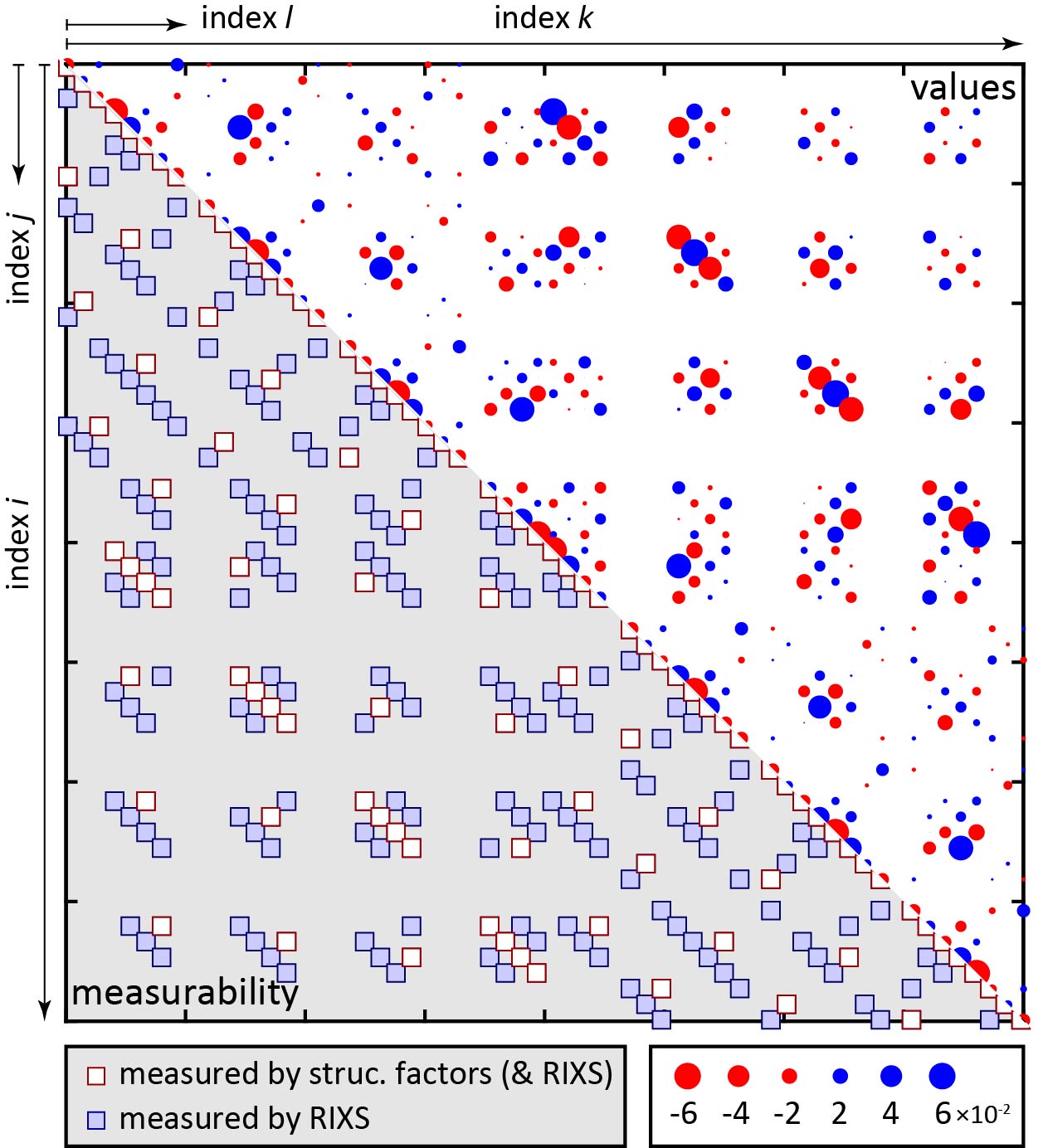}\vspace{-2mm}
    \caption{\label{fig:flattenedMatrixOther} Upper triangle: values and distribution of elements in the flattened matrix $O^{\con}_{(ij)(kl)}$, with point sizes corresponding to the scale of the values. Lower triangle: matrix elements that can be measured by spin or charge structure factors (red) and by RIXS (red and blue). The matrix and its measurability are symmetric when transposing the rows ($i$ and $j$ indices) and columns ($k$ and $l$ indices). This example is drawn from same system as Fig.~\ref{fig:flattenedMatrix}.}    
\end{figure}

In the main text, we choose $O^{\con}_{(ik)(jl)}$ as the form of matrix flattened from the 2CRDM tensor $O^{\con}_{ijkl}$, due to its favorable properties regarding the maximal eigenvalues derived in Sec.~\ref{sec:entanglement:entangleBounds}. Given the symmetry of the tensor, there is an alternative, and nonequivalent convention to flatten 2CRDM into a matrix: grouping the $i$ and $j$ (indices for annihilation operators) as the row index and using the $k$ and $l$ (indices for creation operators) as the column index. As illustrated in Fig.~\ref{fig:flattenedMatrixOther}, this alternative flattened matrix, denoted as $O^{\con}_{(ij)(kl)}$, exhibits different distributions of nonzero elements compared to the $O^{\con}_{(ik)(jl)}$. This difference is mainly reflected by the significant diagonal values in $O^{\con}_{(ij)(kl)}$, corresponding to the situation for $i=k$ and $j=l$. Here, the element $\braket{c_i c_j (c_k c_l)^{\dagger}}$ simplifies to $-\braket{c_i^\dagger c_i  c_j^{\dagger} c_j  }$, which can be measured by spin and charge structure factors (with spin flavors are absorbed in these indices). Despite this structural difference, the accessibility of RIXS spectra to nonzero matrix elements remains unchanged.

\begin{figure}[!t]
    \centering
    \includegraphics[width=\linewidth]{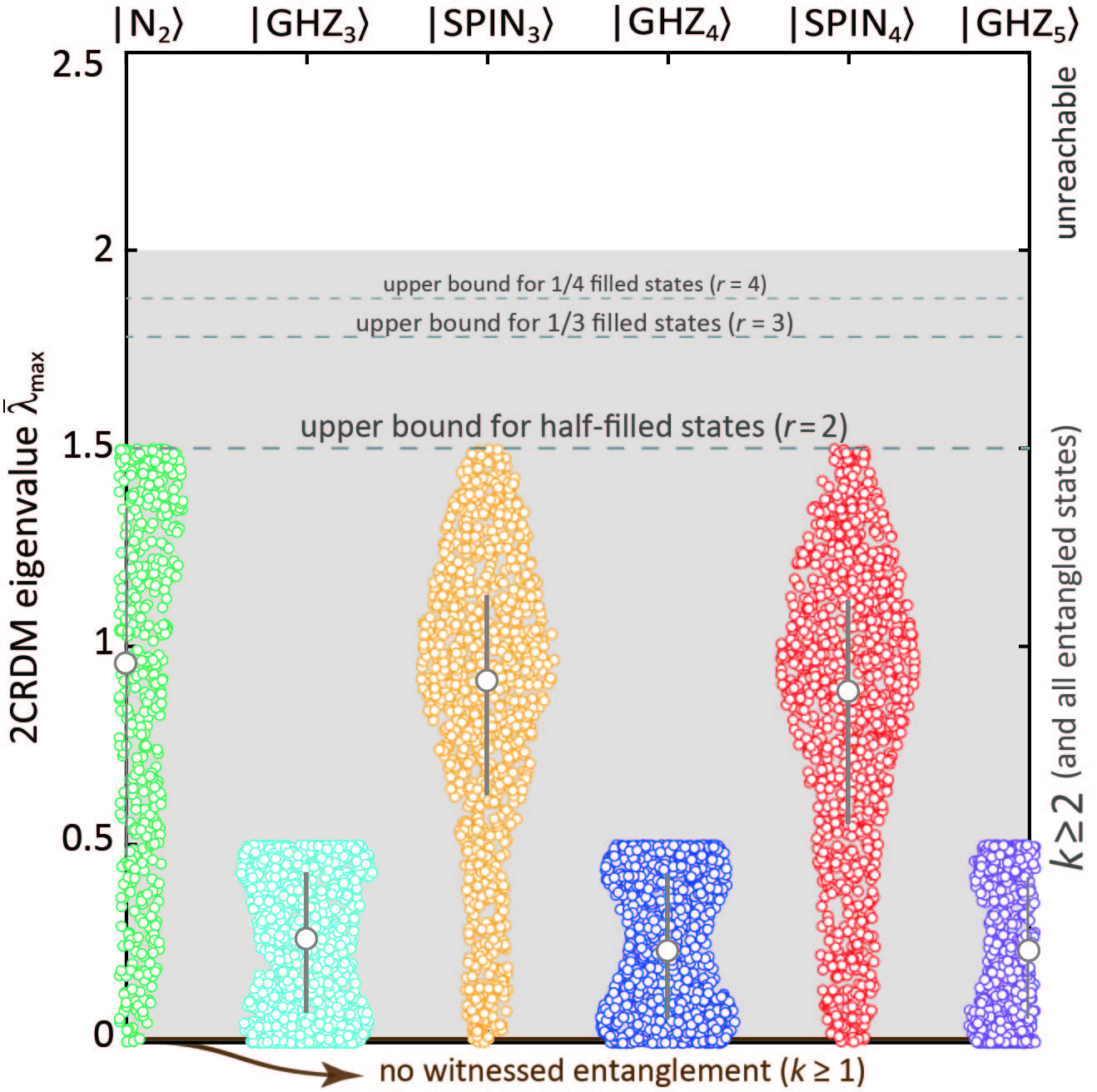}\vspace{-3mm}
    \caption{\label{fig:randomStatesOtherFlatten} The distribution of the maximal eigenvalue $\bar\lambda_{\rm max}$ obtained from $O^{\con}_{(ij)(kl)}$ for random trial states across various classes of wavefunctions. The shaded area indicates the range of eigenvalues for 2-producible states, while the dashed lines denote the upper bounds reached by different fillings.}
\end{figure}

To show why the eigenvalues obtained from this flattened matrix cannot be used as an entanglement witness, we follow the same strategy as Sec.~\ref{sec:entanglement:entangleBounds} and derive their upper bounds. To differ from the $\lambda_{\rm max}$ obtained from the $O^{\con}_{(ik)(jl)}$ matrix in the main text, we denote the eigenvalues obtained from $O^{\con}_{(ij)(kl)}$ as $\bar{\lambda}$.

For a general 2-producible states, representable as the Slater decomposition in Eq.~\eqref{eq:SlaterDecomposition}, the matrix $O^{\con}_{(ij)(kl)}$ has several small block-diagonal subspaces, leading to trivial eigenvalues of $0$ and $-2z_i^2 z_j^2$. The major block that may carry entanglement leads to an $r$th order eigen-equation:
\begin{equation}\label{eq:otherFlattenEigenEq}
    \bar\lambda^r+a_1\bar\lambda_{r-1}+\cdots+a_r=0\,,
\end{equation}
with coefficients
\begin{equation}\label{eq:otherFlattenCoeff}
    \begin{split}
    &a_m=2^m\mkern-10mu\sum_{\mathcal{S}=\{j_1,\ldots,j_m\}\atop\mathcal{S}\subseteq\{1,\ldots,r\}}\mkern-6mu z_{j_1}^2 z_{j_2}^2\ldots z_{j_m}^2
    \\&\times\Bigg[z_{j_1}^2 z_{j_2}^2\ldots z_{j_m}^2-\mkern-6mu\sum_{\{i_1,\ldots,i_{m-1}\}\subseteq\mathcal{S}}z_{i_1}^2\ldots z_{i_{m-1}}^2\Bigg]\,.
    \end{split}
\end{equation}
Similar to the discussion in the main text, the extremum eigenvalues should appear when all $z_j$ are the identical. Thus, we set $z_j^2=1/r$ and the coefficients become
\begin{equation}\label{eq:otherFlattenCoeff2}
\begin{split}
    a_m&=\left(\begin{matrix}r\\m\end{matrix}\right)\frac{2^m}{r^m}\left(\frac{1}{r^m}-\frac{m}{r^{m-1}}\right)\,.
\end{split}
\end{equation}
Substituting Eq.~\eqref{eq:otherFlattenCoeff2} into Eq.~\eqref{eq:otherFlattenEigenEq}, we obtain the equation for extremum eigenvalues:
\begin{equation}
    \left(\bar\lambda_{\rm max/min}-2+\frac{2}{r^2}\right)\left(\bar\lambda_{\rm max/min}+\frac{2}{r^2}\right)^{r-1}=0
\end{equation}
Therefore, the maximal eigenvalue $\bar\lambda_{\max}=2-{2}/{r^2}$ for a 2-producible state. Unlike the property of $\lambda_{\max}$, here the $\bar\lambda_{\max}$ reaches its upper bound (\textit{i.e.}, 2) at $r=\infty$.

To verify this conclusion, we randomly sample 1000 2-producible states $\ket{\text{N}_2}$ and present the obtained maximal eigenvalues $\bar\lambda_{\max}$ in Fig.~\ref{fig:randomStatesOtherFlatten}. With $r=2$ in $\ket{\text{N}_2}$, all simulated $\bar\lambda_{\max}$s fall under 1.5, consistent with the above conclusion for 2-producible states.

For $k>2$, if we still employ the $\ket{\text{GHZ}_k}$ states, \textit{i.e.}, Eq.~\eqref{eq:GHZk}, as references, the eigenvalues exhibit trivial solutions including $0$, $-2z_i^2 z_j^2$ and $2(z_i^2-z_i^4)$ for all $k>2$ and $r>1$. This leads to  $\bar\lambda_{\rm max/min}=\pm1/2$. 

It is likely that the GHZ states are not maximally entangled in the context of the alternative flattened $O^{\con}_{(ij)(kl)}$. Therefore, we further examine the $\ket{\text{SPIN}_3}$ and $\ket{\text{SPIN}_4}$ states. As shown in Fig.~\ref{fig:randomStatesOtherFlatten}, their maximal eigenvalues are all bounded by 1.5 for half-filled systems ($r=2$), equal to that of 2-producible states. Since the state $\ket{\text{N}_2}$ can be regarded as a special case in the class of $\ket{\text{SPIN}_3}$ and $\ket{\text{SPIN}_4}$ states, it is not surprising that the upper bound of $\ket{\text{N}_2}$ can be reached. Notably, these upper bounds exceed those obtained from the  $\ket{\text{GHZ}_k}$ states (\textit{i.e.}, 1/2). 

Therefore, although the eigenvalues of the matrix ${O}^{\con}_{(ij)(kl)}$ are also upper bounded, generally by the total electron number $N_e$\,\cite{raeber2015large}, they are not ideally suited for witnessing entanglement because these bounds do not exhibit a monotonic increase with the entanglement depth $k$, as observed in the GHZ state. While it is possible that a tight and monotonic upper bound could exist for general $k$-producible states in forms significantly different from the GHZ state, deriving a general analytical form similar to Sec.~\ref{sec:entanglement:entangleBounds} for such cases remains a challenge. For example, Ref.~\onlinecite{raeber2015large} has revealed that the large eigenvalues of this alternative flattened 2CRDM matrix are associated with the off-diagonal long-range order (ODLRO), signaling strong electron pair correlations in superconductivity. This connection suggests a potential way to witness entanglement depth using extreme eigenvalues of this alternative form.

\section{ENTANGLEMENT METRICS BASED ON TENSOR SINGULAR VALUES}\label{app:tensorDecomposition}

\begin{figure}[!t]
    \centering
    \includegraphics[width=\linewidth]{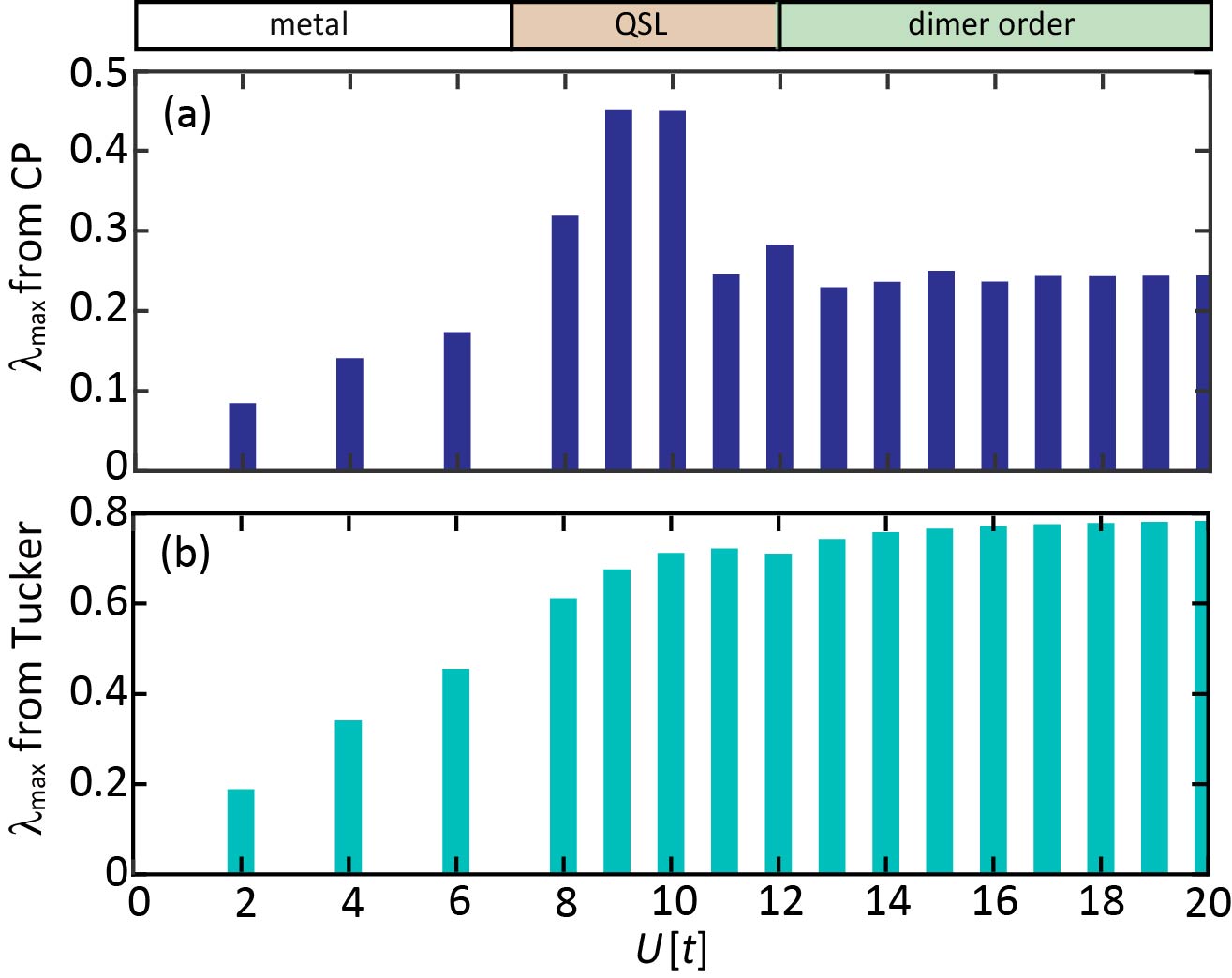}\vspace{-3mm}
    \caption{\label{fig:tensorHOSVDTriangular}
    The maximal tensor singular values $\lambda_{\max}$, calculated using (a) the CP decomposition and (b) the Tucker decomposition for the {Hubbard model on a $72\times 3$ three-leg triangular cylinder}. The model parameters and cluster geometry correspond to those shown in Fig.~\ref{fig:triangularHubbard} in the main text. The upper bar denotes the ground-state phases for different $U$s, adapted from Ref.~\onlinecite{peng2021gapless}.
}    
\end{figure}
As discussed in Sec.\ref{sec:entanglement}, an observable used for witnessing entanglement should be invariant under the unitary basis transformations. We employed the maximal eigenvalue $\lambda_{\rm max}$ of the flattened matrix ${O}^{\con}_{(ik)(jl)}$ as this basis-invariant metric. This metric is particularly advantageous due to its compact analytical upper bound and its linear scaling with entanglement depth, as demonstrated in Sec.~\ref{sec:entanglement:entangleBounds}. More generally, the elementary unitary invariants of the 2CRDM, represented as of the fourth-order tensor $O_{ijkl}^\con$, are its high-order tensor singular values. The $\lambda_{\rm max}$ of the flattened matrix is a function of these singular values, and the extremal singular values can provide a more accurate reflection of the maximally entangled components in a many-body wavefunction. In this section, we numerically investigate tensor singular values of $O_{ijkl}^\con$, using the Tucker decomposition and the canonical polyadic (CP) decomposition.

The CP decomposition is a complete tensor factorization technique that approximates a tensor as a sum of rank-one tensors (vectors):
\begin{equation}
    O_{ijkl}^\con \approx \sum_{r=1}^R \lambda_r U_{ir}  U^\prime_{jr}  V_{kr}  V^\prime_{lr},
\end{equation}
where $R$ denotes the rank of the decomposition, $\lambda_r$ are the high-order singular values, and the columns of the $U$, $U^\prime$, $V$, and $V^\prime$ matrices correspond to the factorized vectors\,\cite{kolda2009tensor}. Without an analytical formula for this decomposition, we employ the alternating least squares (ALS) method to numerically compute the $\lambda_r$, iteratively updating the factorized matrices and weights by solving a series of least-squares problems. The rank $R$ theoretically reaches $N^3$ for the exact CP decomposition, but is usually truncated to reflect the low-rank nature of the highly symmetric tensor and to manage computational complexity\,\cite{haastad1990tensor}. Here, we first unfold $O_{ijkl}^\con$ along the first index and performs SVD to determine the truncated $R$ based on the desired variance preservation. With this fixed $R$, the CP-ALS method iteratively converges to the $R$ sets of vectors and high-order singular values, with the maximal singular value serving as the metric for assessing entanglement depth. 

To evaluate the distribution of CP singular values across different systems and validate the use of the flattened matrix $\lambda_{\max}$ described in the main text, we examine the triangular-lattice Hubbard model, whose  entanglement depth have been analyzed in Sec.~\ref{sec:examples:triangularHubbard}. As shown in Fig.~\ref{fig:tensorHOSVDTriangular}(a), the maximal CP singular value starts at zero when $U=0$ and increases to 0.17 as $U$ reaches $6t$, reflecting the correlations within the metallic state. A further increase in $U$ causes a sharp rise in $\lambda_{\max}$, peaking at 0.45, which corresponds to the highly entangled state in the QSL phase. After this peak, the $\lambda_{\max}$ derived from the CP decomposition drops to 0.23--0.24 for $U>12t$, indicating the transition to the dimer-order phase. This trend in maximal tensor singular values qualitatively matches the behavior of the maximal eigenvalues $\lambda_{\max}$ of the flattened matrix shown in Fig.~\ref{fig:triangularHubbard}(a). While these tensor singular values cannot be translated into an entanglement depth without the analytical upper bounds, the consistent trend between these values and the flattened matrix eigenvalues validates the latter as an effective metric for witnessing entanglement.

We further investigate the Tucker decomposition, a technique known for its lower computational complexity and enhanced numerical stability compared to the CP decomposition. Unlike the CP decomposition, which yields individual singular values, Tucker decomposition factors the tensor into a core tensor with lower rank:
\begin{equation}
    O_{ijkl}^\con = \sum_{p=1}^P \sum_{q=1}^Q \sum_{r=1}^R \sum_{s=1}^S \mathbb{g}_{pqrs} U_{ip}  U^\prime_{jq}  V_{kr}  V^\prime_{ls}\,,
\end{equation}
where the core tensor $\mathbb{g}_{pqrs} \in \mathbb{R}^{P \times Q \times R \times S}$. The ALS method is employed similarly to the CP decomposition. After computing the core tensor, it is flattened into a matrix as outlined in Sec.~\ref{sec:entanglement}, and its maximal eigenvalue $\lambda{\max}$ is extracted. Using again the triangular Hubbard model as an example, we simulate the $\lambda_{\max}$ through the Tucker decomposition across various interactions [see Fig.~\ref{fig:tensorHOSVDTriangular}(b)]. Although $\lambda_{\max}$ generally increases with $U$ in general, the distinctive peak observed in the QSL phase is no longer present. This reduction in the effectiveness of singular values from the Tucker decomposition arises from the incompleteness of the low-rank factorization, which blends multiple singular values, leading to a smoothing of extreme values.

The singular values obtained from both tensor decompositions are not employed for witnessing electronic entanglement depth, primarily because their upper bounds cannot be derived analytically in the manner established in Sec.~\ref{sec:entanglement}. Furthermore, tensor decompositions rely on the symmetries among all four indices, making them particularly sensitive to the full spectrum of tensor elements. In practice, RIXS spectra can access only a subset of elements at specific locations. While these omitted elements are typically small in magnitude, their absence disrupts the symmetry of the tensor. This issue is compounded by the inherent numerical instability of the ALS method, leading to less robust solutions for tensor singular values compared to the more reliable eigenvalues derived from a flattened matrix. A comprehensive exploration of these tensor properties and decomposition methods will require extensive future work.

\bibliography{refs_paper}
\end{document}